\documentclass[preprint,3p,times]{elsarticle}

\usepackage{amssymb}

\usepackage{amsmath,bm}
\usepackage{graphicx}
\usepackage{caption}
\usepackage{subcaption}
\usepackage{xcolor}
\usepackage{amssymb}
\usepackage{booktabs}
\usepackage{multirow}
\usepackage{marvosym}
\usepackage{enumitem}

\newcommand{\PFS}[3]{\ensuremath{{#1}_\mathrm{#2}^\mathrm{#3}}}
\definecolor{brandeisblue}{rgb}{0.0, 0.44, 1.0}
\definecolor{red}{RGB}{0,0,0}

\journal{Journal Name}

\begin{document}

\begin{frontmatter}



\title{Imposing minimum and maximum member size, minimum cavity size, and minimum separation distance between solid members in topology optimization}

 \author[label1]{Eduardo Fern\'{a}ndez}
 \author[label2]{Kai ke Yang}
 \author[label3]{Stijn Koppen}
 \author[label1]{Pablo Alarc\'{o}n}
 \author[label1]{Simon Bauduin}	 
 \author[label1]{Pierre Duysinx}
 
 \address[label1]{Department of Aerospace and Mechanical Engineering, University of Liege, All\'{e}e de la D\'{e}couverte 13A, B52, 4000, Liege, Belgium.}
 \address[label2]{State IJR Center of Aerospace Design and Additive Manufacturing, Northwestern Polytechnical University, Xian 710072, Shaanxi, China.} 
 \address[label3]{Department of Precision and Microsystems Engineering, Delft University of Technology, Mekelweg 2, 2628 CD Delft, The Netherlands.}

\begin{abstract}
{\color{red}
This paper focuses on density-based topology optimization and proposes a combined method to simultaneously impose Minimum length scale in the Solid phase (MinSolid), Minimum length scale in the Void phase (MinVoid) and Maximum length scale in the Solid phase (MaxSolid). MinSolid and MinVoid mean that the size of solid parts and cavities must be greater than the size of a prescribed circle or sphere. This is ensured through the robust design approach based on eroded, intermediate and dilated designs. MaxSolid seeks to restrict the formation of solid parts larger than a prescribed size, which is imposed through local volume restrictions. In the first part of this article, we show that by proportionally restricting the maximum size of the eroded, intermediate and dilated designs, it is possible to obtain optimized designs satisfying, simultaneously, MinSolid, MinVoid and MaxSolid. However, in spite of obtaining designs with crisp boundaries, some results can be difficult to manufacture due to the presence of multiple rounded cavities, which are introduced by the maximum size restriction with the sole purpose of avoiding thick solid members in the structure. To address this issue, in the second part of this article we propose a new geometric constraint that seeks to control the minimum separation distance between two solid members, also called the Minimum Gap (MinGap). Differently from MinVoid, MinGap introduces large void areas that do not necessarily have to be round. 2D and 3D test cases show that simultaneous control of MinSolid, MinVoid, MaxSolid and MinGap can be useful to improve the manufacturability of maximum size constrained designs.
}
\end{abstract}

\begin{keyword}
Length Scale \sep Minimum Gap \sep Manufacturing Constraints \sep Filtering \sep SIMP 
\end{keyword}

\end{frontmatter}


\section{Introduction}
\label{Intro}

Topology optimization is gaining ground as a powerful design tool for developing highly efficient and lightweight components. The achievement owes to the progress of incorporating manufacturing constraints into the topology optimization formulation, which reduces the need of labour-intensive post-processing \citep{Liu2018}. In this context, length scale control has been an active topic of research as it encompasses a wide variety of manufacturing constraints \citep{Lazarov2016}. In addition, controlling the members' sizes allows designers to indirectly include desired properties in the optimization problem which would otherwise be difficult or computationally inefficient to incorporate. For instance, \citet{Amir2018} show that by controlling Minimum size of Solid phase ({\color{red}MinSolid}) and Minimum size of Void phase ({\color{red}MinVoid}), it is possible to obtain optimized designs satisfying strength requirements, thus highly non-linear constraints related to the local control of the stresses can be avoided in the optimization problem. Similarly, in fabrication processes such as casting or additive manufacturing, large thermal gradients can be alleviated implicitly by controlling the Minimum size of Solid phase ({\color{red}MinSolid}) and the Maximum size of the Solid phase ({\color{red}MaxSolid}) \citep{Thompson2016}. Maximum member size control can also increase structural redundancy and implicitly improve resistance to buckling \citep{Clausen2016} or localized damage \citep{Jansen2014}. 

However, in some cases, simultaneous control of {\color{red}MinSolid} and {\color{red}MaxSolid} is not sufficient to improve manufacturability of optimized designs. {\color{red} For instance, it has been shown that maximum size constraints tend to place elongated and very thin cavities or large numbers of small round cavities in the design, which are introduced for the sole purpose of avoiding thick solid members \cite{Fernandez2019}.} These cavities are undesirable features for additive manufacturing as some processes require a minimum distance between structural members to introduce tools to remove the supports, or to allow sufficient flow of powder during its extraction \citep{Thompson2016}. The geometrical complexity introduced by the maximum size restriction can be alleviated by controlling, for example, the Minimum size in the Void phase ({\color{red}MinVoid}).

\begin{figure}[b!]
\centering
 	\begin{subfigure}[b]{0.32\linewidth}
		\centering{}\includegraphics[width=1\linewidth]{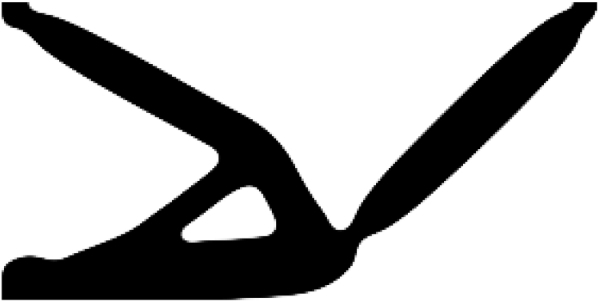}
 		\caption{}
 		\label{fig:Robust_F_a}
	\end{subfigure}
	~ \hspace{5mm}
	\begin{subfigure}[b]{0.32\linewidth}
		\centering{}\includegraphics[width=1\linewidth]{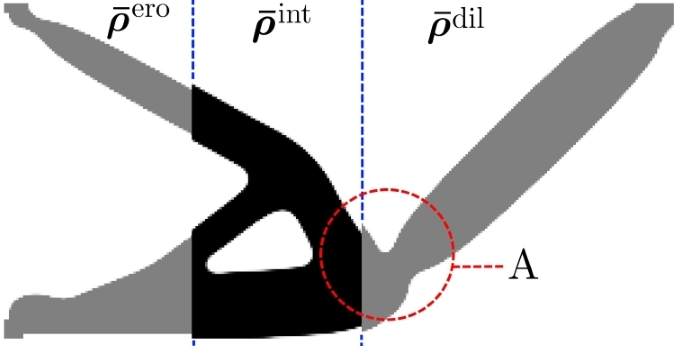}
 		\caption{}
 		\label{fig:Robust_F_b}
	\end{subfigure}
	\\
	\begin{subfigure}[b]{0.35\linewidth}
		\centering{}\includegraphics[width=1\linewidth]{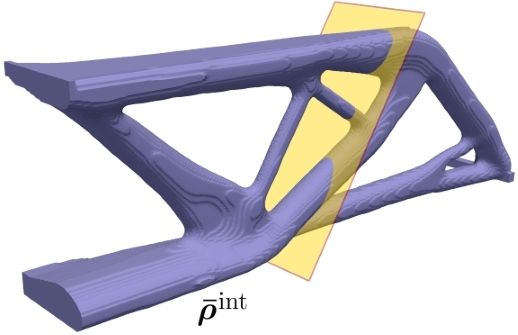}
		\vspace{-4mm}
		\caption{}
		\label{fig:Robust_F_d}
	\end{subfigure}
	~ \hspace{5mm}
	\begin{subfigure}[b]{0.35\linewidth}
		\centering{}\includegraphics[width=1\linewidth]{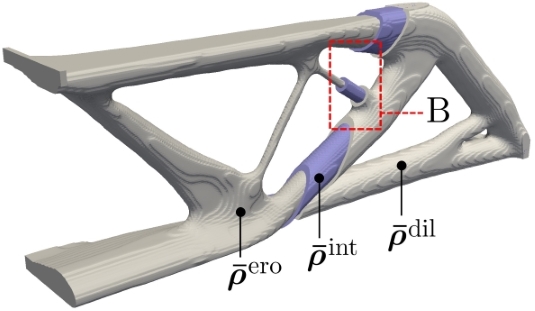}
		\caption{}
		\label{fig:Robust_F_e}
	\end{subfigure}
	\vspace{-2mm}
\caption{Optimized 2D force inverter and 3D MBB-beam using the robust design approach. In (a) and (c) are the blueprint solutions. In (b) and (d) are the three designs involved in the robust design formulation. Labelled zones, A and B, are shown in Fig. \ref{fig:Robust_Forumulation_2}.}
\label{fig:Robust_Forumulation}
\end{figure}

A variety of approaches aiming to control the minimum size of cavities have been proposed in the literature. Among others we can cite \citet{Almeida2009}. They invert the weight of the classical hat density filter \citep{Bruns2001}. The method is rather simple to implement but introduces large amount of gray elements and does not impose minimum length scale on both phases (MinSolid and MinVoid). {\color{red} \citet{Sigmund2007} introduces the erosion and dilation filters, which are based on morphological operators. The erosion filter defines the density of the element equal to the minimum density within the filtering region. Conversely, the dilation filter defines the element density equal to the maximum density within the filtering region. Thus, the erosion filter ensures a Minimum size in the Void phase (MinVoid) while the dilation filter ensures a Minimum size in the Solid phase (MinSolid)}. Later, \citet{Sigmund2009} presents a topology optimization formulation based on morphological operators to include manufacturing uncertainties which may result in uniformly thin (eroded) or uniformly thick (dilated) designs. The author proposes a robust formulation which minimizes the objective function for the worst case among the eroded, intermediate and dilated designs, which are depicted in Fig. \ref{fig:Robust_Forumulation}. The formulation eliminated the long-standing problem of one-node connected hinges in compliant mechanism design and, as shown by \citet{Wang2011}, it allows simultaneous control over the Minimum size in the Solid and Void phases {(\color{red}MinSolid} and {\color{red}MinVoid}), providing designs with crisp boundaries and improved manufacturability. More recently, \citet{Carstensen2018} propose a projection scheme with control over {\color{red}MinSolid and MinVoid}. The principle of the erosion and dilation operators is applied in this method, nonetheless this approach uses different non-linear weighting functions to dictate which filter is applied. The authors report 2D results satisfying minimum size in compliance minimization problems. However, in compliant mechanism problems, the method provides results with intermediate densities impeding direct fabrication of the optimized designs. In contrast to the cited works that rely on filtering techniques, \citet{Zhou2015} propose a constraint formulation to achieve minimum length scale on both phases ({\color{red}MinSolid and MinVoid}). The method introduces one constraint for each phase and utilizes the structural skeleton as switch indicator. The method effectively avoids the one-node connected hinges in the compliant mechanism problem. However, the geometric constraints must be introduced almost at the end of the optimization process as the method requires a well defined structure to detect the skeleton. Given the above, the method serves as a post-processing tool rather than an optimization constraint.

Despite the large number of contributions on the minimum length scale control, few papers impose in addition the Maximum size in the Solid phase (MaxSolid). In density-based topology optimization we can cite \citet{Lazarov2017}, who propose a maximum size restriction based on morphological operators. The constraint seeks to match the eroded design to a field composed entirely of void elements, which guarantees a {\color{red}MaxSolid} in the intermediate (blueprint) design. In addition, \citet{Lazarov2017} include the geometric constraints proposed by \citet{Zhou2015} to obtain, for the first time in the literature devoted to density methods, results with maximum and two-phase minimum length scale control, i.e. {\color{red}MinSolid, MinVoid and MaxSolid}. However, \citet{Lazarov2017} encounter undesirable geometrical features in some designs: a large number of hollow circles in the structure, a wavy pattern in the vicinity of the connections between solid members, and large amounts of intermediate densities. The work of \citet{Carstensen2018}, cited in the preceding paragraph, includes in addition a filtering scheme to impose minimum and maximum length scale control ({\color{red}MinSolid}, {\color{red}MinVoid} and {\color{red}MaxSolid}). However, some results present a large number of intermediate densities which hinders manufacturability. 

In the level set framework, there is also a large amount of work related to length scale control. For instance, we can cite the work of \citet{Allaire2016}, who formulate geometric constraints using local dimensions of solid and void phases obtained by the signed distance function. The authors also present a restriction based on the structural skeleton, resembling that proposed by \citet{Zhou2015} in the density framework. \citet{Allaire2016} report several 2D and 3D results that satisfy {\color{red}MinSolid} and {\color{red}MaxSolid}. The authors mention that one of the major drawbacks is that results are highly dependent on the initial guess. In addition, some designs contain topological features that are typical of a local optimum, such as bars connected by a single extreme to the main structure. However, results are crisp which eases design interpretation and manufacturing. For a more complete review on length scale control within the level set framework, readers are referred to \cite{Liu2016}.

In the density framework, the works focused on simultaneous control of {\color{red}MinSolid, MinVoid and MaxSolid, have not been able to achieve manufacturable solutions mainly due to the presence of intermediate densities and one-node connected hinges}. In addition, the cited methods have only been evaluated on 2D-design domains, so their performance in 3D is unclear. This paper aims to contribute to the above points. We present a novel strategy for density-based topology optimization to control, simultaneously, the Minimum size in the Solid phase ({\color{red}MinSolid}), the Minimum size in the Void phase ({\color{red}MinVoid}), and the Maximum Size in the Solid phase ({\color{red}MaxSolid}). To impose the minimum length scale, we use the robust design approach based on the eroded, intermediate and dilated designs. {\color{red}Following the principle of the method, we minimize the objective function for the worst-performing design. To control the maximum size, we use local volume constraints \cite{Guest2009}, which are gathered into one global constraint using aggregation functions \cite{Fernandez2019}. The maximum size constraint is applied in the eroded, intermediate and dilated designs, therefore, 3 global constraints are included in the robust optimization problem. The challenge of this approach is that the eroded, intermediate and dilated designs feature different dimensions due to the morphological operators, therefore, the maximum size restriction must be scaled according to the design in which it is applied. The method is evaluated using 2D and 3D design problems, considering a linear elastic and isotropic material. Results show that by combining the robust design method with maximum size restrictions, MinSolid, MinVoid and MaxSolid can be achieved simultaneously. Afterwards, it is shown that despite obtaining solutions with crisp boundaries, free of one-node connected hinges and satisfying minimum and maximum length scale, the designs can be difficult to manufacture due to the presence of multiple rounded cavities. An intuitive way to decrease the number of cavities is to enlarge the Minimum size in the Void phase (MinVoid) through the robust design approach. However, MinVoid is ensured by preventing cavities to be smaller than a prescribed circle or sphere, i.e.~the cavities in 2D could be produced by machining tools, and in 3D by Boolean operations using a sphere as a removal pattern. Therefore, radius of curvature at re-entrant corners cannot be smaller than MinVoid. Consequently, decreasing the number of cavities by choosing a larger MinVoid would produce large curvatures at re-entrant corners, which could negatively affect design performance. In order to reduce the amount of cavities, increase their size and do not affect the minimum radius of curvature at re-entrant corners, in this paper we propose a new geometric constraint that imposes the minimum separation distance between two solid members, meaning the Minimum Gap (MinGap). This constraint is integrated in the robust optimization problem aiming to control 4 geometric entities: MinSolid, MinVoid, MaxSolid, and MinGap. Results show that a large MinGap does not affect the radius at re-entrant corners, it implicitly reduces the quantity of closed cavities and increases the minimum size of those remaining, which can contribute to improve manufacturability of maximum size constrained designs.
}

The remainder of the article is organized as follows. {\color{red} In Section \ref{sec:2} we present the chosen test cases and the formulation of the topology optimization problem considering the robust design approach. Section \ref{sec:MaxSizeC} presents the formulation of the maximum size restriction and its implementation in the robust formulation. In Section \ref{sec:MinGap} we introduce the new restriction that limits the minimum distance between solid members. Section \ref{sec:5} gathers the results and discussions while Section \ref{sec:6} summarizes the final conclusions of this work.}

\section{Design problems and topology optimization formulation}
\label{sec:2}

The proposed strategies to control {\color{red} the Minimum size in the Solid phase (MinSolid), the Minimum Size in the Void phase (MinVoid), the Maximum size in the Solid phase (MaxSolid) and the Minimum Gap between solid members (MinGap)}, are presented in a density-based topology optimization framework. The test cases introduced in this section are solved using two open source codes, Polytop \citep{Talischi2012} and TopOpt \citep{Aage2015}. The first is a code written in Matlab aimed at solving 2D problems whereas the latter is a C++ code able to solve large scale 3D problems by means of its parallelized structure based on the PETSc library (Portable, Extensible Toolkit for Scientific Computation). Equations presented in this paper are shown as implemented in these codes, i.e.~using vector and matrix operations in order to vectorize loops in MATLAB \citep{Andreassen2011}, and to use the default routines of PETSc.

\begin{figure}[t!]
\centering    
    \begin{subfigure}[b]{0.35\linewidth}
		\centering{}\includegraphics[width=1\linewidth]{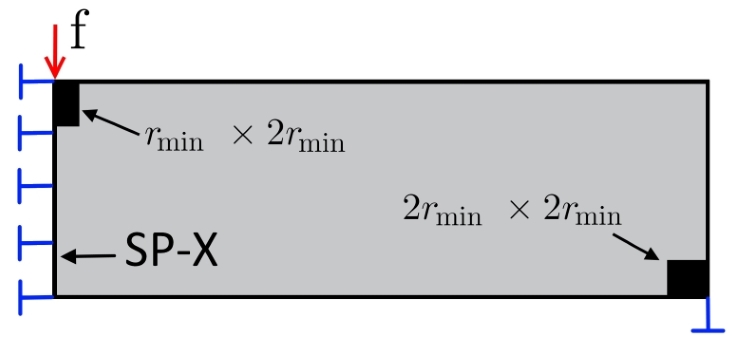}
		\vspace{-5mm}
 		\caption{}
 		\label{fig:2DCB}
	\end{subfigure}
	~ \hspace{15mm} 
 	\begin{subfigure}[b]{0.4\linewidth}
		\centering{}\includegraphics[width=1\linewidth]{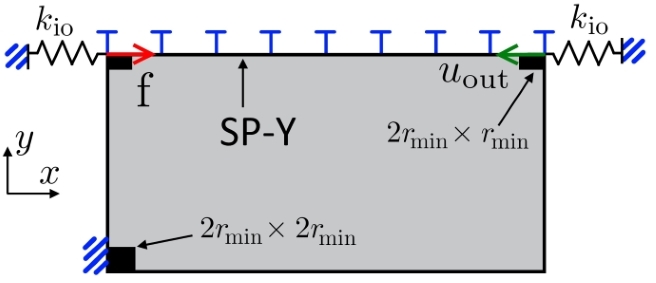}
		\vspace{-5mm}
 		\caption{}
 		\label{fig:2DFI}
	\end{subfigure}
	\\ 
	\begin{subfigure}[b]{0.5\linewidth}
		\centering{}\includegraphics[width=1\linewidth]{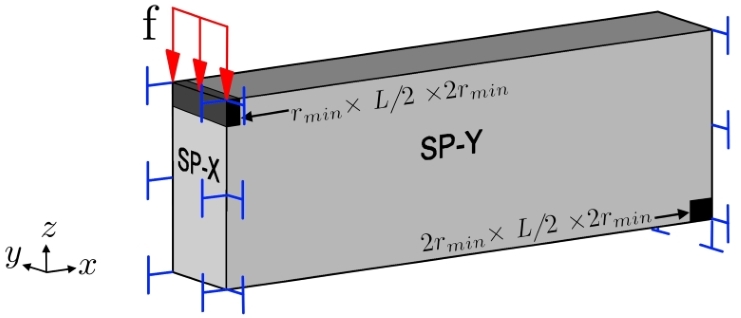}
		\vspace{-8mm}
		\caption{}
		\label{fig:3DCB_Half}
	\end{subfigure}
\caption{Design domains. (a) 2D-MBB beam of size 3L$\times$L. (b) Force inverter of size 2L$\times$L. (c) 3D-MBB beam of size 3L$\times$L/2$\times$L. The Blue lines represent the degrees of freedom that are fixed to zero. The Black areas represent the non-optimizable zones defined as solid material. (a) and (b) are implemented in PolyTop \cite{Talischi2012} while (c) in TopOpt \cite{Aage2015}.}
\label{fig:Design_Domains}
\end{figure}

{\color{red}
\subsection{Test Cases}
}

Two well known problems in the literature are addressed in this work, the compliance minimization and the compliant mechanism design, both in linear elasticity. The so-called MBB (Messerschmitt{-{–}}B{\"o}lkow{-}Blohm) beam and the compliant force inverter are chosen as test cases. The MBB beam design problem is solved in 2D and 3D to analyse the performance of the proposed methodologies in both cases. The force inverter is chosen because its complexity promotes small local features such as one-node connected hinges \cite{Sigmund2009}, which makes length scale control more difficult than in compliance minimization. 

{\color{red} In order to reduce computational requirements and simulation time, we model only a half of the 2D-MBB beam and force interter, and only a quarter of the 3D-MBB beam by imposing symmetry conditions, as shown in Fig. \ref{fig:Design_Domains}}. The design domains are discretized into 4-node quadrilateral and 8-node hexahedral elements in 2D and 3D, respectively. The 2D-MBB beam and the force inverter are discretized using 30000 and 33600 elements, while the 3D-MBB is discretized into 1.33 million elements. 

{\color{red}
\subsection{Three field scheme for robust design approach}
}

{ \color{red} Due to numerical instabilities resulting from a poor finite element modeling, structural topology optimization based on the density approach needs restriction methods to ensure well-posed and mesh-independent solutions \cite{Sigmund1998}. The method implemented here is the three-field scheme \cite{Sigmund2013}}. The first field $\bm{\rho}$ represents the design variables of the optimization problem, the second field $\bm{\tilde{\rho}}$ is obtained through a filtering process of the design variables, and the third field $\bm{\bar{\rho}}$ is obtained by a smoothed Heaviside projection of the filtered field, thus $\bm{\bar{\rho}} \left( \bm{\tilde{\rho}} ( \bm{\rho} ) \right)$. Since the eroded, intermediate and dilated designs are built in the three-field scheme, the filtering and projection processes are described in detail. From now on, various geometrical parameters will be introduced, which are summarized in Table \ref{tab:Parameters} for the sake of legibility.

\begin{table}[t!]
		\caption{List of parameters related to the length scale control. Within square brackets it is shown the parameter with reference to a specific design. The superscripts ero, int and dil refer to the eroded, intermediate and dilated designs, respectively.}
		\vspace{-3mm}
		\centering
		\begin{tabular}{p{0.1\linewidth} p{0.8\linewidth}}
				 \toprule
				 Parameter & Description \\
				 \cmidrule(r){1-1} \cmidrule(r){2-2} 
				 $r_\mathrm{fil}$ & Radius of the circular or spherical filtering region.
				 \vspace{1mm}\\				 
				 $\mu$            & Smoothed Heaviside function threshold. [$\mu^\mathrm{ero}$, $\mu^\mathrm{int}$, $\mu^\mathrm{dil}$].
				 \vspace{1mm}\\				 
				 $r_\mathrm{min,Solid}$ & Minimum size in the Solid phase (MinSolid). [$r_\mathrm{min,Solid}^\mathrm{ero}$, $r_\mathrm{min,Solid}^\mathrm{int}$, $r_\mathrm{min,Solid}^\mathrm{dil}$].  
				 \vspace{1mm}\\
				 $r_\mathrm{min,Void}$ & Minimum size in the Void phase (MinVoid). [$r_\mathrm{min,Void}^\mathrm{ero}$, $r_\mathrm{min,Void}^\mathrm{int}$, $r_\mathrm{min,Void}^\mathrm{dil}$].  
				 \vspace{1mm}\\
				 $t_\mathrm{ero}$ & Offset distance between the intermediate and eroded designs. 
				 \vspace{1mm}\\
				 $t_\mathrm{dil}$ & Offset distance between the intermediate and dilated designs. 
				 \vspace{1mm}\\		
				 $r_\mathrm{o}$ & External radius of the region $\Omega$. [$r_\mathrm{o}^\mathrm{ero}$, $r_\mathrm{o}^\mathrm{int}$, $r_\mathrm{o}^\mathrm{dil}$]. 			 
				 \vspace{1mm}\\	
				 $r_\mathrm{max}$ & Maximum size in the Solid phase (MaxSolid). [$r_\mathrm{max}^\mathrm{ero}$, $r_\mathrm{max}^\mathrm{int}$, $r_\mathrm{max}^\mathrm{dil}$].
				 \vspace{1mm}\\
				 $h$ 			  & Minimum Gap (MinGap). [$h^\mathrm{ero}$, $h^\mathrm{int}$, $h^\mathrm{dil}$]. 				 
				 \\
				 \bottomrule
		\end{tabular}
	\label{tab:Parameters}
\end{table}

\subsubsection{Filtered field}
\label{sec:2.1}

\begin{figure}[t!]
\centering
 	\begin{subfigure}[b]{0.216\linewidth}
		\centering{}\includegraphics[width=1\linewidth]{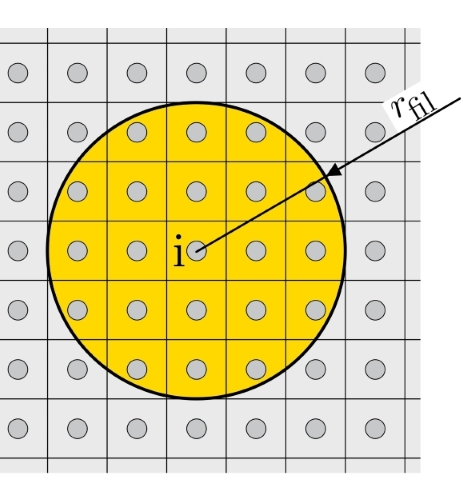}
 		\caption{}
 		\label{fig:Filter_1}
	\end{subfigure}
	~
	\begin{subfigure}[b]{0.22\linewidth}
		\centering{}\includegraphics[width=1\linewidth]{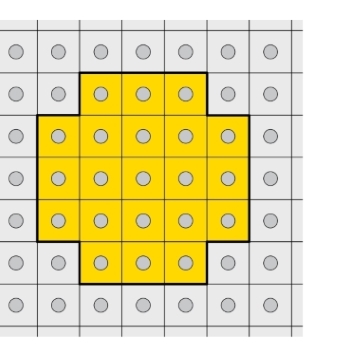}		
		\caption{}
		\label{fig:Filter_2}
	\end{subfigure}
	~
	\begin{subfigure}[b]{0.25\linewidth}
		\centering{}\includegraphics[width=1\linewidth]{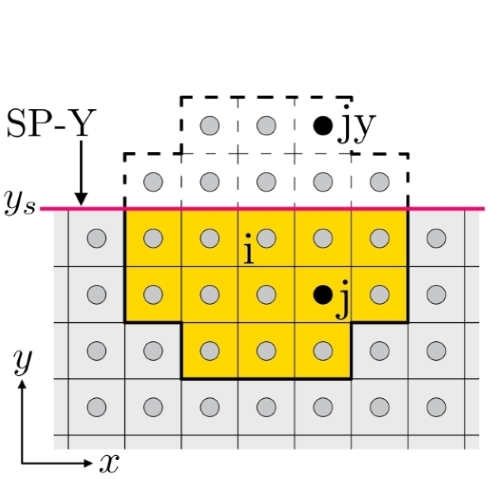}
		\caption{}
		\label{fig:Filter_3}
	\end{subfigure}
	~
	\begin{subfigure}[b]{0.25\linewidth}
		\centering{}\includegraphics[width=1\linewidth]{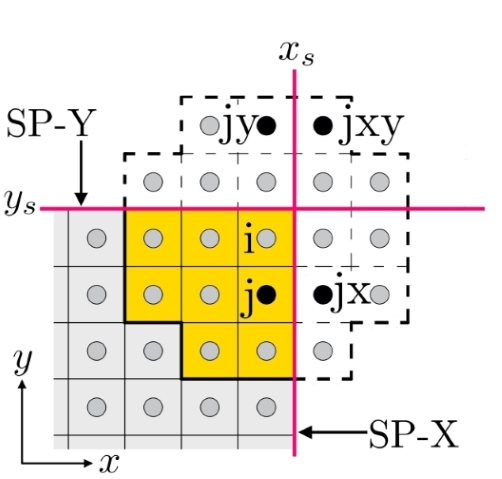}
		\caption{}
		\label{fig:Filter_4}
	\end{subfigure}
	\vspace{-2mm}
\caption{Filtering process. In (a) is the region from which the weighted average is obtained. In (b) is the discretized filtering region. In (c) is shown the element $j \mathrm{y}$ which is the reflection of $j$ with respect to SP-Y. In (d), the element $j$ can be mirrored 3 times in the quarter 3D-MBB beam.}
\label{fig:Filter}
\end{figure}

{\color{red}
The filtering of the design variables is a restriction method that avoids the mesh-dependency and checkerboard issues. For that, the filtered variable associated to element $i$ is defined as the weighted average of the design variables in a neighborhood of radius $r_\mathrm{fil}$ \cite{Bruns2001}, as shown in Fig. \ref{fig:Filter_1}. The filtered variable $\tilde{\rho}_i$ can be written as: 
}
\begin{equation} \label{eq:FilterNC}
	\tilde{\rho}_i = \frac{\displaystyle\sum_{j=1}^{N} \rho_j \: v_j \: w\left( X_i , X_j \right)}{\displaystyle\sum_{j=1}^{N} v_j \: w \left( X_i , X_j \right)} \; ,
\end{equation}
{\color{red}
\noindent where $\rho_j$ and $v_j$ are the design variable and volume of element $j$, $N$ is the total number of elements, and $w$ is the weighting function that depends on the distance between the centroid of the elements $i$ and $j$. A centroid is denoted by $X$ and is defined by its Cartesian coordinates ($x,y,z$). Therefore, we have $X_i(x_i,y_i,z_i)$ and $X_j(x_j,y_j,z_j)$. The weighting function $w$ is defined as follows:
}
\begin{equation} \label{eq:wij}
	w \left( X_i , X_j \right) = \text{max} \left(  0,1 - \frac{\lvert \rvert {X_i-X_j} \lvert \rvert}{r_{\mathrm{fil}}} \right) \;.
\end{equation}

{\color{red}
The filtering of design variables \eqref{eq:FilterNC} is included in the PolyTop \cite{Talischi2012} and TopOpt \cite{Aage2015} codes, however, we slightly change the definition of $\tilde{\rho}_i$ in order to consider boundary effects.} The filtering procedure conducted at the boundaries of the design domain can have a strong effect on the optimized solution. As shown by \citet{Clausen2017}, boundary effects can even affect the topology beyond the border by decreasing the structural performance or affecting the imposed length scale. An effective treatment is to extend the design domain at the boundaries \citep{Sigmund2007,Clausen2017}. This approach, however, brings an additional computational cost due to the extra amount of finite elements that must be stored in memory, which can have a big impact on large 3D problems. Given the above, the filtering scheme is efficiently adjusted to incorporate the boundary effect without entailing a real extension of the finite element mesh. In addition, unlike the vast majority of works in literature, here the filter is corrected for problems modelled with symmetric boundary conditions. The adjustments of the filter are explained in detail as they will also be of importance for the proposed strategy.

Since we use a regular mesh and the same filter radius throughout the design domain, the volume of all filtering neighbourhoods would also be the same if the finite element mesh were extended beyond the boundaries of the design domain, i.e.~the denominator in \eqref{eq:FilterNC} would be the same for all filtered variables inside the design space. The denominator of \eqref{eq:FilterNC} is denoted by ${V_\mathrm{fil}}$ and computed as:
\begin{equation}
{V_\mathrm{fil}} = \displaystyle \sum_{j=1}^N w\left( X_k , X_j \right) \: v_j  \; ,
\end{equation}
\noindent where $k$ is the index of any element whose discretized filtering region does not reach the boundaries of the design domain, as shown in Fig. \ref{fig:Filter_2}. {\color{red} Then, to emulate a mesh extension throughout the boundaries of the design space, the filter \eqref{eq:FilterNC} is replaced by:}
\begin{equation} \label{eq:FilterWC}
	\tilde{\rho}_i = \frac{\displaystyle\sum_{j=1}^{N} \rho_j \: v_j \: \hat{w}_{i,j}}{V_\mathrm{fil}} \; .
\end{equation}

{\color{red}
For an element $i$ whose filtering region reaches the boundaries, equation \eqref{eq:FilterWC} assumes void elements outside the design space. For an element $i$ whose filtering region reaches a plane of symmetry, equation \eqref{eq:FilterWC} reflects design variables $\rho_j$ with respect to the plane of symmetry.} For this, we denote as $\hat{w}_{i,j}$ the weighting function that includes in addition the weight of the mirrored variables, i.e.~the total contribution of $\rho_j$ over the filtered field $\tilde{\rho}_i$. For instance, if we consider $j{\mathrm{y}}$ as the reflection of the variable $j$ with respect to SP-Y (see Fig.~\ref{fig:Filter_3}), the total weight $\hat{w}_{i,j}$ is computed as: 
\begin{equation} \label{eq:wij_total}
	\hat{w}_{i,j} = w \left( X_i , X_j \right) +w \left( X_i , X_{j\mathrm{y}} \right) \; .
\end{equation}  

The centroid of the element $j\mathrm{y}$ can be obtained as $X_{j{\mathrm{y}}} = (x_j \; , \; y_j+2(y_\mathrm{s}-y_j) \; , \; z_j)$, where $y_\mathrm{s}$ is the coordinate that defines SP-Y, as shown in Fig. \ref{fig:Filter_3}. In the case of the quarter 3D-MBB beam, the variable $j$ can be mirrored up to 3 times in the vicinity of SP-X and SP-Y, as shown in Fig. \ref{fig:Filter_4}. In this case $w \left( X_i , X_{j\mathrm{x}} \right)$, $w \left( X_i , X_{j\mathrm{y}} \right)$ and $w \left( X_i , X_{j\mathrm{xy}} \right)$ are the additional contributions of variable $j$ in the filtering process. 

Subsequently, the weights and constant parameters are preallocated in a matrix $\mathrm{\mathbf{D}_I}$ as follows: 
\begin{equation} \label{eq:DI_matrix}
	\mathrm{{D}_I}_{(i,j)} = \frac{v_j \: \hat{w}_{i,j} }{V_{\mathrm{fil}}} \;.
\end{equation}

Finally, the filtering process including the treatment on the boundaries can be carried out as:
\begin{equation} \label{eq:FilteredField}
	\bm{\tilde{\rho}}= \mathrm{\mathbf{D}_I} \: \bm{\rho} \;.
\end{equation}

\subsubsection{Projected field and sensitivity analysis}
\label{sec:2.2}

{\color{red}
To impose minimum length scale (MinSolid and MinVoid), we use the robust design approach \cite{Sigmund2009}, which considers manufacturing uncertainties that can result in an uniformly thin (eroded) or uniformly thick (dilated) structure with respect to the intended (intermediate) design. The designs involved in this formulation are denoted by $\bm{\bar{\rho}}^{\mathrm{ero}}$, $\bm{\bar{\rho}}^{\mathrm{int}}$ and $\bm{\bar{\rho}}^{\mathrm{dil}}$. Superscripts ${\mathrm{ero}}$, ${\mathrm{int}}$ and ${\mathrm{dil}}$ refer to the eroded, intermediate and dilated fields, respectively.} These three designs are obtained from the filtered field using the following smoothed Heaviside function:
\begin{equation} \label{eq:Heaviside}
\bar{\rho}_i = H(\tilde{\rho}_i,\: \beta,\: \mu) = 
\frac
	{\mathrm{tanh}(\beta\:\mu)+\mathrm{tanh}(\beta\:(\tilde{\rho}_i-\mu))}
	{\mathrm{tanh}(\beta\:\mu)+\mathrm{tanh}(\beta\:(        1     -\mu))} \;.
\end{equation}

\begin{figure}[t!]
	\centering
	\includegraphics[width=1\linewidth]{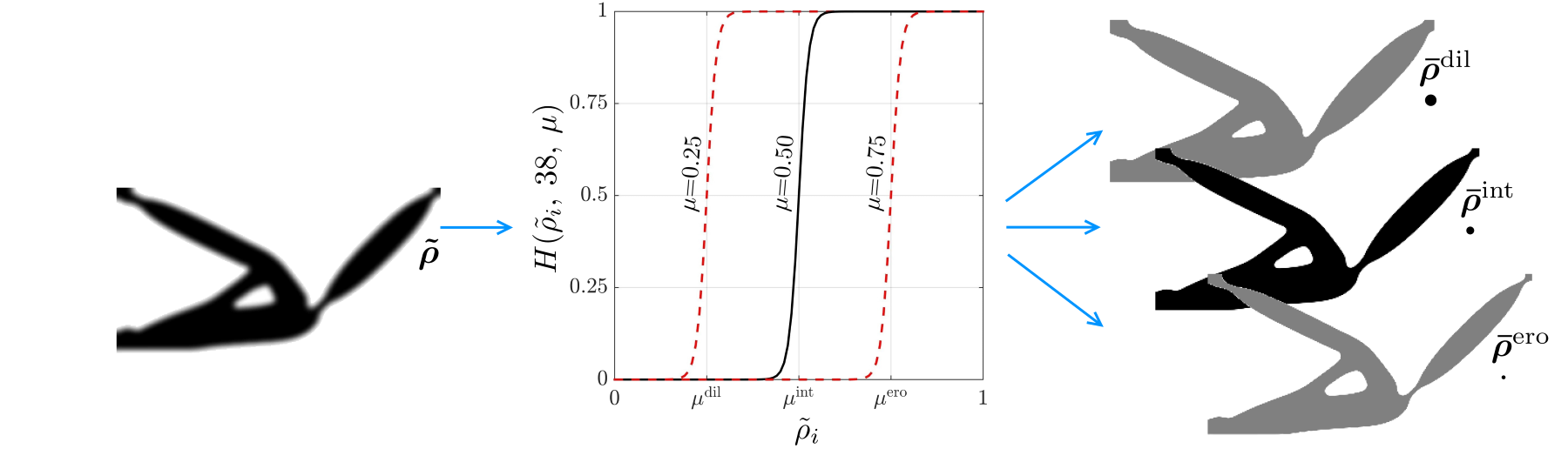}
	\caption{{\color{red}From left to right, the figure shows a filtered field, the smoothed Heaviside function at $\beta=38$, and the eroded, intermediate and dilated designs. The eroded and dilated designs are faded to emphasize that the intermediate design is the one intended for manufacturing. The black dot below the projected field $\bm{\bar{\rho}}$ shows the minimum member size measured graphically from the design.}}
\label{fig:F_Robust}
\end{figure}

{\color{red}
This Heaviside function plays two roles. First, it reduces the amount of grey elements coming from the filtering process, and second, it produces the uniform manufacturing error for the robust formulation \citep{Wang2011}. The parameter $\beta$ controls the steepness of the Heaviside function and, for values greater than 0, reduces the amount of grey elements. The parameter $\mu$ controls the threshold of the projection and provides the uniform manufacturing error. That is, the eroded, intermediate and dilated designs are obtained using the same steepness $\beta$ but different thresholds $\mu$. Fig. \ref{fig:F_Robust} shows the eroded, intermediate and dilated designs obtained from a filtered field $\bm{\tilde{\rho}}$ using $\mu^\mathrm{ero}$=0.75, $\mu^\mathrm{int}$=0.50, $\mu^\mathrm{dil}$=0.25 and $\beta$=38. The chosen parameters will be explained in the following sections. 
}

Considering column arrays and denominator layout \cite{Xing2019}, the sensitivity analysis of the three-field-scheme can be obtained by applying the chain rule as follows:
\begin{equation} \label{eq:derivProjection}
	\frac{\mathrm{d}\bm{\bar{\rho}}}{\mathrm{d}\bm{\rho}} = \frac{\mathrm{d} \bm{\tilde{\rho}}}{\mathrm{d} \bm{\rho}} \frac{\mathrm{d} \bm{\bar{\rho}}}{\mathrm{d} \bm{\tilde{\rho}}} = \mathrm{\mathbf{D}_I^{\intercal}} \mathrm{\mathbf{J}}_{\bar{\rho}} \;.
\end{equation}

\noindent were $\mathrm{\mathbf{J}}_{\bar{\rho}}$ is the Jacobian matrix of the smoothed Heaviside function defined as $\mathrm{\mathbf{J}}_{\bar{\rho}}=\mathrm{diag}(H'(\tilde{\rho}_1),...,H'(\tilde{\rho}_N))$. Here $H'(\tilde{\rho}_i)$ is the derivative of $H$ with respect to $\tilde{\rho}_i$. For efficiency purposes, the matrix $\mathrm{\mathbf{J}}_{\bar{\rho}}$ is not created in practice, but here it is written for the sake of completeness. A more efficient implementation involves the use of the Hadamard product for which the reader can refer to \citet{Talischi2012}.

\subsection{Topology optimization problem}
\label{sec:2.3}

{\color{red}
As the compliance minimization and the compliant mechanism design problems share a common basis, a generic density-based topology optimization problem considering the robust formulation is given as follows: 
}
{\color{red}
\begin{align} \label{eq:OPTI_General}
	\begin{split}
  		{\min_{\rho}} &\quad \max \left( \bm{t}^{\intercal} \bm{u}^{\mathrm{ero}} \:,\: \bm{t}^{\intercal} \bm{u}^{\mathrm{int}} \:,\: \bm{t}^{\intercal} \bm{u}^{\mathrm{dil}}\right)				\\
	  	{s.t.:} &\quad \bm{v}^{\intercal} \bm{\bar{\rho}}^{\mathrm{ero}} \leq V^*_{\mathrm{ero}} 	\\
	  			&\quad \bm{v}^{\intercal} \bm{\bar{\rho}}^{\mathrm{int}} \leq V^*_{\mathrm{int}}    \\
	  			&\quad \bm{v}^{\intercal} \bm{\bar{\rho}}^{\mathrm{dil}} \leq V^*_{\mathrm{dil}}    \\
	  			&\quad 0 \leq {\rho_i} \leq1 \; ,
	\end{split}
\end{align}
}
\noindent where $\bm{u}^\mathrm{ero}$, $\bm{u}^\mathrm{int}$ and $\bm{u}^\mathrm{dil}$ are the nodal displacements obtained after solving the state equations:
\begin{subequations}
\begin{eqnarray}
\mathbf{K}(\bm{\bar{\rho}}^{\mathrm{ero}})\:\bm{u}^{\mathrm{ero}}=\bm{f} \\
\mathbf{K}(\bm{\bar{\rho}}^{\mathrm{int}})\:\bm{u}^{\mathrm{int}}=\bm{f} \\
\mathbf{K}(\bm{\bar{\rho}}^{\mathrm{dil}})\:\bm{u}^{\mathrm{dil}}=\bm{f}
\end{eqnarray}
\end{subequations}

Here, $\bm{f}$ is the array containing the external force. $\mathbf{K}$ is the stiffness matrix that is assembled from the element stiffness matrices $\mathbf{k}_i$, which are computed as $\mathbf{k}_i$=$E_i\:\mathbf{k}_0$, where $\mathbf{k}_0$ is the stiffness matrix of unit stiffness and $E_i$ is the Young's modulus defined by the modified SIMP \cite{Sigmund2007} interpolation scheme: 
\begin{equation}
	E_i = E_{\mathrm{min}} + \bar{\rho}_i^\eta(E_0-E_{\mathrm{min}}) \; ,
\end{equation}
\noindent where $\eta$ is the penalization power, $E_0$ is the Young's modulus of the solid phase, and $E_\mathrm{min}$ is the Young's modulus of the void phase. {\color{red} In this work, we set $E_0$=1 and $E_\mathrm{min}$=$10^{-6}$.} 

The optimization problem \eqref{eq:OPTI_General} minimizes the objective function of the worst performing design. For the compliance minimization problem, $\bm{t}$ represents the input force over the domain, i.e.~it is equal to the array $\bm{f}$. For the force inverter problem, $\bm{t}$ is an array that contains value $1$ at the output degree of freedom and 0 otherwise. {\color{red} In the general formulation of the robust approach \eqref{eq:OPTI_General}, the volume restriction is applied in all 3 designs, where $\bm{v}$ is the array containing element volumes and $V^*$ is the upper bound of the restriction. However, since the dilated design is always the most voluminous, the volume constraints of the eroded and intermediate designs can be removed from the optimization problem. On the other hand, the intermediate design is the one intended for manufacturing, therefore the desired volume constraint is $V^*_\mathrm{int}$.} To impose $V^*_\mathrm{int}$ through $V^*_\mathrm{dil}$, we use the strategy adopted in \citep{Amir2018}, i.e.~every 10 iterations $V^*_{\mathrm{dil}}$ is updated according to the desired $V^*_\mathrm{int}$ as follows: 
\begin{equation} \label{eq:VolumeDilated}
	V^*_{\mathrm{dil}} = V^*_{\mathrm{int}} \frac{\bm{v}^\intercal \bm{\bar{\rho}}^{\mathrm{dil}}}{\bm{v}^\intercal \bm{\bar{\rho}}^{\mathrm{int}}} \;.
\end{equation}

{\color{red}
Regarding the robust formulation, we also emphasize that in compliance minimization, the eroded design is always the one with the largest compliance, therefore the intermediate and dilated designs can be removed from the objective function without compromising robustness in terms of length scale control \cite{Amir2018}. This alleviates computational cost by avoiding the state equations that solve $\bm{u}^{\mathrm{int}}$ and $\bm{u}^{\mathrm{dil}}$, and it reduces the optimization problem to the following:
}
{\color{red} 
\begin{align} \label{eq:OPTI_Compliance}
	\begin{split}
  		{\min_{\rho}} &\quad \bm{f}^{\intercal} \bm{u}^{\mathrm{ero}} 				\\
	  	{s.t.:}	&\quad \bm{v}^{\intercal} \bm{\bar{\rho}}^{\mathrm{dil}} \leq V^*_{\mathrm{dil}}    \\
	  			&\quad 0 \leq {\rho_i} \leq1 
	\end{split}
\end{align}
}

Concerning the force inverter design problem, results show that either the eroded or dilated field is the worst performer in most iterations. For this reason, we do not consider the intermediate field in the objective function. This simplification, according to our studies, does not compromise robustness in terms of length scale control. Thus, the force inverter design problem becomes:
\begin{align} \label{eq:OPTI_ForceInverter}
	\begin{split}
  		{\min_{\rho}} &\quad \max \left( \bm{t}^{\intercal} \bm{u}^{\mathrm{ero}} \:,\: \bm{t}^{\intercal} \bm{u}^{\mathrm{dil}}\right)				\\
	  	{s.t.:} &\quad \bm{v}^{\intercal} \bm{\bar{\rho}}^{\mathrm{dil}} \leq V^*_{\mathrm{dil}}    \\
	  			&\quad 0 \leq {\rho_i} \leq1 
	\end{split}
\end{align}

{\color{red}
\subsection{The Minimum size in the Solid phase (MinSolid) and the Minimum size in the Void phase (MinVoid)}
}

{\color{red}
As \citet{Wang2011} show, the robust formulation \eqref{eq:OPTI_General} imposes a Minimum size in the Solid phase (MinSolid) and a Minimum size in the Void phase (MinVoid). The minimum length scales, MinSolid and MinVoid, are defined by a prescribed circle or sphere of radius $r_\mathrm{min}$. For example, in 2D, MinSolid could be defined by the radius $r_\mathrm{min,Solid}$ of a deposition tool, and MinVoid by the radius $r_\mathrm{min,Void}$ of a milling tool \cite{Sigmund2009,Wang2011,Qian2013}. In 3D, MinSolid and MinVoid can be defined by the radius $r_\mathrm{min,Solid}$ and $r_\mathrm{min,Void}$ of a sphere used in a Boolean operation of union and subtraction, respectively. 
}

{\color{red}
Due to the morphological operators, the eroded, intermediate and dilated designs feature different minimum length scales, as can be seen from the black dots in Fig. \ref{fig:F_Robust}. Therefore, to refer to a specific design, we include the corresponding superscripts. Thus, MinSolid is defined by $r_\mathrm{min,Solid}^\mathrm{ero}$, $r_\mathrm{min,Solid}^\mathrm{int}$ and $r_\mathrm{min,Solid}^\mathrm{dil}$, and MinVoid by $r_\mathrm{min,Void}^\mathrm{ero}$, $r_\mathrm{min,Void}^\mathrm{int}$ and $r_\mathrm{min,Void}^\mathrm{dil}$.
}

{\color{red}
In \cite{Wang2011}, it is shown that MinSolid and MinVoid depend on the projection thresholds $\mu$ and on the filter size $r_\mathrm{fil}$. Therefore, the minimum length scale intended by the user ($r_\mathrm{min,Solid}^\mathrm{int}$ and $r_\mathrm{min,Void}^\mathrm{int}$) must be implicitly imposed through 4 parameters: $\mu^\mathrm{ero}$, $\mu^\mathrm{int}$, $\mu^\mathrm{dil}$ and $r_\mathrm{fil}$. To determine these 4 unknowns for a desired length scale, we use the results from \cite{Wang2011}. There, graphical solutions are provided that relate $r_\mathrm{min,Solid}$ and $r_\mathrm{min,Void}$ with $\PFS{\mu}{}{ero}$, $\PFS{\mu}{}{int}$, $\PFS{\mu}{}{dil}$ and $\PFS{r}{fil}{}$. However, we will need additional information that is not provided in \cite{Wang2011}, so the graphical solutions are reconstructed in this paper. The numerical procedure for constructing the graphs is summarized as follows. A one-dimensional field $\bm{\rho}_\mathrm{1D}$ must be chosen; then, a set of thresholds [$\PFS{\mu}{}{ero}$, $\PFS{\mu}{}{int}$, $\PFS{\mu}{}{dil}$] is selected and the three-field scheme is applied on $\bm{\rho}_\mathrm{1D}$; the one-dimensional size of the eroded, intermediate and dilated fields is measured; finally, the obtained length scale (MinSolid and MinVoid) is placed on a graph. To construct a curve, the procedure is repeated several times for different values of $\PFS{r}{fil}{}$, $\PFS{\mu}{}{ero}$, $\PFS{\mu}{}{int}$ and $\PFS{\mu}{}{dil}$. 
}

\begin{figure}[t!]
\centering
    \begin{tabular}{c c}
	\begin{subfigure}[b]{0.45\linewidth}
		\centering{}\includegraphics[width=1\linewidth]{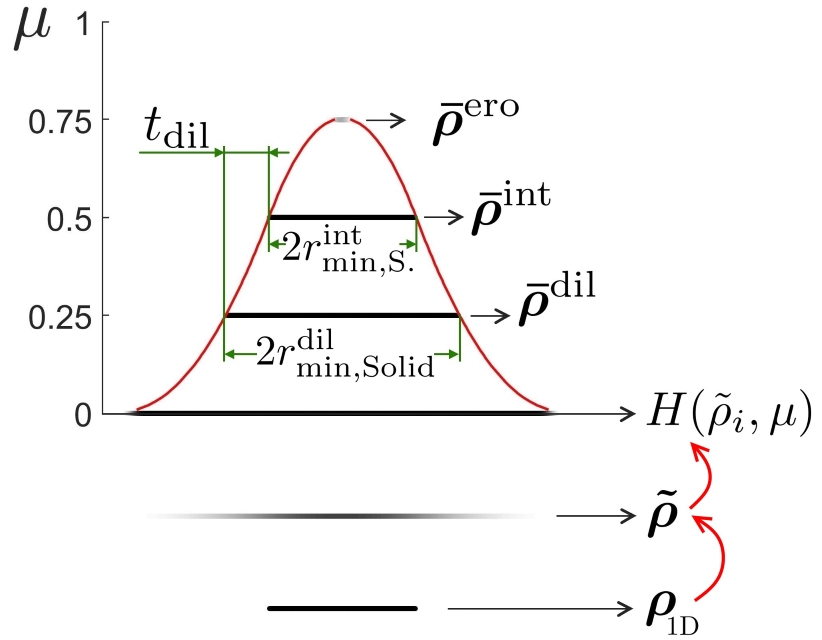}
		\caption{}
		\label{fig:MeasureGraph_a}
	\end{subfigure}
	& 
	\begin{subfigure}[b]{0.45\linewidth}
		\centering{}\includegraphics[width=1\linewidth]{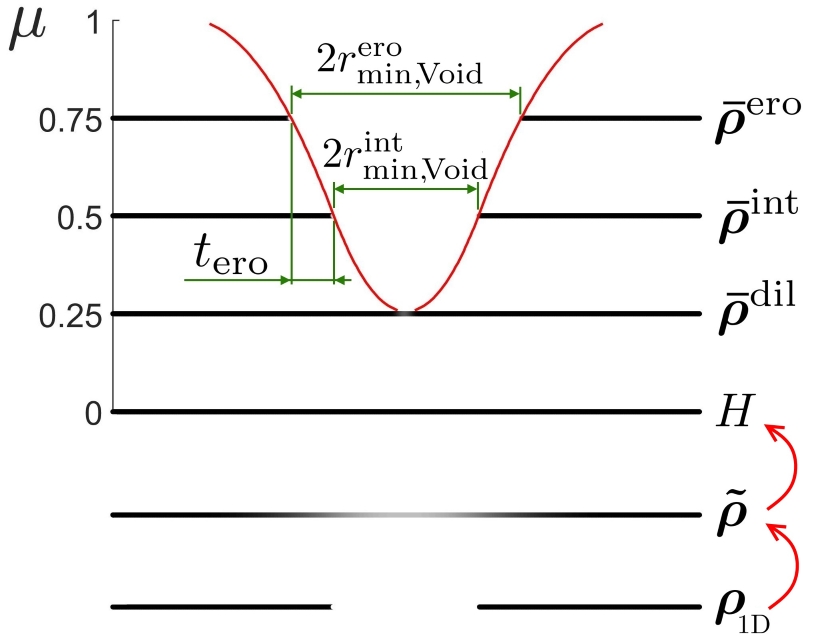}
		\caption{}
		\label{fig:MeasureGraph_b}
	\end{subfigure}    
    \vspace{1mm}\\
 	\begin{subfigure}[b]{0.45\linewidth}
		\centering{}\includegraphics[width=1\linewidth]{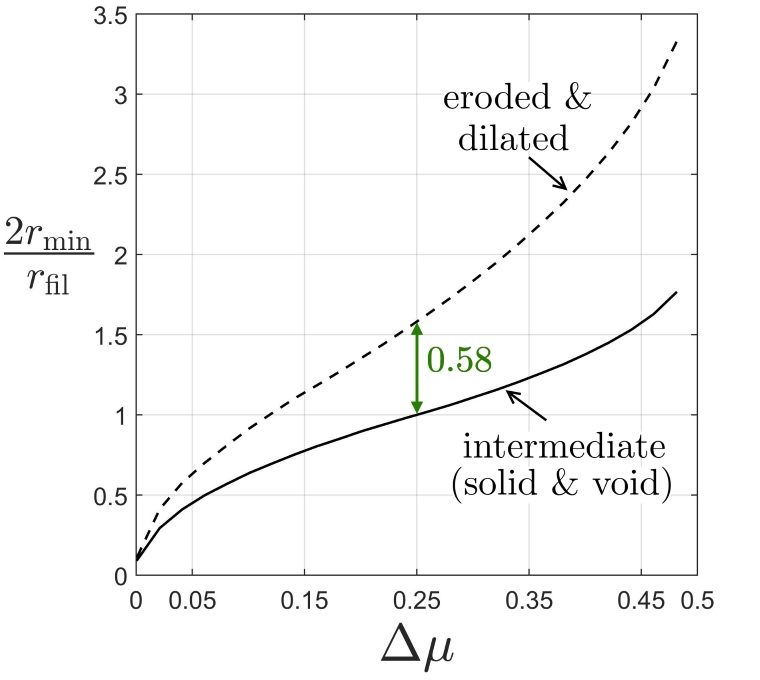}
		\caption{}
		\label{fig:MeasureGraph_c}
	\end{subfigure}
	&
	\begin{subfigure}[b]{0.45\linewidth}
		\centering{}\includegraphics[width=1\linewidth]{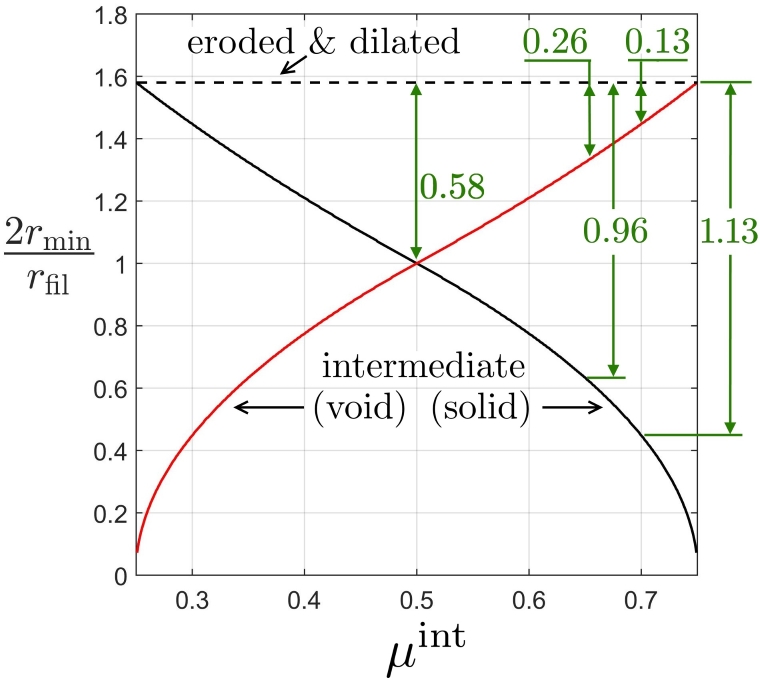}
		\caption{}
		\label{fig:MeasureGraph_d}
	\end{subfigure}
	\end{tabular} 	
\caption{Geometric relationships between the eroded, intermediate and dilated designs. In (a), filtering process to obtain the relation between $\mu$ and the MinSolid. In (b), the filtering process to obtain the relation between $\mu$ and the MinVoid. In (c), the graph that relates $\PFS{r}{min,Solid}{}$ and $\PFS{r}{min,Void}{}$ with $\PFS{r}{fil}{}$ and $\Delta\mu$, considering $\mu_\mathrm{int}=0.5$. In (d), the graph that relates $\PFS{r}{min,Void}{}$ and $\PFS{r}{min,Solid}{}$ with $\PFS{r}{fil}{}$ and $\mu_\mathrm{int}$ considering $\mu^\mathrm{ero}=0.75$ and $\mu^\mathrm{dil}=0.25$.}
\label{fig:MeasureGraph}
\end{figure}

In \cite{Wang2011}, the curve is constructed by setting the threshold $\PFS{\mu}{}{int}$ to 0.5, and by defining the erosion threshold greater than 0.5 and the dilation threshold smaller than 0.5. In this work, as in \cite{Amir2018,Wang2011,Qian2013}, the thresholds  $\PFS{\mu}{}{ero}$ and $\PFS{\mu}{}{dil}$ are defined symmetrical with respect to  $\PFS{\mu}{}{int}$, as follows:
\begin{equation}
   \mu^\mathrm{ero}-\mu^\mathrm{int}=\mu^\mathrm{int}-\mu^\mathrm{dil}=\Delta\mu \;\; ,
\end{equation}
\noindent where $\Delta\mu$ is the symmetric distance with respect to $\mu^\mathrm{int}$. Therefore, the graphical solution can be built for $\PFS{\mu}{}{int}$ fixed at 0.5 and $\Delta\mu$ variable between 0 and 0.5. {\color{red} To choose $r_\mathrm{fil}$ and to finally recreate the graphical solution, the following must be taken into consideration. {\textit{The intermediate design can be seen as a dilation of the eroded design, therefore, a solid member in the intermediate design will exist only if the solid member is projected by at least one solid element in the eroded design. Similarly, the intermediate design can be interpreted as an erosion of the dilated design, therefore, a cavity in the intermediate design will exist only if this cavity is projected by at least one void element in the dilated design.}} Knowing this, we describe below the process to construct the graphical solution that relates the desired length scale and the 4 unknown parameters.}

To obtain the curve that relates $\PFS{r}{min,Solid}{int}$ and $\PFS{r}{min,Void}{int}$ with $r_\mathrm{fil}$, $\PFS{\mu}{}{int}=0.5$ and $0<\Delta\mu<0.5$, a 1D field containing solid elements only at the center of the design domain is created, as $\bm{\rho}_\mathrm{1D}$ shown at the bottom of Fig. \ref{fig:MeasureGraph_a}. Then, a filter size is chosen such that the eroded design results in a set composed of one or few solid elements. The MinSolid is obtained from the intermediate and dilated designs by measuring the length of the 1D designs, as shown in Fig. \ref{fig:MeasureGraph_a}. To obtain the MinVoid, the procedure described above is repeated but with a field $\bm{\rho}_\mathrm{1D}$ containing void elements at the center of the design domain, as shown in Fig. \ref{fig:MeasureGraph_b}. The filter size in this case is chosen such that the dilated design results in a field composed of one or few void elements. These procedures are repeated for several values of $\Delta\mu$ between 0 and 0.5. Then, the geometrical parameters obtained for each $\Delta\mu$ are plotted in Fig. \ref{fig:MeasureGraph_c}. In that graph, the MinSolid and MinVoid curves overlap, meaning that $r_\mathrm{min,Solid}^\mathrm{int} = r_\mathrm{min,Void}^\mathrm{int}$ and $\PFS{r}{min,Void}{ero} = \PFS{r}{min,Solid}{dil}$. For further details on the production of these graphs the reader is referred to \cite{Wang2011,Qian2013}.

\vspace{3mm}
{\color{red}
\subsection{Sets of thresholds and offset distances}
}

{\color{red}
We recall that the minimum length scale defined by the user, $r_\mathrm{min,Solid}^\mathrm{int}$ and $r_\mathrm{min,Void}^\mathrm{int}$, must be imposed indirectly through $r_\mathrm{fil}$, $\PFS{\mu}{}{ero}$, $\PFS{\mu}{}{int}$ and $\PFS{\mu}{}{dil}$. As the system that relates these parameters is indeterminate, there is more than one combination of parameters that leads to the desired length scale \cite{Wang2011}. For instance, using Fig. \ref{fig:MeasureGraph_c}, the minimum length scale $r_\mathrm{min,Solid}^\mathrm{int} = r_\mathrm{min,Void}^\mathrm{int} = 4$ elements can be imposed through $r_\mathrm{fil}=8.88$ elements and [$\PFS{\mu}{}{ero}$, $\PFS{\mu}{}{int}$, $\PFS{\mu}{}{dil}$] = [0.70, 0.50, 0.30], or through $r_\mathrm{fil}=7.27$ elements and [$\PFS{\mu}{}{ero}$, $\PFS{\mu}{}{int}$, $\PFS{\mu}{}{dil}$] = [0.80, 0.50, 0.20].

In this paper we choose the set of thresholds [$\PFS{\mu}{}{ero}$, $\PFS{\mu}{}{int}$, $\PFS{\mu}{}{dil}$] = [0.75, 0.50, 0.25], simply because the filter radius is defined by an integer, since $r_\mathrm{fil} = 2 r_\mathrm{min,Solid}^\mathrm{int} = 2 r_\mathrm{min,Void}^\mathrm{int}$. This allows to reduce rounding errors and impose the desired length scale more accurately. In addition, the minimum length scale can be easily verified from the optimized design by counting the number of elements, which is done in the following section to assess whether or not the minimum length scale is met when including maximum size constraints in the robust optimization formulation. However, it is worth mentioning that the methodology presented in the following sections can be carried out with another set of thresholds without affecting the findings and conclusions of this work. 
}

The chosen set of thresholds imposes $r_\mathrm{min,Void}^\mathrm{int} = r_\mathrm{min,Solid}^\mathrm{int}$. For comparative purposes to be discussed in the following sections, we choose 2 other sets of parameters to impose $r_\mathrm{min,Void}^\mathrm{int} \approx 2 r_\mathrm{min,Solid}^\mathrm{int}$ and $r_\mathrm{min,Void}^\mathrm{int} \approx 3 r_\mathrm{min,Solid}^\mathrm{int}$. The unknown parameters are obtained from Fig. \ref{fig:MeasureGraph_d}, which is constructed considering $\PFS{\mu}{}{ero}$=0.75, $\PFS{\mu}{}{dil}$=0.25, and $\PFS{\mu}{}{int}$ between 0.25 and 0.75. The chosen thresholds and filter radius for the desired minimum length scale are summarized in Table \ref{tab:MuRelation}.    

\begin{table}[t!]
	\caption{Thresholds and filter radius for user defined $\PFS{r}{min,Solid}{int}$ and $\PFS{r}{min,Void}{int}$. In addition, the minimum length scale in the eroded and dilated designs is provided.}
		\centering
		\begin{tabular}{c p{1.8cm} c p{1.8cm} p{1.8cm} p{1.8cm} p{1.8cm} p{1.8cm}}
				 \toprule
				 \multicolumn{2}{c}{Parameters} & & \multicolumn{5}{c}{Minimum length scale} \\
				 \cmidrule(r){1-2} \cmidrule(r){4-8}
				 $[\mu^\mathrm{ero},\: \mu^\mathrm{int},\: \mu^\mathrm{dil}]$ & $\PFS{r}{fil}{}$ & &   $\PFS{r}{min,Void}{int}$ & $\PFS{r}{min,Void}{ero}$ & $\PFS{r}{min,Solid}{dil}$ & $\PFS{t}{ero}{}$ & $\PFS{t}{dil}{}$   \\
				 \cmidrule(r){1-2} \cmidrule(r){4-8} 
				 \small{$[0.75,\: 0.50,\: 0.25]$} & {\small 2.0} $r_\mathrm{min,Solid}^\mathrm{int}$ & & {\small 1.0} $r_\mathrm{min,Solid}^\mathrm{int}$ & {\small 1.6} $r_\mathrm{min,Solid}^\mathrm{int}$ & {\small 1.6} $r_\mathrm{min,Solid}^\mathrm{int}$ & {\small 0.6} $r_\mathrm{min,Solid}^\mathrm{int}$ & {\small 0.6} $r_\mathrm{min,Solid}^\mathrm{int}$     
				 \vspace{2mm}\\
				 \small{$[0.75,\: 0.65,\: 0.25]$} & {\small 3.2} $r_\mathrm{min,Solid}^\mathrm{int}$ & & {\small 2.1} $r_\mathrm{min,Solid}^\mathrm{int}$ & {\small 2.5} $r_\mathrm{min,Solid}^\mathrm{int}$ & {\small 2.5} $r_\mathrm{min,Solid}^\mathrm{int}$ & {\small 0.4} $r_\mathrm{min,Solid}^\mathrm{int}$ & {\small 1.5} $r_\mathrm{min,Solid}^\mathrm{int}$
				 \vspace{2mm}\\
				 \small{$[0.75,\: 0.70,\: 0.25]$} & {\small 4.4} $r_\mathrm{min,Solid}^\mathrm{int}$ & & {\small 3.2} $r_\mathrm{min,Solid}^\mathrm{int}$ & {\small 3.5} $r_\mathrm{min,Solid}^\mathrm{int}$ & {\small 3.5} $r_\mathrm{min,Solid}^\mathrm{int}$ & {\small 0.3} $r_\mathrm{min,Solid}^\mathrm{int}$ & {\small 2.5} $r_\mathrm{min,Solid}^\mathrm{int}$
				 \\
				 \bottomrule
		\end{tabular}
	\label{tab:MuRelation}
\end{table}

\begin{figure}[t!]
\centering
 	\begin{subfigure}[b]{0.19\linewidth}
		\centering{}\includegraphics[width=1\linewidth]{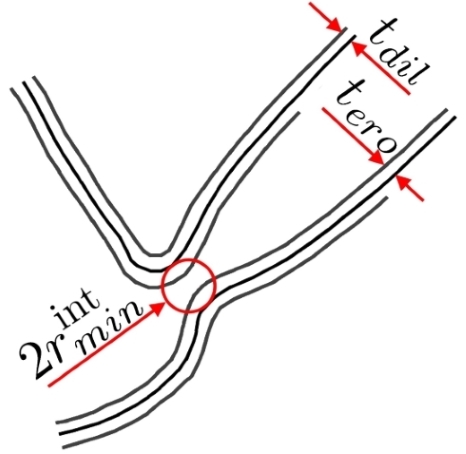}
		\caption{}
		\label{fig:Robust_F_c}
	\end{subfigure}
	~
	\begin{subfigure}[b]{0.22\linewidth}
		\centering{}\includegraphics[width=1\linewidth]{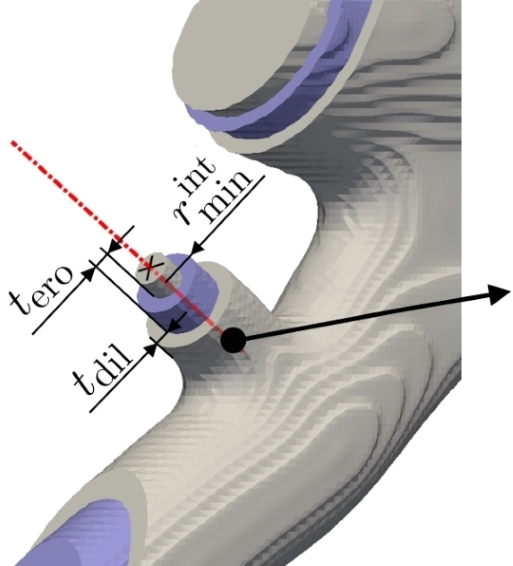}
		\caption{}
		\label{fig:Robust_F_f}
	\end{subfigure}
	~
	\begin{subfigure}[b]{0.24\linewidth}
		\centering{}\includegraphics[width=1.0\linewidth]{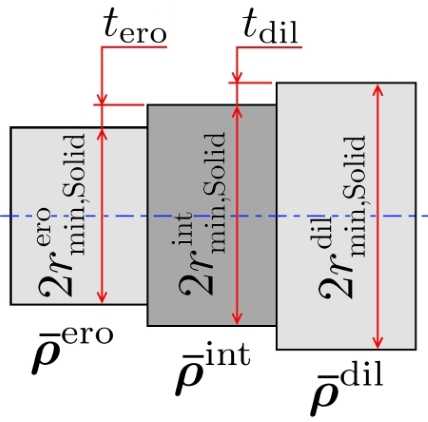}
		\caption{MinSolid}
		\label{fig:Robust_F_Fd}
	\end{subfigure}
	~ \hspace{2mm}
	\begin{subfigure}[b]{0.24\linewidth}
		\centering{}\includegraphics[width=1.0\linewidth]{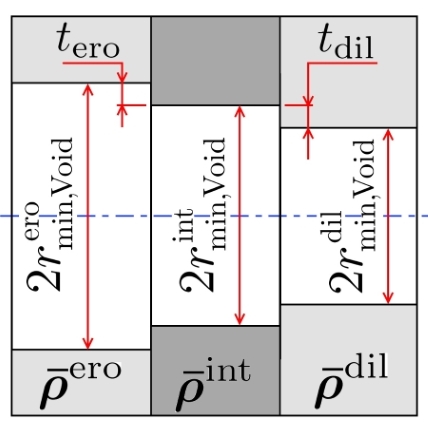}
		\caption{MinVoid}
		\label{fig:Robust_F_Fdd}
	\end{subfigure}
	\vspace{-2mm}
\caption{Geometric relationships between the eroded, intermediate and dilated designs. (a) and (b) show the labelled zones of Fig. \ref{fig:Robust_Forumulation}. (a) shows the contour lines of the 3 designs, and (b) shows the 3 designs in a cross section. (c) and (d) show sketches of the minimum length scales.}
\label{fig:Robust_Forumulation_2}
\end{figure}

To impose Minimum size in the Solid phase (MinSolid), Minimum size in the Void phase (MinVoid) and Maximum size in the Solid phase (MaxSolid), it will be necessary to consider the relations of minimum length scales between the eroded, intermediate and dilated designs. For this purpose we use the offset distances between the designs. The offset distances are defined with respect to the surface of the intermediate design. As shown in Fig.\:\ref{fig:Robust_Forumulation_2}, $t_\mathrm{ero}$ represents the offset distance between the intermediate and eroded designs, and $t_\mathrm{dil}$ represents the offset distance between the intermediate and dilated designs. These can be obtained from Figs. \ref{fig:MeasureGraph_c} and \ref{fig:MeasureGraph_d} by taking the vertical distance between the corresponding curves, as shown by the green lines in Figs. \ref{fig:MeasureGraph_c} and \ref{fig:MeasureGraph_d}. The measured values of $\PFS{t}{ero}{}$ and $\PFS{t}{dil}{}$ are summarized in Table \ref{tab:MuRelation}. The procedure for obtaining $\PFS{t}{ero}{}$ and $\PFS{t}{dil}{}$ is explained for [$\PFS{\mu}{}{ero}$, $\PFS{\mu}{}{int}$, $\PFS{\mu}{}{dil}$] = [0.75, 0.50 , 0.25] and $r_\mathrm{fil}=2r_\mathrm{min}$, as follows:
\begin{subequations}
\begin{align} 
	t_\mathrm{ero} = r_\mathrm{min,Void}^\mathrm{ero}  - r_\mathrm{min,Void}^\mathrm{int} = 0.58 \: \frac{r_\mathrm{fil}}{2}= 0.58 \: r_\mathrm{min,Solid}^\mathrm{int} \; , \\
	t_\mathrm{dil} = r_\mathrm{min,Solid}^\mathrm{dil} - r_\mathrm{min,Solid}^\mathrm{int} = 0.58 \: \frac{r_\mathrm{fil}}{2}= 0.58 \: r_\mathrm{min,Solid}^\mathrm{int} \; .
\end{align}
\end{subequations}

{\color{red}
\section{Imposing Maximum size in the Solid phase (MaxSolid)}
\label{sec:MaxSizeC}
}

{\color{red}
In this section we present the combined method to impose Minimum size in the Solid phase (MinSolid), Minimum size in the Void phase (MinVoid) and Maximum size in the Solid phase (MaxSolid). To this end, we incorporate maximum size constraints into the robust optimization problem \eqref{eq:OPTI_General}. Before detailing the method, we briefly describe the implemented maximum size constraint.
}

\vspace{3mm}
\subsection{The maximum size constraint}

The maximum size constraint used in this work is based on the formulation proposed by \citet{Guest2009}, which consists of a local volume restriction applied in a region $\Omega$ around each element $i$. Following the notation in \cite{Fernandez2019}, we define the local volume restriction as follows: 
\begin{equation} \label{eq:MS_Local_Constraint}
	g_{i} = \varepsilon - \frac{\displaystyle\sum_{j \in \Omega_i} v_j\:(1-\bar{\rho}_j)^q} {\displaystyle\sum_{j \in \Omega_i} v_j} \leq 0 \quad ,
\end{equation}
{\color{red}
\noindent were $\varepsilon$ is the minimum fraction of void to be placed within $\Omega_i$ and $q$ is a parameter that penalizes intermediate densities. Thus, by placing a cavity (of size proportional to the void fraction $\varepsilon$) inside the local region $\Omega_i$, the local restriction $g_i$ prevents the formation of solid parts larger than the prescribed size of the region $\Omega_i$.
}

\begin{figure}[t]
\centering
 	\begin{subfigure}[b]{0.252\linewidth}
		\centering{}\includegraphics[width=1\linewidth]{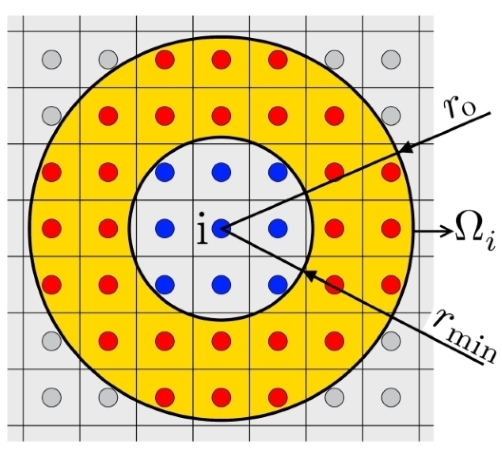}
 		\caption{}
 		\label{fig:Test_Region_MaxSize_2D}
	\end{subfigure}
	~ \hspace{5mm}
	\begin{subfigure}[b]{0.200\linewidth}
		\centering{}\includegraphics[width=1\linewidth]{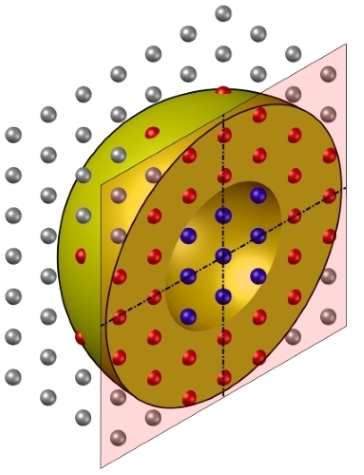}
		\caption{}
		\label{fig:Test_Region_MaxSize_3D}
	\end{subfigure}
\caption{ Local region $\Omega_i$ in (a) 2D and in (b) 3D where the maximum size constraint is applied.}
\label{fig:Test_Region_MaxSize}
\end{figure}

\begin{figure}[t!]
\centering
    \begin{tabular}{c c c c}
 	\begin{subfigure}[b]{0.22\linewidth}
		\centering{}\includegraphics[width=1\linewidth]{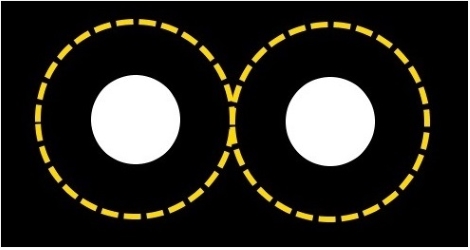}
		\caption{}
		\label{fig:MaxSize_Circular_Annular_a}
	\end{subfigure}
	&
	\begin{subfigure}[b]{0.22\linewidth}
		\centering{}\includegraphics[width=1\linewidth]{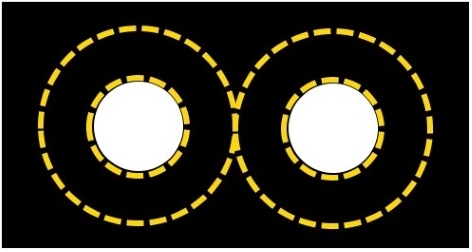}
		\caption{}
		\label{fig:MaxSize_Circular_Annular_b}
	\end{subfigure}
	&
	\begin{subfigure}[b]{0.22\linewidth}
		\centering{}\includegraphics[width=1\linewidth]{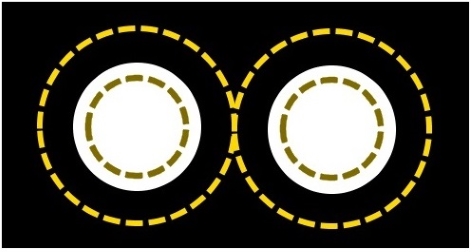}
		\caption{}
		\label{fig:MaxSize_Circular_Annular_c}
	\end{subfigure}
	&
	\begin{subfigure}[b]{0.22\linewidth}
		\centering{}\includegraphics[width=1\linewidth]{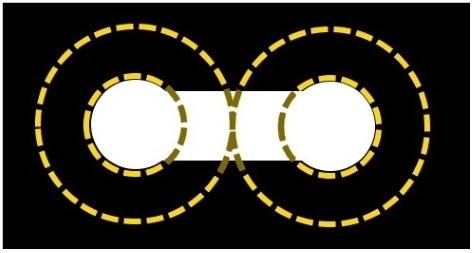}
		\caption{}
		\label{fig:MaxSize_Circular_Annular_d}
	\end{subfigure}
	\end{tabular} 	
	\vspace{-2mm}
\caption{{\color{red}An illustration of two cavities within a design. The yellow dashed lines show the contour of the local region $\Omega_i$.  In (a), the maximum size constraint is satisfied when using a circular region. In (b), the maximum size constraint is violated when using an annular region. In (c) and (d) are two possible solutions that satisfy the maximum size restriction when considering an annular region.}}
\label{fig:MaxSize_Circular_Annular}
\end{figure}

\begin{figure}[t!]
\centering
 	\begin{subfigure}[b]{0.32\linewidth}
		\centering{}\includegraphics[width=1\linewidth]{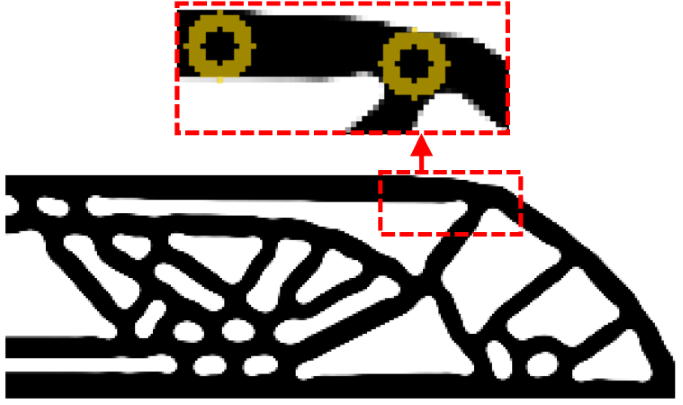}
		\caption{Annular $\Omega_i$, $\varepsilon=5\%$ and $p=100$.}
		\label{fig:F_CAI_Annular_a}
	\end{subfigure}
	~
	\begin{subfigure}[b]{0.32\linewidth}
		\centering{}\includegraphics[width=1\linewidth]{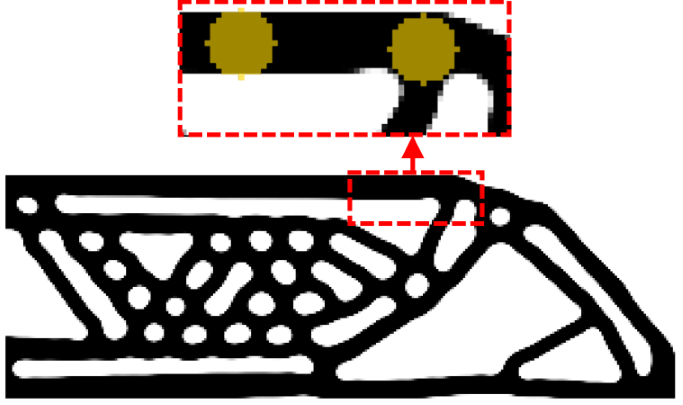}
		\caption{Circular $\Omega_i$, $\varepsilon=5\%$ and $p=100$.}
		\label{fig:F_CAI_Annular_b}
	\end{subfigure}
	~
	\begin{subfigure}[b]{0.32\linewidth}
		\centering{}\includegraphics[width=1\linewidth]{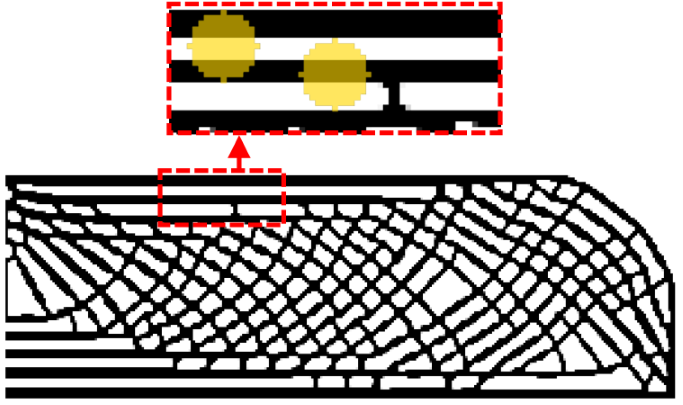}
		\caption{Circular $\Omega_i$, $\varepsilon=50\%$ and $p=10$.}
		\label{fig:F_CAI_Annular_c}
	\end{subfigure} 	
	\vspace{-2mm}
\caption{{\color{red}2D-MBB beam for compliance minimization and maximum size constraints.}}
\label{fig:F_CAI_Annular}
\end{figure}

We define the region $\Omega_i$ as a ring of outer radius $r_\mathrm{o}$ and inner radius $r_\mathrm{min}$, as shown in Fig. \ref{fig:Test_Region_MaxSize_2D}. In 3D, $\Omega_i$ is a spherical shell of thickness $r_{\mathrm{o}} - r_{\mathrm{min}}$, as depicted in Fig. \ref{fig:Test_Region_MaxSize_3D}. {\color{red} The annular region is chosen over the circular one for 3 reasons. First, an annular region and a circular region of external radius $r_\mathrm{o}$ impose the same maximum size on the solid part \cite{Fernandez2019}. Second, in comparison to a circular region, the annular one use less computer memory to store the neighborhoods $\Omega$. Third and most important, compared to a circular region, an annular one introduces less round cavities into the topology \cite{Fernandez2019}. To explain the latter reason, we make use of Fig. \ref{fig:MaxSize_Circular_Annular}, which shows a design with two circular cavities of radius $r_\mathrm{min,Void}^\mathrm{int}$, with distance between centers of $2r_\mathrm{o}$. In this configuration, the maximum size restriction is satisfied for a circular region (Fig. \ref{fig:MaxSize_Circular_Annular_a}), though it is violated for an annular one (Fig. \ref{fig:MaxSize_Circular_Annular_b}). To satisfy the constraint, among other possibilities, the optimizer can increase the size of the cavities to reach the annular region (Fig. \ref{fig:MaxSize_Circular_Annular_c}), or it can merge the two holes (Fig. \ref{fig:MaxSize_Circular_Annular_d}). In the latter case, a single channel-shaped hole is obtained. Therefore, the annular region reduces the number of closed cavities by a geometric condition that forces the round cavities to enlarge or merge. Figs. \ref{fig:F_CAI_Annular_a} and \ref{fig:F_CAI_Annular_b} show two results obtained with the strategy proposed in the following subsections, using an annular region in Fig. \ref{fig:F_CAI_Annular_a} and a circular region in Fig. \ref{fig:F_CAI_Annular_b}. It can be seen that the solution obtained with the annular region features less closed cavities than the one obtained with the circular region, which is desired in this work. For a more detailed analysis on the annular region, readers are referred to \cite{Fernandez2019}. 
}

{\color{red}
The desired Maximum size in the Solid phase (MaxSolid) is defined by a circle or sphere of radius $r_\mathrm{max}$. It is noteworthy that the maximum size $r_\mathrm{max}$ depends on the external radius $r_\mathrm{o}$ and on the void fraction $\varepsilon$. This can be observed in Fig. \ref{fig:F_CAI_Annular}. There, solutions are obtained using the same external radius $r_\mathrm{o}=6$ elements, but Figs. \ref{fig:F_CAI_Annular_a} and \ref{fig:F_CAI_Annular_b} are obtained with $\varepsilon=5\%$, while Fig. \ref{fig:F_CAI_Annular_c} with $\varepsilon=50\%$. In the same figures, a detailed view shows the $\Omega_i$ region in front of the imposed maximum size. It can be seen that for a small $\varepsilon$, the maximum size imposed on the solid phase is nearly equal to the size of the $\Omega_i$ region, meaning that $r_\mathrm{max} \approx r_\mathrm{o}$. On the contrary, a large value of $\varepsilon$ reduces significantly the Maximum size in the Solid phase, as shown in Fig. \ref{fig:F_CAI_Annular_c}. In the latter case, thin structural branches extend across the design domain, which is known as infill topology optimization \cite{Clausen2016}. Although it is true that both small and large $\varepsilon$ values impose a Maximum size in the Solid phase, using a large value $\varepsilon$ does not allow to control the Minimum size in the Solid phase (see \cite{Fernandez2019} for further details). For this reason, we use $\varepsilon=5\%$ in all the examples of this work and, as will be argued in Section \ref{sec:MinGap}, we can consider $r_\mathrm{max} \approx r_\mathrm{o}$. 
}

As the constraint $g_i$ inserts a low amount of void ($\varepsilon = 5\%$) within the region $\Omega_i$, intermediate densities must be penalized, otherwise the constraint would be easily satisfied by few elements with intermediate densities. For this reason, the void measure ($1-\bar{\rho}_i$) is penalized using a exponent $q$ in \eqref{eq:MS_Local_Constraint}. As in \cite{Fernandez2019,Guest2009}, this penalization parameter is defined as $q=\eta$. 

To reduce computational burden on the optimization problem, the $N$ local restrictions $g_i$ are replaced by a single global constraint $G_\mathrm{ms}$, which takes the value of the most critical constraint, i.e.~$G_\mathrm{ms} = \max(\bm{g})$. To resort to gradient-based optimizers,  the max function is replaced by a smooth approximation using a \textit{p-mean} function, thus the global constraint reads: 
\begin{equation} \label{eq:Gms}
 	{G}_{\mathrm{ms}}= \varepsilon - 1 + \left( \frac{1}{N} \sum_{i=1}^N (g_{i}+1-\varepsilon)^p \right)^{1/p} \leq 0\;.
\end{equation}
{\color{red}
\noindent where the term $1-\varepsilon$ is to ensure real and positives values under the $p$ root. The parameter $p$ controls the  accuracy of the aggregation function so that the larger $p$ is, the smaller the error between \textit{p-mean}($\bm{g}$) and $\max(\bm{g})$ is. However, a large value of $p$ introduces a high non-linearity that can compromise the convergence of the optimization problem \cite{Fernandez2019}. On the other hand, given that \textit{p-mean}($\bm{g}$) $<$ $\max(\bm{g})$, if the value of $p$ is too small, the global restriction $G_\mathrm{ms}$ could be satisfied ($G_\mathrm{ms} \leq 0$) even if several local restrictions are violated ($g_i>0$). Therefore, a good compromise is to use a $p$ small enough to avoid introducing a high non-linearity into the optimization problem, but large enough to capture the violated constraints. In \cite{Fernandez2019} it is shown that the above compromise is influenced by the value of $\varepsilon$, such that the smaller the value of $\varepsilon$, the larger the value of $p$ is. For example, solution in Fig. \ref{fig:F_CAI_Annular_c} is obtained with $\varepsilon=50\%$ and $p=10$, whilst Figs. \ref{fig:F_CAI_Annular_a} and \ref{fig:F_CAI_Annular_b} are obtained with $\varepsilon=5\%$ and $p=100$. From now on, in all the examples where MaxSolid is imposed we use $\varepsilon=5\%$ and $p=100$. 
}

The high non-linearity introduced by the aggregation process causes oscillations that compromise the convergence of the optimization process \cite{Fernandez2019}. Here we adopt two common strategies to reduce oscillations: a continuation method \cite{Sigmund2013,Rojas2015} and moving limits on designs variables \cite{Yang1996,Fernandez2019}. In the continuation method, the SIMP penalty $\eta$ is initialized at $1.0$ and is increased in increments of $0.25$ to a maximum of $3.0$. The Heaviside parameter $\beta$ is initialized at $1.5$ and is increased by a factor $1.5$ to a maximum of $38$. The continuation procedure includes $40$ iterations between parameters increment. The stopping criteria of the complete topology optimization process is met either when the maximum number of iterations of the continuation method is reached, or when the maximum change in the design variables between two consecutive iterations is smaller than $0.001$. To impose move limits, the maximum change of the design variables is controlled as follows:
\begin{equation}
\mathrm{max}(0,\rho_i-m_\mathrm{l}) \leq \rho_i \leq \mathrm{\mathrm{min}}(1,\rho_i+m_\mathrm{l})
\end{equation}
\noindent where $m_\mathrm{l}$ is the move limit value. As the non-linearity of the optimization problem rises as $\eta$ and $\beta$ increase, oscillations intensify as the optimization progresses. For this reason, the maximum allowable change of design variables is decreased as $\eta$ and $\beta$ increase. At the beginning of the optimization process $m_\mathrm{l}=0.5$ and at the end $m_\mathrm{l}=0.05$. A linear interpolation function defines $m_\mathrm{l}$ during the optimization as follows:
\begin{equation} \label{eq:move_limit}
	m_\mathrm{l} = \frac{0.5-0.05}{3.0-1.0} (3.0-\eta)+0.05 \;.
\end{equation}

\clearpage
\subsubsection{Adjustment of the maximum size constraint due to boundary effects}
\label{sec:MaxSizeC.Adjustment}
\begin{figure}[t!]
\centering
 	\begin{subfigure}[b]{0.30\linewidth}
		\centering{}\includegraphics[width=1\linewidth]{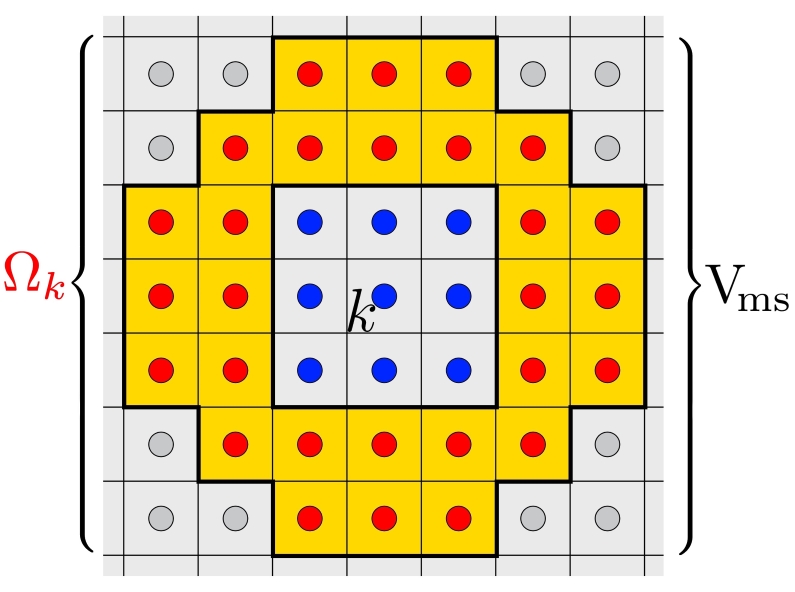}
 		\caption{}
 		\label{fig:Test_Region_MaxSize_2D_2}
	\end{subfigure}
	~ \hspace{0mm}
	\begin{subfigure}[b]{0.3\linewidth}
		\centering{}\includegraphics[width=1\linewidth]{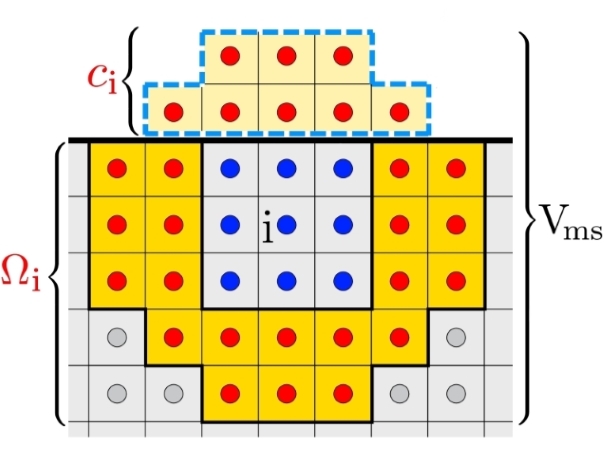}
		\caption{}
		\label{fig:Test_Region_MaxSize_Symmetry}
	\end{subfigure}
	~ \hspace{0mm}
	\begin{subfigure}[b]{0.3\linewidth}
		\centering{}\includegraphics[width=1\linewidth]{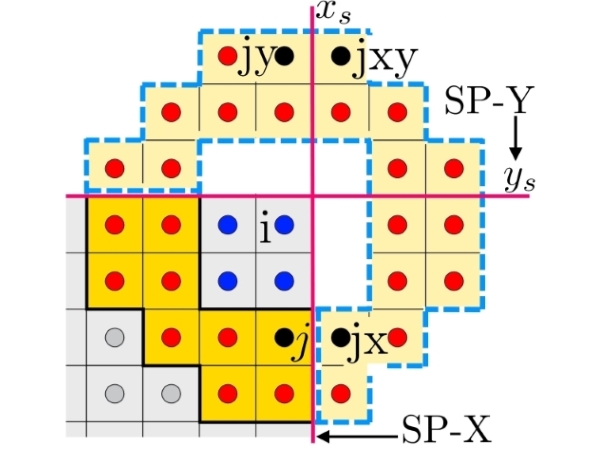}
		\caption{}
		\label{fig:Test_Region_MaxSize_Discrete}
	\end{subfigure}
\caption{In (a) is a discretized local region $\Omega_k$ that does not reach the borders of the design domain . In (b), the $\Omega_i$ region reaches the boundary of the design domain. In (c), the $\Omega_i$ region reaches to two planes of symmetry.}
\label{fig:Test_Region_MaxSize_2}
\end{figure}

{\color{red}
Same as the filtering process, the maximum size constraint \eqref{eq:MS_Local_Constraint} must be adapted at the boundaries of the design domain, otherwise boundary effects would influence the imposed maximum size, e.g.~by reducing MaxSolid by half at the borders of the design space. To remedy this, we simulate an extension of the finite element mesh beyond the boundaries of the design space. For instance, if the mesh were extended beyond the design space, the denominator in \eqref{eq:MS_Local_Constraint} would be the same for all constraints $g_i$ within the design space. This denominator is denoted as $\mathrm{V_{ms}}$ and calculated as follows:
}
{\color{red}
\begin{equation}
\mathrm{V}_\mathrm{ms} = \displaystyle \sum_{j \in \Omega_k} v_j \; , 
\end{equation}
}
{\color{red}
\noindent where $k$ is the index of an element whose region $\Omega_k$ does not reach the boundaries of the design domain, as shown in Fig.~\ref{fig:Test_Region_MaxSize_2D_2}. That means, $\mathrm{V}_\mathrm{ms}$ represents the volume of the region $\Omega_k$. 
}

For practical reasons, we define the local region $\Omega_i$ using a weighting factor $l\left(X_i , X_j\right)$ as follows:
\begin{equation} \label{eq:DII_lij}
 	l \left(X_i , X_j\right)=
 	\left\{    \begin{matrix}
 										1 \;\; , & \text{if} \;\; r_{\mathrm{min}} \leq \lvert \rvert {X_i-X_j} \lvert \rvert \leq r_{\mathrm{o}} \vspace{3mm} \\ 
 										0 \;\; , & \text{otherwise}
 	           \end{matrix}
 	\right.
\end{equation} 

To take into account the planes of symmetry, we define as $\hat{l}_{i,j}$ the $\Omega_i$ region that considers the reflection of $j$ with respect to the planes of symmetry. Therefore, for an element $i$ whose region $\Omega_i$ does not reach symmetry planes, $\hat{l}_{i,j} = l \left(X_i , X_j\right)$. However, if the region $\Omega_i$ reaches the symmetry planes SP-X and SP-Y, as in Fig. \ref{fig:F_CAI_Annular_c}, the total weight $\hat{l}_{i,j}$ is as follows:
\begin{equation} \label{eq:DII_lij_tot}
	\hat{l}_{i,j} = {l} \left(X_i , X_j               \right) + 
	                {l} \left(X_i , X_{j \mathrm{x}}  \right) + 
	                {l} \left(X_i , X_{j \mathrm{y}}  \right) + 
	                {l} \left(X_i , X_{j \mathrm{xy}} \right) \;.
\end{equation}

{\color{red}
As the maximum size constraint $g_i$ measures the fraction of void within the region $\Omega_i$, it is necessary to consider the void elements that are outside the design domain in order to simulate an extension of the mesh, i.e.~the fraction of void that would be gained by applying an extension of the mesh. For this, we define as $c_i$ the fraction of $\Omega_i$ that remains outside the design domain and does not reach planes of symmetry, as shown in Fig. \ref{fig:Test_Region_MaxSize_Symmetry}. Thus, the additional void fraction resulting from an extension of the mesh can be calculated as:
}
\begin{equation} \label{eq:ci_factor}
	c_{i} = 1 - \frac{\displaystyle\sum_{j=1}^N \: v_j \: \hat{l}_{i,j}} {V_{\mathrm{ms}}} \quad .
\end{equation}

Then, to emulate the extension of the mesh considering the planes of symmetry, the maximum size restriction is calculated as follows:
\begin{equation} \label{eq:MS_Local_Constraint_corrected}
	g_{i} = \varepsilon - c_i - \frac{\displaystyle\sum_{j=1}^N \hat{l}_{i,j} \: v_j \: (1-\bar{\rho}_j)^q} {V_{\mathrm{ms}}} \leq 0 \quad ,
\end{equation}

The constraint $g_i$ must be evaluated at every element, therefore it is convenient to vectorize the computation to avoid loops. To this end, the constant parameters $\hat{l}_{i,j}$, $V_{\mathrm{ms}}$ and $v_j$ are assembled in a matrix $\mathrm{\mathbf{D}}_{\mathrm{II}}$ as follows:
\begin{equation} \label{eq:DII}
	\mathrm{D_{II}}_{(i,j)} = \frac{v_j\:l_{i,j}}{V_{\mathrm{ms}}} \;,
\end{equation}
\noindent and the $N$ local constraints $g_i$ can be obtained as:
\begin{equation} \label{eq:gms}
 	\bm{g}= \bm{\varepsilon} - \bm{c} - \mathrm{\mathbf{D}}_{\mathrm{II}} \bm{\delta} \;,
\end{equation}
\noindent where $\bm\delta$ is the array that contains a measure of void for each element $i$, calculated as:
\begin{equation} \label{eq:delta}
 	\delta_i = (1-\bar\rho_i)^q \;.
\end{equation}

Considering column arrays and denominator layout \cite{Xing2019}, the sensitivities of the global constraint $G_{\mathrm{ms}}$ are obtained by applying the chain rule as:
\begin{equation} \label{eq:chainSensitivity}
	\frac{\mathrm{d} G_{\mathrm{ms}}}{\mathrm{d}\bm{\rho}} = \frac{\mathrm{d}\bm{\tilde{\rho}}}{\mathrm{d}\bm{\rho}} 
							\: \frac{\mathrm{d}\bm{\bar{\rho}}}{\mathrm{d}\bm{\tilde{\rho}}} 
	                      	\: \frac{\mathrm{d}\bm{\delta}}{\mathrm{d}\bm{\bar{\rho}}} 
	                     	\: \frac{\mathrm{d}\bm{g}}{\mathrm{d}\bm{\delta}} 
	                      	\: \frac{\mathrm{d}G_{\mathrm{ms}}}{\mathrm{d}\bm{g}}
	                      	=
	                      	-\mathrm{\mathbf{D}_I^{\intercal}} 
	                      	\: \mathrm{\mathbf{J}}_{\bar{\rho}}
	                      	\: \mathrm{\mathbf{J}}_{\delta}
	                      	\: \mathrm{\mathbf{D}_{II}^{\intercal}} 
	                      	\: \frac{(\bm{g}+\bm{1}-\bm{\varepsilon})^{(p-1)}}{N}
	                      	\: \left( \frac{1}{N} \sum_{i=1}^N ({g}_{i}+{1}-{\varepsilon})^p \right)^{1/p-1} \;.
\end{equation}
\noindent where $\mathrm{\mathbf{J}}_{\delta}$ is the Jacobian matrix of the voids vector with penalization defined as $\mathrm{\mathbf{J}}_{\delta}=\mathrm{diag}(\delta'(\bar{\rho}_1),...,\delta'(\bar{\rho}_N))$. Here, $\delta'(\bar{\rho}_i)$ is the derivative of $\delta$ with respect to $\bar{\rho}_i$.

\subsection{Minimum and Maximum length scale control (MinSolid, MinVoid and MaxSolid)}
\label{sec:Robust}

{\color{red}
Aiming to impose minimum and maximum length scale control (MinSolid, MinVoid and MaxSolid), the maximum size restriction $G_\mathrm{ms}$ is introduced in the robust formulation based on the eroded, intermediate and dilated design. Since the intermediate design is the one intended for manufacturing, the desired maximum size is defined by the prescribed radius $\PFS{r}{max}{int}$ of a circle or sphere. We recall that the maximum size restriction $g_i$ is imposed using an annular $\Omega_i$ region of external radius $r_\mathrm{o}$ and with a void fraction $\varepsilon=5\%$.
}

{\color{red}
As the manufacturing error resulting in the eroded, intermediate and dilated designs is uniform throughout the design, the offset distances between designs are considered constant all over the structural surface. Therefore, if a solid part features a maximum size $\PFS{r}{max}{int}$ in the intermediate design, the same part in the eroded and dilated designs will feature a maximum size: 
}
\begin{subequations} \label{eq:MaxRelation}
\begin{align}
	r_\mathrm{max}^\mathrm{ero} &  = r_\mathrm{max}^\mathrm{int} - t_\mathrm{ero} \; ,\\
	r_\mathrm{max}^\mathrm{dil} &  = r_\mathrm{max}^\mathrm{int} + t_\mathrm{dil} \; . 
\end{align}
\end{subequations}

{\color{red}
These relations of maximum length scales between the eroded, intermediate and dilated designs points out that it is theoretically possible to impose a maximum size $r_\mathrm{max}^\mathrm{int}$ through the eroded and dilated designs, as long as the $\Omega_i$ regions are scaled accordingly. For example, a maximum size constraint applied in the intermediate design $g_i(\Omega_i^\mathrm{int})$ may be geometrically equivalent to a constraint applied in the dilated design $g_i(\Omega_i^\mathrm{dil})$ and to a constraint applied in the eroded design $g_i(\Omega_i^\mathrm{ero})$, if the regions $\Omega_i^\mathrm{dil}$ and $\Omega_i^\mathrm{ero}$ are scaled according to the offset distances. However, to check whether the three constraints $g_i(\Omega_i^\mathrm{ero})$, $g_i(\Omega_i^\mathrm{int})$ and $g_i(\Omega_i^\mathrm{dil})$ are geometrically compatible and to observe the effect of these on the topology, we solve a set of compliance minimization problems with variations in the field(s) on which the maximum size constraint acts. In the general case, the optimization problem is given as follows:  
}
{\color{red}
\begin{align} \label{eq:OPTI_MaxSize_Robust}
	\begin{split}
  		{\min_{\rho}} & \quad \bm{f}^\intercal \bm{u}^\mathrm{ero} 	\\
	  	{s.t.:} & \quad  \bm{v}^\intercal \bm{\bar{\rho}}^\mathrm{dil} \leq V^*_\mathrm{dil}	\\  		
	  		& \quad G_\mathrm{ms} ( \Omega^\mathrm{ero} ) \leq 0 \vspace{1mm}\\ 
	  		& \quad G_\mathrm{ms} ( \Omega^\mathrm{int} ) \leq 0 \vspace{1mm} \\ 
	  		& \quad G_\mathrm{ms} ( \Omega^\mathrm{dil} ) \leq 0 \\
			& \quad 0 \leq \rho_i \leq1
	\end{split}
\end{align}
}

{\color{red}
The 2D-MBB beam with a volume constraint of $50\%$ is chosen as test case. The minimum length scale (MinSolid and MinVoid) is set to  $\PFS{r}{min,Solid}{int} = \PFS{r}{min,Void}{int} = 3$ elements. The Maximum size in the Solid phase (MaxSolid) is set to $\PFS{r}{max}{int}=5$ elements. The minimum length scale is indirectly imposed through the set of thresholds [0.75, 0.50, 0.25] and filter radius $r_\mathrm{fil}=10$ elements. According to Table \ref{tab:MuRelation}, the offset distances are $t_\mathrm{ero}=t_\mathrm{dil}=1.8$ elements. Therefore, the regions $\Omega^\mathrm{ero}$, $\Omega^\mathrm{int}$ and $\Omega^\mathrm{dil}$ are respectively defined by the outer radii $\PFS{r}{o}{ero}=3.2$, $\PFS{r}{o}{int}=5$ and $\PFS{r}{o}{dil}=6.8$ elements, and by the inner radii $\PFS{r}{min,Void}{ero}=1.2$, $\PFS{r}{min,Void}{int}=3$ and $\PFS{r}{min,Void}{dil}=4.8$ elements.      
}

\begin{figure}[t!]
\vspace{4mm}
\centering
	\begin{subfigure}[b]{0.18\linewidth}
		\centering{}\includegraphics[width=1\linewidth]{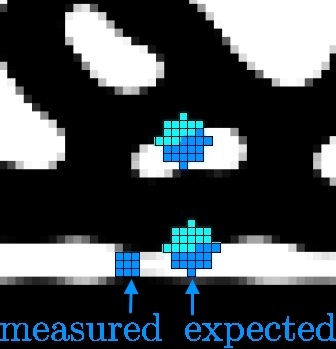}
		\caption{Zone A}
		\label{fig:Robust_Details_a}
	\end{subfigure}
	~\hspace{6mm}
	\begin{subfigure}[b]{0.18\linewidth}
		\centering{}\includegraphics[width=1\linewidth]{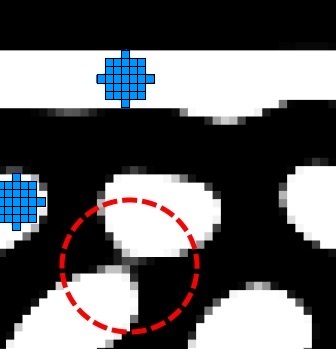}
		\caption{Zone B}
		\label{fig:Robust_Details_b}
	\end{subfigure}
	~\hspace{6mm}
	\begin{subfigure}[b]{0.18\linewidth}
		\centering{}\includegraphics[width=1\linewidth]{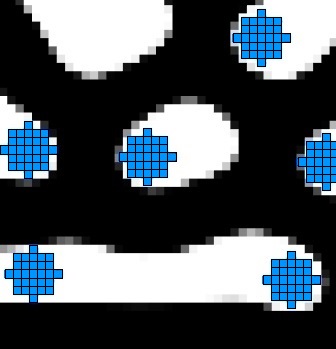}
		\caption{Zone C}
		\label{fig:Robust_Details_c}
	\end{subfigure}
	~\hspace{6mm}
	\begin{subfigure}[b]{0.18\linewidth}
		\centering{}\includegraphics[width=1\linewidth]{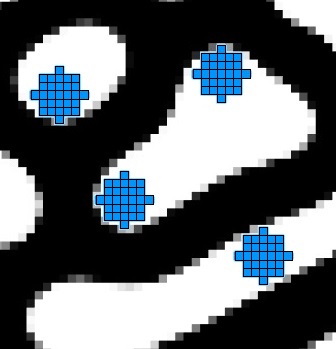}
		\caption{Zone D}
		\label{fig:Robust_Details_d}
	\end{subfigure}
	\vspace{-2mm}
\caption{Labelled areas of Table \ref{tab:Effect_MaxSize_Robust}. Blue circles represent the theoretical minimum size of the void phase. In (a), the minimum size region changes color when touching solid elements.}
\label{fig:Robust_Details}
\end{figure}

Table \ref{tab:Effect_MaxSize_Robust} summarizes the results. The first column specifies the design on which the maximum size restriction is applied and the value of the objective function. The eroded, intermediate and dilated designs are reported in columns 2, 3 and 4, respectively. Some blueprints include labelled areas which are shown in detail in Fig.~\ref{fig:Robust_Details}. In addition, the intended length scale is graphically reported next to the blueprint designs. A black circle {\color{black}\textbullet} represents the desired Minimum size in the Solid phase. A blue circle {\color{brandeisblue}\textbullet} represents the desired Minimum size of the Void phase, and a black ring {\small \CircPipe} represents the maximum size region $\Omega$. From Table 3, the following observations are drawn:

\begin{table}[t!]
	\caption{Results for compliance minimization with maximum size constraints applied in the different fields. A black ring {\small \CircPipe} next to the field $\bm{\bar{\rho}}$ represents the size of the test region $\Omega$. The objective value is given in parentheses. } 
		\begin{tabular}{c c c c}
				 \toprule
				 {\footnotesize Constraint} & $\bm{\bar{\rho}}^{\mathrm{ero}}$ & $\bm{\bar{\rho}}^{\mathrm{int}}$ (Blueprint) & $\bm{\bar{\rho}}^{\mathrm{dil}}$   \\
				 \cmidrule(r){1-4}
				 & 
				 \multirow{4}{*}{\includegraphics[width=0.265\linewidth]{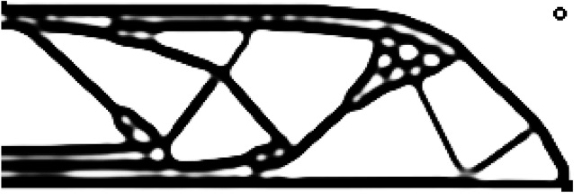} }
				 & 
				 \multirow{4}{*}[0 mm]{\includegraphics[width=0.265\linewidth]{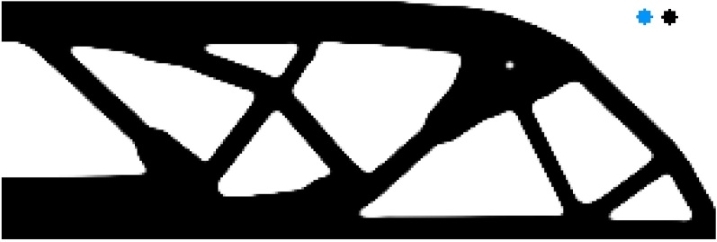} }
				 &
				 \multirow{4}{*}[0 mm]{\includegraphics[width=0.265\linewidth]{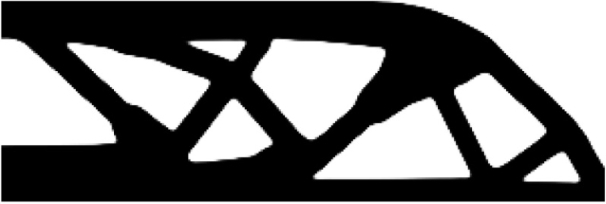} }
				 \\				 
				 \footnotesize $G_{\mathrm{ms}}(\Omega^{\mathrm{ero}})$
				 \\
				 \footnotesize $(21.324)$
				 \\ \\
				 \cmidrule(r){1-4}
				 & 
				 \multirow{4}{*}[0 mm]{\includegraphics[width=0.265\linewidth]{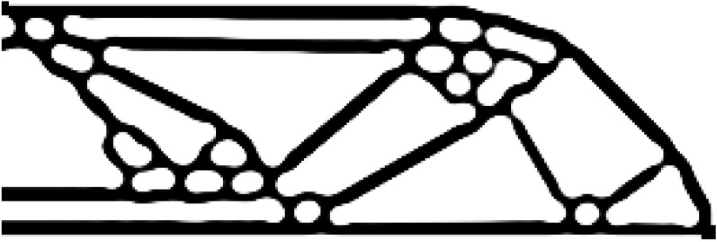}}
				 &
				 \multirow{4}{*}[0 mm]{\includegraphics[width=0.265\linewidth]{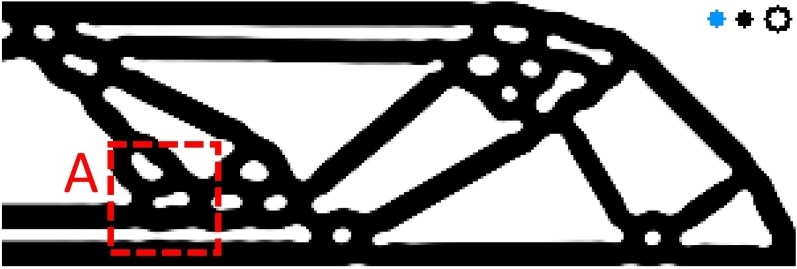}}
				 & 
				 \multirow{4}{*}[0 mm]{\includegraphics[width=0.265\linewidth]{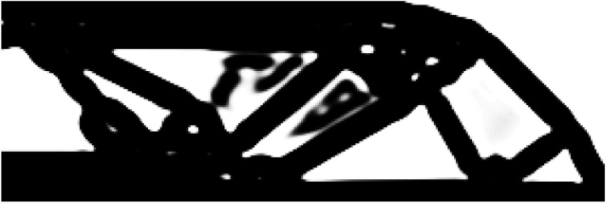}}
				 \\ 
				 \footnotesize $G_{\mathrm{ms}}(\Omega^{\mathrm{int}})$
				 \\ 
				 \footnotesize $(26.729)$				 
				 \\ \\
				 \cmidrule(r){1-4}
				 & 
				 \multirow{4}{*}[0 mm]{\includegraphics[width=0.265\linewidth]{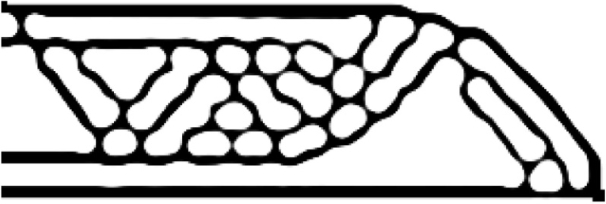}}
				 & 
				 \multirow{4}{*}[0 mm]{\includegraphics[width=0.265\linewidth]{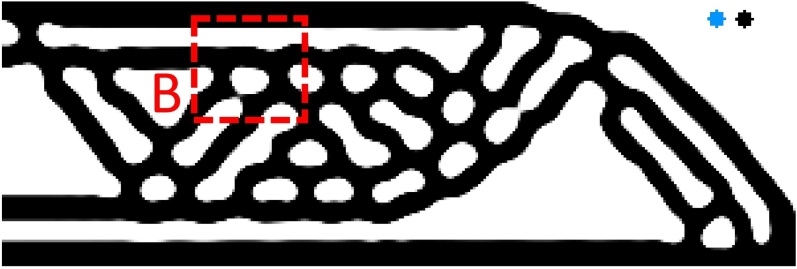}}
				 &
				 \multirow{4}{*}[0 mm]{\includegraphics[width=0.265\linewidth]{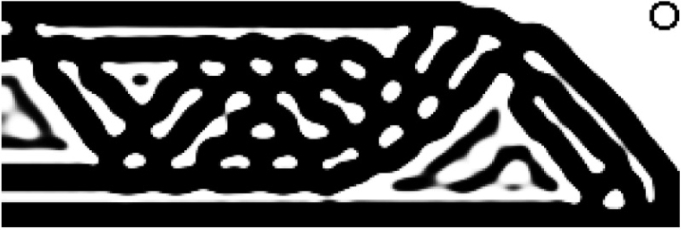}}
				 \\ 
				 \footnotesize $G_{\mathrm{ms}}(\Omega^{\mathrm{dil}})$
				 \\
				 \footnotesize $(40.259)$
				 \\ \\
				 \cmidrule(r){1-4}
				 \footnotesize  $G_{\mathrm{ms}}(\Omega^{\mathrm{ero}}) $
				 & 
				 \multirow{5}{*}[0 mm]{\vspace{3mm}\includegraphics[width=0.265\linewidth]{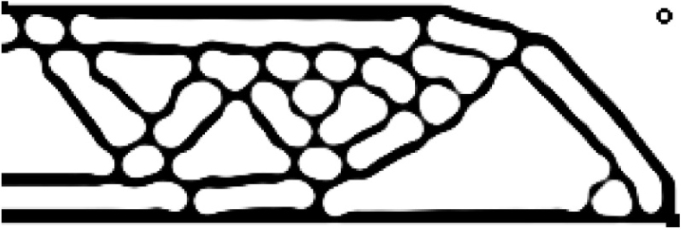}}
				 &  
				 \multirow{5}{*}[0 mm]{\vspace{3mm}\includegraphics[width=0.265\linewidth]{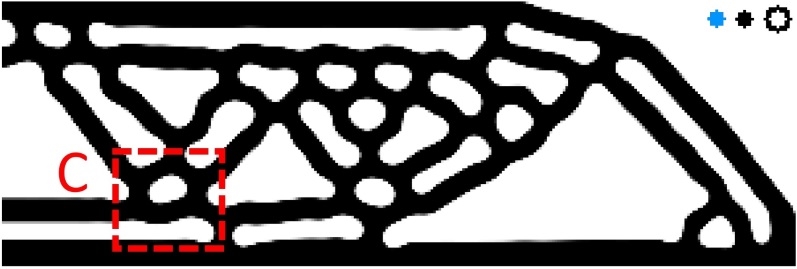}}
				 & 
				 \multirow{5}{*}[0 mm]{\vspace{3mm}\includegraphics[width=0.265\linewidth]{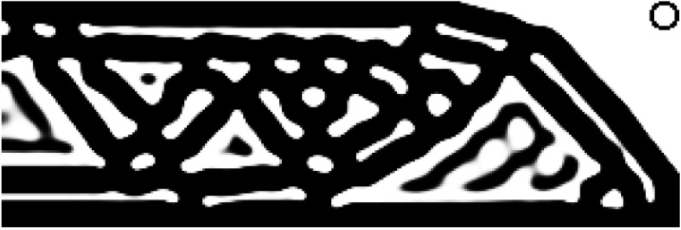}}
  			 	 \\ 
  			 	 \footnotesize $G_{\mathrm{ms}}(\Omega^{\mathrm{int}}) $
  			 	 \\
  			 	 \footnotesize $G_{\mathrm{ms}}(\Omega^{\mathrm{dil}}) $
  			 	 \\
  			 	 \footnotesize $(34.460)$
  			 	 \\ 
				 \cmidrule(r){1-4}
				 & 
				 \multirow{5}{*}[0 mm]{\vspace{3mm}\includegraphics[width=0.265\linewidth]{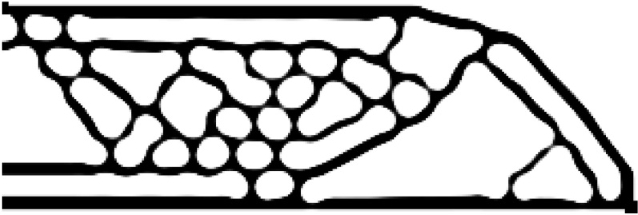}}
				 &  
				 \multirow{5}{*}[0 mm]{\vspace{3mm}\includegraphics[width=0.265\linewidth]{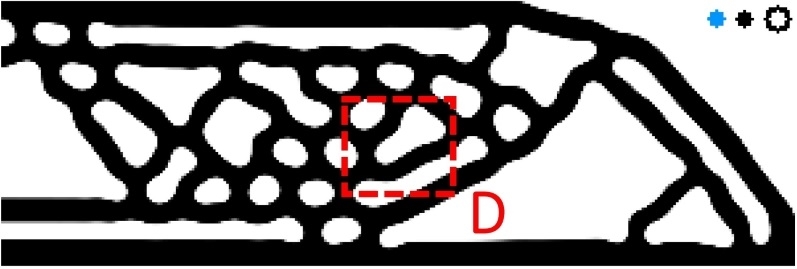}}
				 & 
				 \multirow{5}{*}[0 mm]{\vspace{3mm}\includegraphics[width=0.265\linewidth]{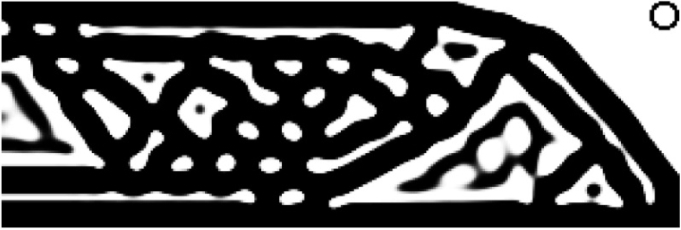}}
  			 	 \\ 
  			 	 \footnotesize $G_{\mathrm{ms}}(\Omega^{\mathrm{int}}) $
  			 	 \\
  			 	 \footnotesize $G_{\mathrm{ms}}(\Omega^{\mathrm{dil}}) $
  			 	 \\
  			 	 \footnotesize  $(35.209) $
  			 	 \\ \vspace{-2mm} \\
				 \bottomrule
		\end{tabular}
	\label{tab:Effect_MaxSize_Robust}
\end{table}

\begin{enumerate}[label=(\roman*)]
{\color{red}
\item By including only the constraint $\PFS{G}{ms}{}(\Omega^\mathrm{ero})$ in the robust optimization problem, the maximum size $\PFS{r}{max}{int}$ in the intermediate design is not imposed. Furthermore, the small cavity in the intermediate design suggests that there is no control over the Minimum size in the Void phase (MinVoid). 
}

{\color{red}
\item By including only the constraint $\PFS{G}{ms}{}(\Omega^\mathrm{int})$ in the robust optimization problem, a maximum size is imposed on the eroded and intermediate designs. The maximum length scale even coincides with the theoretical values of $\PFS{r}{max}{ero}$ and $\PFS{r}{max}{int}$. However, the desired Minimum size in the Void phase (MinVoid) is not reached, since cavities in the intermediate design are smaller than expected, as seen in Fig.~\ref{fig:Robust_Details_a}. These cavities are not present in the dilated design, therefore the dilation projection closes them, which means, the minimum width of the cavities in the intermediate design is less than $2t_\mathrm{dil}$. Interestingly, $2t_\mathrm{dil} = 3.6$ elements, which is rather close to the measured size, as can be seen in Fig. \ref{fig:Robust_Details_a}. 
}

{\color{red}
\item By including only the constraint $\PFS{G}{ms}{}(\Omega^\mathrm{dil})$ in the robust optimization problem, the Maximum size in the Solid phase (MaxSolid) is imposed on all 3 designs and matches the predicted values $\PFS{r}{max}{ero}$, $\PFS{r}{max}{int}$ and $\PFS{r}{max}{dil}$. In addition, it can be seen in Fig. \ref{fig:Robust_Details_b} that the Minimum size in the Void phase (MinVoid) is achieved. Therefore, in order to impose MinVoid and MaxSolid in the intermediate design, the maximum size restriction must be applied at least in the dilated design. The reason can be explained by considering the following. {\textit{The intermediate design can be seen as an erosion of the dilated design. Thus, if a small cavity is present in the dilated design, the intermediate design will project the cavity larger}}. We recall that the same idea was applied to construct the graph of Fig. \ref{fig:MeasureGraph_b}. However, from Fig. \ref{fig:Robust_Details_b} it can be seen that there are small hinges, which suggests that the Minimum size in the Solid phase (MinSolid) is not guaranteed.  
}

\item Including the three restrictions in the optimization problem results in a design that satisfies the desired length scale (MinSolid, MinVoid and MaxSolid), as can be seen in \ref{fig:Robust_Details_c}. So, although $G_\mathrm{ms}(\Omega^\mathrm{dil})$ alone imposes a maximum size on all three designs, the fact of restricting all three designs separately has a positive effect on avoiding the presence of small hinges. Probably, this is due to the fact that the presence of a small hinge in the intermediate design slightly thickens the structure. Therefore, $G_\mathrm{ms}(\Omega^\mathrm{int})$ could avoid or eliminate the small hinges by avoiding thickening. If that were the case, it would be enough to add $G_\mathrm{ms}(\Omega^\mathrm{dil})$ and $G_\mathrm{ms}(\Omega^\mathrm{int})$ in the optimization problem, as can be seen in the last row of Table \ref{tab:Effect_MaxSize_Robust}. However, our experience is that the three restrictions collectively lead to a better optimum. For this reason, in this work we apply the maximum size restriction on all 3 designs whenever possible, as in equation \eqref{eq:OPTI_MaxSize_Robust}.
\end{enumerate}

The observations (i)-(iv) are based on the results of a single test case, therefore they cannot be generalized. However, the results are sufficient to understand that at least the dilated design should be restricted consistently with the desired maximum size $\PFS{r}{max}{int}$, {\color{red} as this ensures the introduction of small cavities in the dilated design, which are projected with size $\PFS{r}{min,Void}{int}$ in the intermediate design.} 

To corroborate the above observations and verify that the theoretical length scale can be imposed using a different set of thresholds, we now solve a topology optimization problem considering the Minimum size in the Solid phase (MinSolid) different from the Minimum size in the Void phase (MinVoid). For this, the minimum and maximum size of solid phase (MinSolid and MaxSolid) are defined as in the previous example, but now the desired MinVoid is, at least, 2 and 3 times larger than MinSolid. This is achieved with the sets of thresholds [0.75,0.65,0.25] and [0.75,0.70,0.25], as shown in Table \ref{tab:MuRelation}. Since $t_\mathrm{ero}$ results in a number smaller than the size of a finite element, only $G_\mathrm{ms}(\Omega^\mathrm{dil})$ and $G_\mathrm{ms}(\Omega^\mathrm{int})$ are included in the optimization problem. The results for $\PFS{r}{min,Void}{int}=3.2 \PFS{r}{min,Solid}{int}$ and $\PFS{r}{min,Void}{int}= 2.1 \PFS{r}{min,Solid}{int}$ are shown in the first and second rows of Table \ref{tab:MaxSize_Robust_065_070}, respectively. There, the dilated design and a detailed view of the cavities size are shown next to the intermediate design. 

Table \ref{tab:MaxSize_Robust_065_070} shows that minimum and maximum length scales (MinSolid, MinVoid and MaxSolid) are achieved for $\PFS{r}{min,Void}{int}= 2.1 \PFS{r}{min,Solid}{int}$, which confirms that that the relations of lengths scales based on the offset distances is correct. However, it can be seen that for $\PFS{r}{min,Void}{int}= 3.2 \PFS{r}{min,Solid}{int}$ there is no precise control of the Minimum size in the Void phase (MinVoid), as the radius of curvature at the re-entrant corners is less than the desired $\PFS{r}{min,Void}{int}$. This issue is mainly due to the fact that the manufacturing error imposed on the eroded design is too small, since $\mu^\mathrm{ero}-\mu^\mathrm{int}=0.05$. This reduces robustness of the approach based on manufacturing errors, and therefore the control over the MinVoid is not guaranteed.

\begin{table}[t!]
	\caption{2D-MBB beam for compliance minimization with maximum size constraints. The intended minimum length scale and the utilized thresholds $[\mu^\mathrm{ero},\mu^\mathrm{int},\mu^\mathrm{dil}]$ are noted under each solution.}
\centering		\begin{tabular}{c c c}
				 \toprule
				 Detail & $\bm{\bar{\rho}}^{\mathrm{int}}$ (Blueprint) & $\bm{\bar{\rho}}^{\mathrm{dil}}$   \\
				 \cmidrule(r){1-3}
				 {\includegraphics[width=0.19\linewidth]{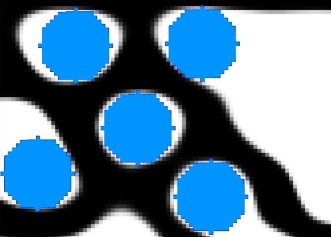} }
				 & 
				 {\includegraphics[width=0.32\linewidth]{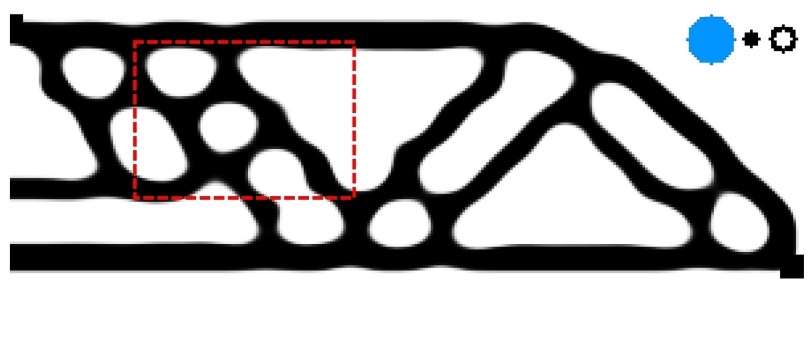} }
				 &
				 {\includegraphics[width=0.32\linewidth]{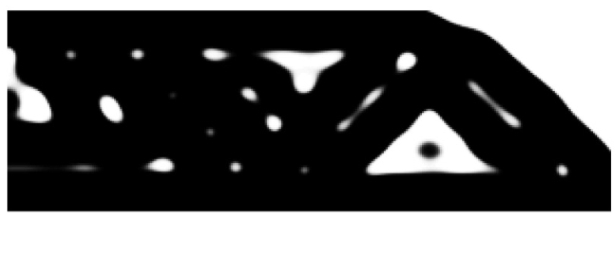} }
				 \vspace{-4mm}\\
				 & \multicolumn{2}{l}{\small{\hspace{5mm} $\PFS{r}{min,Void}{int} = 3.2 \PFS{r}{min,Solid}{int}$ \; , \quad$[0.75,0.70,0.25]$} }
				 \\
				 \cmidrule(r){1-3} 
				 {\includegraphics[width=0.19\linewidth]{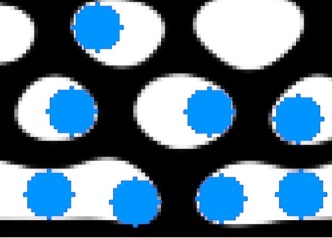}}
				 & 
				 {\includegraphics[width=0.32\linewidth]{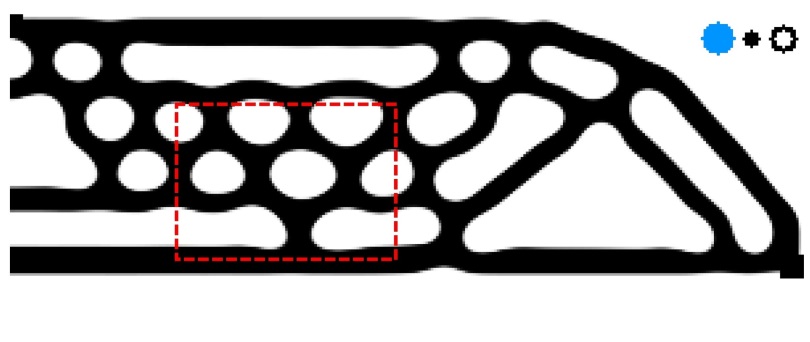}}
				 &
				 {\includegraphics[width=0.32\linewidth]{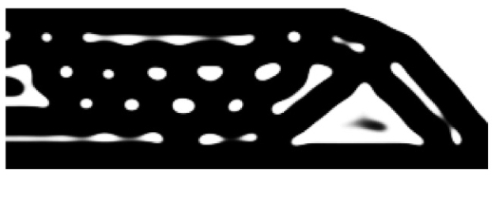}}
				 \vspace{-4mm}\\
				 & \multicolumn{2}{l}{\small{\hspace{5mm} $\PFS{r}{min,Void}{int} = 2.1 \PFS{r}{min,Solid}{int}$ \; , \quad$[0.75,0.65,0.25]$} }
				 \\				 
				 \cmidrule(r){1-3} 
				 {\includegraphics[width=0.19\linewidth]{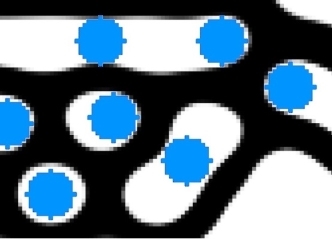}}
				 & 
				 {\includegraphics[width=0.32\linewidth]{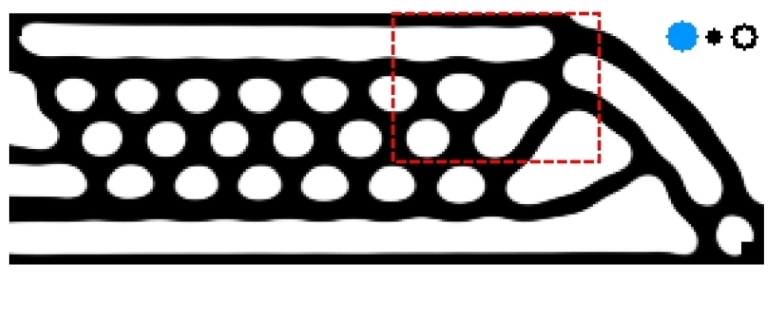}}
				 &
				 {\includegraphics[width=0.32\linewidth]{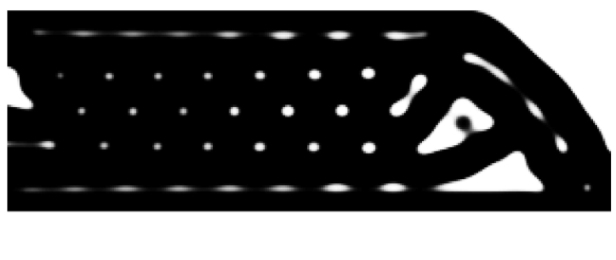}}
				 \vspace{-4mm}\\
				 & \multicolumn{2}{l}{\small{\hspace{5mm} $\PFS{r}{min,Void}{int} = 2.1 \PFS{r}{min,Solid}{int}$ \; , \quad $[0.75,0.65,0.25]$} \;, \; (Without boundary treatment)}						 
				 \\
				 \bottomrule
		\end{tabular}
	\label{tab:MaxSize_Robust_065_070}
\end{table}     

To show that the treatment of the filter and of the maximum size restriction at the boundaries of the design space is effective, we include in the third row of Table \ref{tab:MaxSize_Robust_065_070} a solution obtained without any treatment. That is, the filtering of the design variables is done as in \eqref{eq:FilterNC} and not as in \eqref{eq:FilterWC}, and the maximum size restriction is \eqref{eq:MS_Local_Constraint} and not \eqref{eq:MS_Local_Constraint_corrected}. The comparable design including boundary treatment is shown in the second row of Table \ref{tab:MaxSize_Robust_065_070}. It can be seen that although these two results are obtained with exactly the same optimization parameters, the designs differ significantly. For example, it can be seen in the solution without boundary treatment that the bars at the top and bottom of the domain are thin compared to those at the inside. This is because the region $\Omega$ is half fractionated at the borders of the design space, therefore the imposed maximum size in those zones is $0.5r_\mathrm{max}^\mathrm{int}$. The same applies to the MinSolid and the MinVoid, since the filtering region is also fractionated. This subtle boundary effect produces a significant change in the topology, hence the importance of correcting them \cite{Clausen2017}.

It is worth mentioning that the dilated designs in Tables \ref{tab:Effect_MaxSize_Robust} and \ref{tab:MaxSize_Robust_065_070} contain floating material that is not connected to the structure. We believe that the reason is the volume difference between the intermediate and dilated designs. Due to the dilation filter, the volume of the dilated design is always larger than that of the intermediate design. Furthermore, since the dilation occurs in the normal direction of the surface, the larger the surface of the intermediate design, the larger the volume difference between designs. Therefore, the scaling factor ${v^\intercal \bm{\bar{\rho}}^{\mathrm{dil}}}/{v^\intercal \bm{\bar{\rho}}^{\mathrm{int}}}$ in equation \eqref{eq:VolumeDilated} is expected to be larger in maximum size constrained problems. For instance, the scaling factor ${v^\intercal \bm{\bar{\rho}}^{\mathrm{dil}}}/{v^\intercal \bm{\bar{\rho}}^{\mathrm{int}}}$ is approximately $1.4$ in the last 3 rows of Table \ref{tab:Effect_MaxSize_Robust}, which means that $\PFS{V}{dil}{*} = 70\%$. If the volume constraint in the dilated design is not restrictive enough, then the optimizer will attempt to place the excess of material wherever the maximum size restriction is not violated, resulting in the introduction of solid floating parts. This observation can be confirmed in Table \ref{tab:MaxSize_Robust_30VOL}, where the volume constraint is 30$\%$, which is small enough to avoid the floating material in the dilated design.   

\clearpage

\begin{table}[t!]
	\caption{MBB beam for compliance minimization. The desired length scale is the same as in Table \ref{tab:Effect_MaxSize_Robust}, but here the volume restriction is 30$\%$.}
		\begin{tabular}{c c c}
				 \toprule
				 $\bm{\bar{\rho}}^{\mathrm{ero}}$ & $\bm{\bar{\rho}}^{\mathrm{int}}$ (Blueprint) & $\bm{\bar{\rho}}^{\mathrm{dil}}$   \\
				 \cmidrule(r){1-3}
				 {\includegraphics[width=0.29\linewidth]{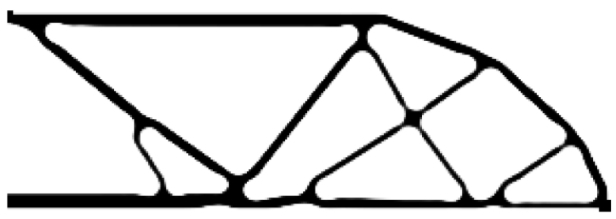} }
				 & 
				 {\includegraphics[width=0.29\linewidth]{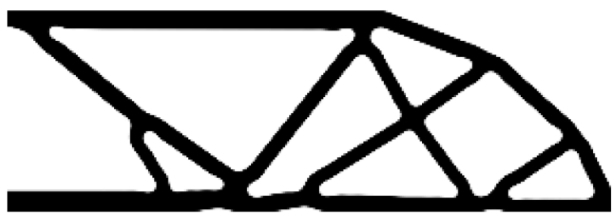} }
				 &
				 {\includegraphics[width=0.29\linewidth]{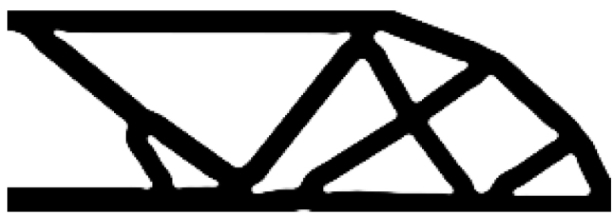} }
				 \\
				 \bottomrule
		\end{tabular}
	\label{tab:MaxSize_Robust_30VOL}
\end{table}

\section{The minimum gap constraint}
\label{sec:MinGap}

\begin{figure}[b!]
\vspace{4mm}
\centering
	\begin{subfigure}[b]{0.4\linewidth}
		\centering{}\includegraphics[width=1\linewidth]{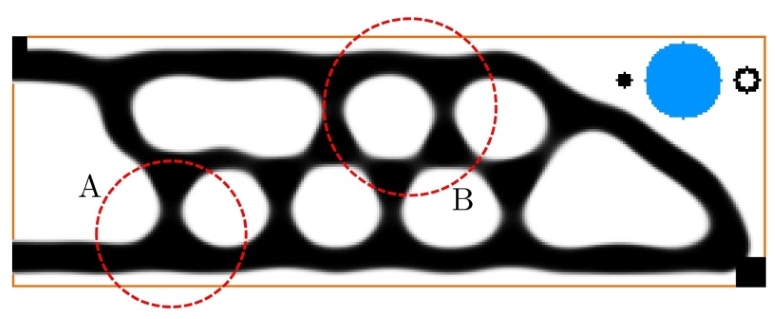}
		\caption{}
		\label{fig:MinVoid_Large_a}
	\end{subfigure}
	~\hspace{0mm}
	\begin{subfigure}[b]{0.15\linewidth}
		\centering{}\includegraphics[width=1\linewidth]{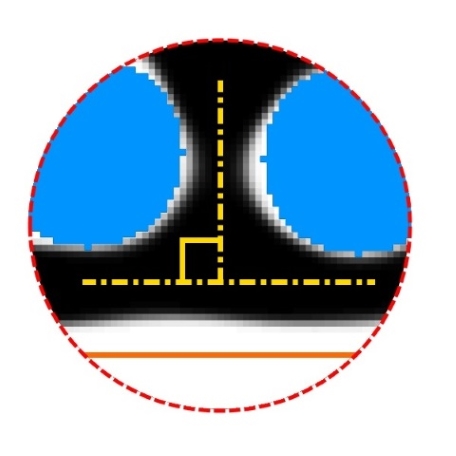}
		\caption{Zone A}
		\label{fig:MinVoid_Large_b}
	\end{subfigure}
	~\hspace{0mm}
	\begin{subfigure}[b]{0.15\linewidth}
		\centering{}\includegraphics[width=1\linewidth]{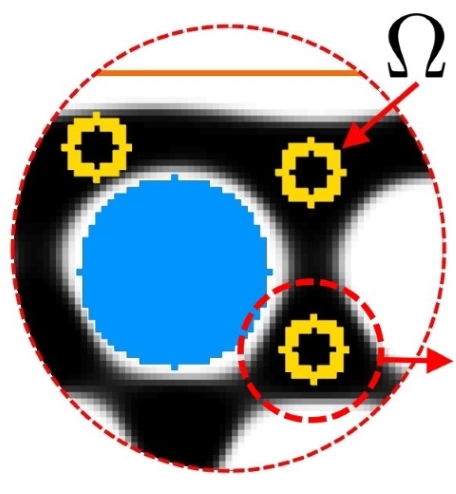}
		\caption{Zone B}
		\label{fig:MinVoid_Large_c}
	\end{subfigure}
	~\hspace{0mm}
	\begin{subfigure}[b]{0.23\linewidth}
		\centering{}\includegraphics[width=1\linewidth]{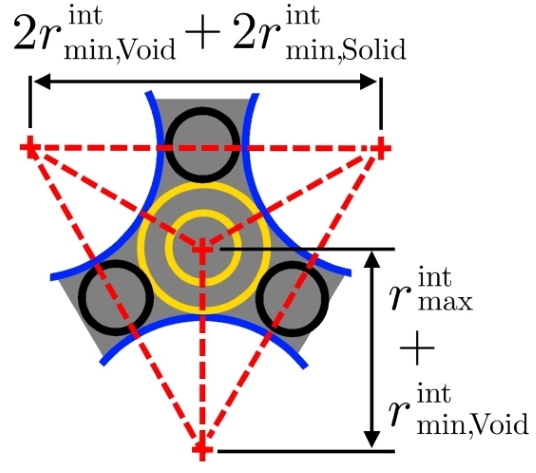}
		\caption{}
		\label{fig:MinVoid_Large_d}
	\end{subfigure}
	\vspace{-2mm}
\caption{{\color{red} In (a) is the optimized 2D-MBB beam for compliance minimization with control over MinSolid, MinVoid and MaxSolid ($\PFS{r}{min,Solid}{int}=3$, $\PFS{r}{min,Volid}{int}=15$ and $\PFS{r}{max}{int}=5$ elements). In (b) it is highlighted that the connection between two bars is perpendicular. In (c) it is shown that the maximum size is not met at the intersection of bars. (d) shows a geometric representation of the length scale at the intersection of bars.}}
\label{fig:MinVoid_Large}
\end{figure}

{\color{red}
As shown in the previous section, by adding maximum size constraints in the robust topology optimization formulation, it is possible to achieve designs satisfying Minimum size in the Solid phase (MinSolid), Minimum size in the Void phase (MinVoid) and Maximum size in the Solid phase (MaxSolid). However, the optimized designs exhibit some features that may affect the performance or manufacturability of the component. For instance, it can be seen that maximum size constrained designs can contain large numbers of closed cavities, which can hinder manufacturability. An intuitive way to reduce the number of closed cavities is to enlarge the Minimum size in the Void phase (MinVoid). This strategy, however, could greatly affect the performance of the optimized design. For example, as a result of the large offset distance between the intermediate and dilated designs, the structure is separated from the borders of the design space, as shown in Fig. \ref{fig:MinVoid_Large_a}. In addition, as $\PFS{r}{min,Void}{int}$ defines the radius of curvature at re-entrant corners, an excessively large value of $\PFS{r}{min,Void}{int}$ forces the structural members to connect perpendicularly, as shown in Fig. \ref{fig:MinVoid_Large_b}. Furthermore, due to geometrical contradictions between minimum and maximum length scales, it could be difficult or even impossible to satisfy the maximum size restriction at the intersection of two or more members. For instance, at the intersection of 3 solid members, as shown in Fig. \ref{fig:MinVoid_Large_c} and as outlined in Fig. \ref{fig:MinVoid_Large_d}, it can be easily demonstrated that to avoid geometrical contradictions, the desired maximum size in the solid phase must meet the following:
}
{\color{red} 
\begin{equation} \label{eq:MaxSize_contradiction_2}
\PFS{r}{max}{int} \geq \left( \frac{2}{\sqrt{3}} - 1 \right) \PFS{r}{min,Void}{int} + \frac{2}{\sqrt{3}} \PFS{r}{min,Solid}{int} \:.
\end{equation}
}

{\color{red}
If the condition \eqref{eq:MaxSize_contradiction_2} is not met, the maximum size restriction $G_\mathrm{ms}$ will never be satisfied, leading to numerical instabilities in the optimizer that may result in the divergence of the optimization problem or in the introduction of undesirable topological features.
}

{\color{red}
Aiming to increase the size of the and reduce the complexity of maximum size constrained designs, in this section we propose a new geometric constraint that seeks to control the minimum distance between two solid members, also called the Minimum Gap (MinGap). It will be shown that for large separation distances, the MinGap constraint considerably increases the size of the cavities and indirectly reduces the number of them, notably, without affecting the radius of curvature at the re-entrant corners.}

{\color{red}
The Minimum Gap constraint is incorporated into the robust optimization problem in order to simultaneously impose minimum length scale (MinSolid and MinVoid), maximum length scale (MaxSolid) and Minimum Gap (MinGap). Similar to the maximum size constraint, the Minimum Gap (MinGap) constraint is based on local volume restrictions. In the following we will show that it is possible to impose MinGap through the maximum size constraint $g_i$ by only changing 2 parameters, the void fraction $\varepsilon$ and the local region $\Omega_i$.  
}

\subsection{The $\varepsilon$ parameter and the minimum separation distance between members}

As described in the previous section, the local volume constraint $g_i$ limits the Maximum size in the Solid phase (MaxSolid) by placing $\varepsilon$ fraction of void in the region $\Omega_i$. Interestingly, the local volume constraint $g_i$ also imposes a minimum separation distance between solid members. To show this, we solve the compliance minimization problem subject to maximum size constraints but with no control over the Minimum size in the Void phase, i.e. $\PFS{r}{min,Void}{}=0$ elements, which is achieved by using the thresholds $[0.25,0.25,0.25]$. Since $\bm{\bar{\rho}}^\mathrm{int}=\bm{\bar{\rho}}^\mathrm{ero}=\bm{\bar{\rho}}^\mathrm{dil}$, a single restriction $G_\mathrm{ms}$ is included in the compliance minimization problem, as follows:
{\color{red}
\begin{align} \label{eq:OPTI_MaxSize_No_Robust}
	\begin{split}
  		{\min_{\rho}} & \quad \bm{f}^\intercal \bm{u}^\mathrm{int} 	\\
	  	{s.t.:} & \quad  \bm{v}^\intercal \bm{\bar{\rho}}^\mathrm{int} \leq V^*_\mathrm{int}	\\  	
	  		& \quad G_\mathrm{ms} ( \Omega^\mathrm{int} ) \leq 0 \vspace{1mm}\\ 
			& \quad 0 \leq \rho_i \leq1
	\end{split}
\end{align}
}

\begin{figure}[b!]
\centering
	\begin{subfigure}[b]{0.33\linewidth}
		\centering{}\includegraphics[width=1\linewidth]{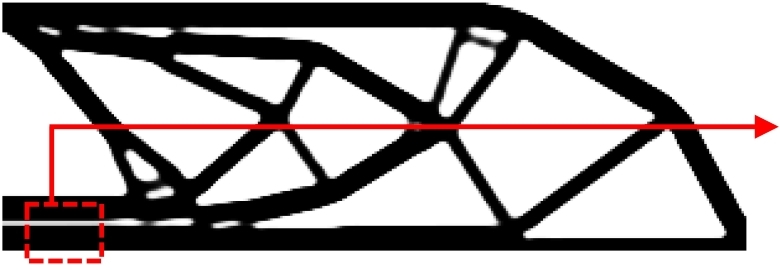}
		\caption{}
		\label{fig:MinGap_Analysys_a}
	\end{subfigure}
	~\hspace{2mm}
	\begin{subfigure}[b]{0.18\linewidth}
		\centering{}\includegraphics[width=1\linewidth]{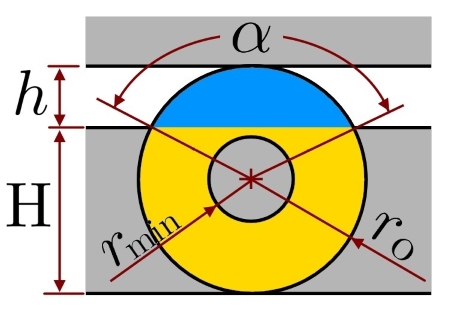}
		\caption{}
		\label{fig:MinGap_Analysys_b}
	\end{subfigure}
	~\hspace{2mm}
	\begin{subfigure}[b]{0.35\linewidth}
		\centering{}\includegraphics[width=1\linewidth]{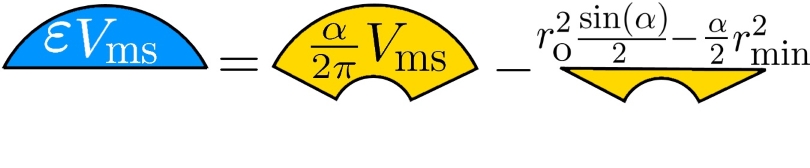}
		\caption{}
		\label{fig:MinGap_Analysys_c}
	\end{subfigure}
	\\
	\begin{subfigure}[b]{0.33\linewidth}
		\centering{}\includegraphics[width=1\linewidth]{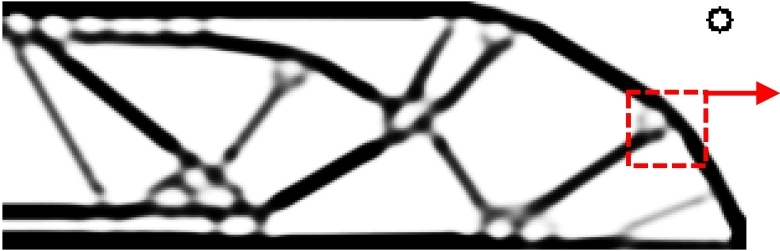}
		\caption{}
		\label{fig:MinGap_Analysys_d}
	\end{subfigure}
	~\hspace{2mm}
	\begin{subfigure}[b]{0.15\linewidth}
		\centering{}\includegraphics[width=1\linewidth]{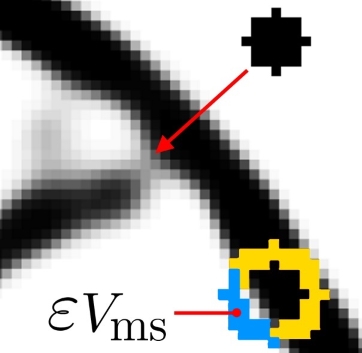}
		\caption{}
		\label{fig:MinGap_Analysys_e}
	\end{subfigure}
\caption{MBB beam with maximum size constraints and simple projection strategy ($\mu^{\mathrm{ero}}$=$\mu^{\mathrm{int}}$=$\mu^{\mathrm{dil}}$=$0.25$). In (a) is the optimized solution using $\varepsilon=0.05$. In (b) is the assumption that allows to relate $\varepsilon$ and the gap distance $h$. In (c) is the equation that relates the geometrical entities. In (d) is an iteration using $\varepsilon=0.3$. In (e) is the contradiction between the maximum size constraint and the minimum size projection.}
\label{fig:MinGap_Analysys}
\end{figure}

The optimization problem \eqref{eq:OPTI_MaxSize_No_Robust} is applied to the 2D-MBB beam. The volume constraint is set to $40\%$ and the filter radius is set to $\PFS{r}{fil}{}=6$ elements. The maximum size constraint is defined with a fraction of void $\varepsilon=5\%$ and with an annular region $\Omega_i^\mathrm{int}$ of outer radius $r_\mathrm{o}=5$ elements and inner radius $r_\mathrm{min}=3$ elements. The result is shown in Fig. \ref{fig:MinGap_Analysys_a}. It can be seen that the structure contains very small cavities, which are introduced by the maximum size restriction to prevent thick solid members. The geometry of such cavities can be related to the fraction of void $\varepsilon$ and to the local region $\Omega_i$ by making some assumptions. In this work we consider that the cavity is flat and extended, and it separates a bar of maximum thickness from another, as outlined in Fig. \ref{fig:MinGap_Analysys_a}. In such a case, the distribution of solid and void within $\Omega_i$ looks as shown in Fig. \ref{fig:MinGap_Analysys_b}. For that configuration, it is possible to obtain the system of equations that relates $\varepsilon$, $h$ and the size of $\Omega_i$ using geometric relationships, as follows: 
\begin{subequations} \label{eq:MinGap_Relation_2D}
\begin{align}
\label{eq:MinGap_Relation_2D_a}
\alpha - \text{sin}(\alpha) &= 2 \varepsilon \pi (1-(r_{\mathrm{min}}/r_\mathrm{o})^2)  \\
\label{eq:MinGap_Relation_2D_b}
h &=r_\mathrm{o}-r_o\text{cos}(\alpha/2) \\
\label{eq:MinGap_Relation_2D_c}
H &=2r_\mathrm{o}-h
\end{align} 
\end{subequations}

Equation \eqref{eq:MinGap_Relation_2D_a} is obtained by equating the areas shown in Fig. \ref{fig:MinGap_Analysys_c}. Equation \eqref{eq:MinGap_Relation_2D_b} is obtained by relating the height $h$ with the external radius $r_o$, and equation \eqref{eq:MinGap_Relation_2D_c} is obtained by equating the diameter of $\Omega$ with $h+\mathrm{H}$. {\color{red} Here, $h$ could be interpreted as the distance between 2 members and $\mathrm{H}$ as the effective maximum size imposed on the solid, i.e.~$\mathrm{H}=2r_\mathrm{max}$. For instance, the system of equations applied to the above  problem (Fig.~\ref{fig:MinGap_Analysys_a}) yields $h=0.5$ elements and $r_\mathrm{max}=4.75$ elements, which is rather close to $r_\mathrm{o}=5.0$ elements. For this reason, $\PFS{r}{max}{} \approx \PFS{r}{o}{}$ was considered in the previous section.
}

We could think in opposite sense, to calculate the fraction of void $\varepsilon$ for a desired separation distance $h$. However, if a fraction of void $\varepsilon$ much higher than $5\%$ is obtained, a geometric conflict could arise that compromises the convergence of the optimization problem. To explain this conflict, we consider the optimized 2D-MBB beam shown in Fig. \ref{fig:MinGap_Analysys_d}, which is obtained using $\varepsilon=30\%$. It can be seen that the separation distance $h$ between bars is bigger than in Fig. \ref{fig:MinGap_Analysys_a}. However, there are lots of intermediate densities at the intersection of bars. This is due to a conflict between the maximum size restriction and the Heaviside projection. As shown in Fig. \ref{fig:MinGap_Analysys_e}, the $G_\mathrm{ms}$ constraint attempts to introduce $\varepsilon$ fraction of void into $\Omega_i$, but the Heaviside function attempts to introduce solid elements into the same region. As the optimization progresses and the penalty parameters rise, the material connectivity is compromised and the optimization problem may diverge.

\subsection{The local region for minimum gap}

\begin{figure}[b]
\centering
	\begin{subfigure}[b]{0.24\linewidth}
		\centering{}\includegraphics[width=1\linewidth]{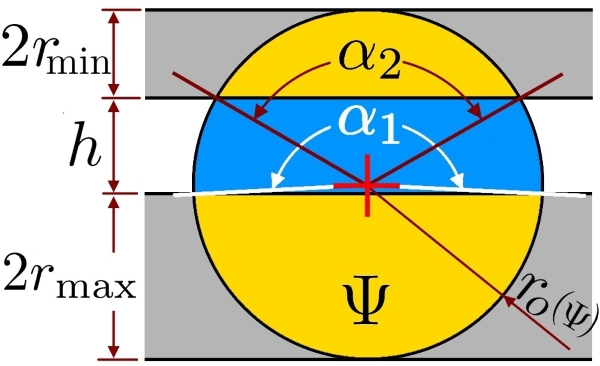}
		\caption{}
		\label{fig:MinGap_Test_Regions_a}
	\end{subfigure}
	~\hspace{2mm}
	\begin{subfigure}[b]{0.192\linewidth}
		\centering{}\includegraphics[width=1\linewidth]{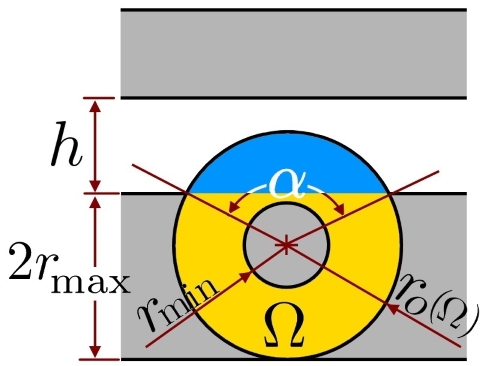}
		\caption{}
		\label{fig:MinGap_Test_Regions_b}
	\end{subfigure}
	~\hspace{2mm}
	\begin{subfigure}[b]{0.28\linewidth}
		\centering{}\includegraphics[width=1\linewidth]{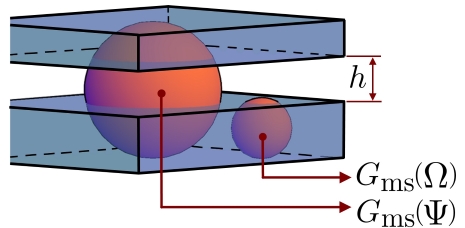}
		\caption{}
		\label{fig:MinGap_Test_Regions_c}
	\end{subfigure}
	\vspace{-3mm}
\caption{(a) The test region to impose the gap distance. (b) The maximum size radius $r_{\mathrm{o}}$ depends on $t_{\mathrm{max}}$ and $\varepsilon$. (c) Test regions in 3D.}
\label{fig:MinGap_Test_Regions}
\end{figure}

To mitigate the geometric conflict mentioned above and exploit the existing relation between $\varepsilon$ and $h$, we propose to replace the region $\Omega$ by another region $\Psi$. The main idea is to select $\Psi$ such that $\varepsilon$ is low. There are several geometric shapes that serve this purpose. The test region used in this work is shown in Fig. \ref{fig:MinGap_Test_Regions_a}. The chosen $\Psi$ is bigger than $\Omega$ in order to reduce the fraction of void. The main drawback is that, depending on the desired length scale, memory allocation to create the matrix $\mathrm{\mathbf{D}}_{\mathrm{II}}\left( \Psi \right)$ can be considerably higher than for creating $\mathrm{\mathbf{D}}_{\mathrm{II}}\left(\Omega\right)$ and $\mathrm{\mathbf{D}}_{\mathrm{I}}$. 

The aim is to use the constraint $g_i(\Psi)$ to impose a minimum distance ($h$) between 2 solid members by introducing a void fraction ($\varepsilon$) within the local region $\Psi$. Therefore, the minimum separation distance $h$ must be imposed indirectly through the void fraction $\varepsilon$ and the size of $\Psi$. To find the geometric relationship between $h$, $\varepsilon$ and $\Psi$, we assume that the two solid members are parallel and that one of them has a maximum thickness $\mathrm{H} = 2\PFS{r}{max}{}$, as shown in Fig.~\ref{fig:MinGap_Test_Regions_a}. That is, if one of these two conditions is not met, the minimum distance between two bars will not be accurately imposed. In 3D, we make the same assumptions but considering plates instead of bars, as shown in Fig. \ref{fig:MinGap_Test_Regions_c}. In 2D and 3D, the outer diameter of the $\Psi$ region is defined as:
\begin{equation}\label{eq:MinGap_rb}
r_\mathrm{o} = r_{min} + r_{max} + h/2  \; .
\end{equation}

The system of equations that yields $\varepsilon$ is obtained by equating geometric entities in the same way as was done for the $\Omega$ region in \eqref{eq:MinGap_Relation_2D}. For the region $\Psi$, the system of equations is:
\begin{subequations} \label{eq:MinGap_Test_Region}
\begin{align}
2 \varepsilon \pi &= \alpha_2 - \alpha_1 + \text{sin} (\alpha_2) - \text{sin} (\alpha_1) \; ,
\\
2r_{\mathrm{max}} + h &= r_\mathrm{o} + r_\mathrm{o} \: \text{cos}(\alpha_2 / 2) \; ,
\\
2r_{\mathrm{max}} &= r_{\mathrm{o}} + r_\mathrm{o} \: \text{cos}(\alpha_1 / 2) \; .
\end{align}
\end{subequations}

In 3D, the equation that solves $\varepsilon$ can be obtained from the volume equation of a spherical cap, which leads to the following expression:
\begin{equation} \label{eq:MinGap_Test_Region_3D}
	\varepsilon 4 r_\mathrm{o}^3 = 3 r_\mathrm{o} (h+2r_{\mathrm{min}})^2 - (h+2r_{\mathrm{min}})^3 - 12 h r_\mathrm{o} r_{\mathrm{min}}^3 + 8 r_{\mathrm{min}}^3   \; .
\end{equation}

Thus, the proposed approach to control the distance between solid structural members has an exact formulation to the maximum size constraint but changing $\varepsilon$ and the local region $\Omega$ by $\Psi$. 

\begin{figure}[t!]
	\centering{}\includegraphics[width=0.75\linewidth]{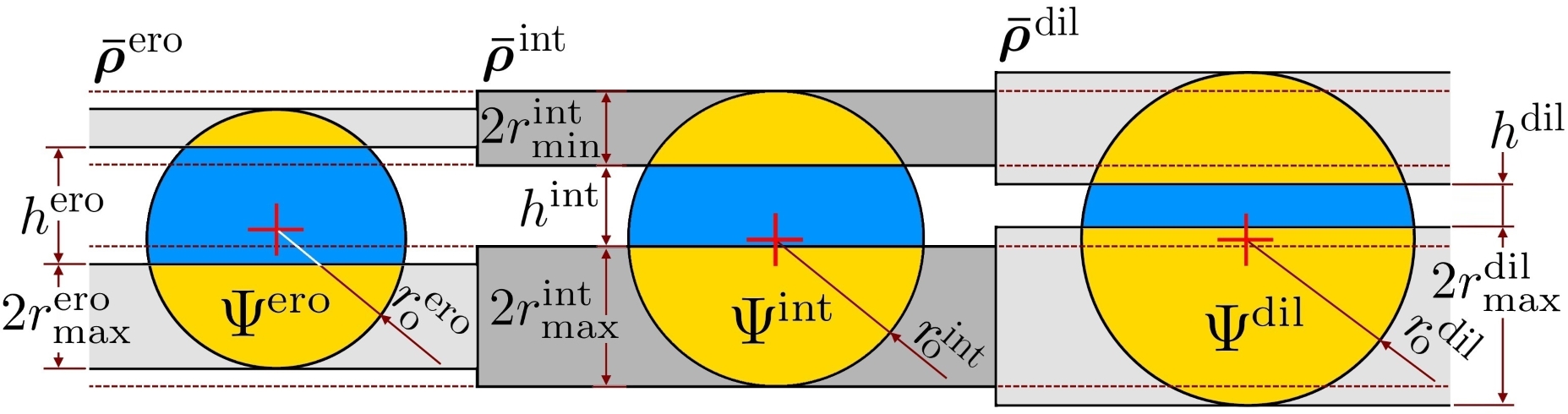}
	\vspace{-1mm}
\caption{The region $\Psi$ for the eroded, intermediate and dilated designs.}
\label{fig:MinGap_Test_Regions_f}
\end{figure}

In the robust design formulation, the desired length scale is defined in the intermediate design. Therefore, the intended separation distance is denoted as $h^\mathrm{int}$. Then, the external radius of $\Psi^\mathrm{int}$ can be computed using \eqref{eq:MinGap_rb} and the void fraction $\varepsilon$ can be obtained by solving \eqref{eq:MinGap_Test_Region} in 2D, and \eqref{eq:MinGap_Test_Region_3D} in 3D. As the local volume restriction $g_i(\Psi)$ must be applied in the eroded, intermediate and dilated designs, the local region $\Psi$ and the separation distance $h$ must be scaled accordingly, as shown in Fig. \ref{fig:MinGap_Test_Regions_f}. For this, the offset distances can be used. Thus, the separation distances $h$ in the eroded and dilated designs can be obtained as:
\begin{subequations}
\begin{align}
h^\mathrm{ero} =& \; h^\mathrm{int} + 2t_\mathrm{ero} \vspace{1mm}\\
h^\mathrm{dil} =& \; h^\mathrm{int} - 2t_\mathrm{dil}
\end{align}
\end{subequations}

Then, the external radius of the regions $\Psi^\mathrm{ero}$ and $\Psi^\mathrm{dil}$ can be obtained using \eqref{eq:MinGap_rb}, considering $h$, $r_\mathrm{min}$ and $r_\mathrm{max}$ of the corresponding design. Finally, the void fraction $\varepsilon$ of $g_i(\Psi^\mathrm{ero})$ and $g_i(\Psi^\mathrm{dil})$ can be obtained by solving \eqref{eq:MinGap_Test_Region} in 2D, and \eqref{eq:MinGap_Test_Region_3D} in 3D, considering $h$, $r_\mathrm{o}$ and $r_\mathrm{max}$ of the corresponding design.

\begin{table}[b!]
	\caption{Set of parameters used in this work for the 2D problems. The user-defined parameter is indicated as UD in parentheses. The considered thresholds are $[\mu^\mathrm{ero},\mu^\mathrm{int},\mu^\mathrm{dil}]=[0.75,0.50,0.25]$, therefore, $\PFS{r}{min}{int}=\PFS{r}{min,Solid}{int}=\PFS{r}{min,Void}{int}$ and $t = \PFS{t}{ero}{} = \PFS{t}{dil}{}$.}
	\centering
		\begin{tabular}{c c c c c c c c}
				 \toprule
				  Constraint & $\varepsilon$ & $r_{\mathrm{min}}$ & $r_{\mathrm{max}}$ & $r_{\mathrm{o}}$ & $h$ & $q$ & $p$\\
				 \cmidrule(r){1-8} 
				 $G_{\mathrm{ms}} (\Omega^{\mathrm{int}})$ & $0.05$ & $r_\mathrm{min}^\mathrm{int}$ {\small(UD)} & $r_{\mathrm{max}}^\mathrm{int}$ {\small(UD)} & eq. \eqref{eq:MinGap_Relation_2D} &  eq. \eqref{eq:MinGap_Relation_2D} & $\eta$ & $100$
				 \vspace{1mm}\\
				 $G_{\mathrm{ms}} (\Omega^{\mathrm{ero}})$ &  $0.05$ & $r_\mathrm{min}^\mathrm{int}-t$ & $r_{\mathrm{max}}^\mathrm{int}-t$ &  eq. \eqref{eq:MinGap_Relation_2D} &  eq. \eqref{eq:MinGap_Relation_2D} & $\eta$ & $100$
				 \vspace{1mm}\\
				 $G_{\mathrm{ms}} (\Omega^{\mathrm{dil}})$ & $0.05$ & $r_\mathrm{min}^\mathrm{int}+t$ & $r_{\mathrm{max}}^\mathrm{int}+t$ &  eq. \eqref{eq:MinGap_Relation_2D} &  eq. \eqref{eq:MinGap_Relation_2D} & $\eta$ & $100$
				 \vspace{1mm}\\
				 $G_{\mathrm{ms}} (\Psi^\mathrm{int})$ &  eq. \eqref{eq:MinGap_Test_Region} &  $r_\mathrm{min}^\mathrm{int}$ &  $r_{\mathrm{max}}^\mathrm{int}$ &  eq. \eqref{eq:MinGap_rb} & $h^\mathrm{int}$ {\small(UD)}  & $1.0$ & $60$
				 \vspace{1mm}\\
				 $G_{\mathrm{ms}} (\Psi^{\mathrm{ero}})$ &  eq. \eqref{eq:MinGap_Test_Region} & $r_\mathrm{min}^\mathrm{int}-t$ & $r_{\mathrm{max}}^\mathrm{int}-t$ &  eq. \eqref{eq:MinGap_rb} & $h^\mathrm{int}+t$ & $1.0$ & $60$
				 \vspace{1mm}\\
				 $G_{\mathrm{ms}} (\Psi^{\mathrm{dil}})$ &  eq. \eqref{eq:MinGap_Test_Region} & $r_\mathrm{min}^\mathrm{int}+t$ & $r_{\mathrm{max}}^\mathrm{int}+t$ &  eq. \eqref{eq:MinGap_rb} & $h^\mathrm{int}-t$ & $1.0$ & $60$
				 \vspace{1mm}\\ 
				 \bottomrule
		\end{tabular}
	\label{tab:Opti_Parameters}
\end{table}

Since the void fraction $\varepsilon$ in $g_i({\Psi})$ is usually greater than $5\%$, it is not necessary to penalize the intermediate densities, therefore we use $q=1$ in \eqref{eq:MS_Local_Constraint_corrected}. In addition, as discussed in the previous section, a large value of $\varepsilon$ allows to reduce the aggregation exponent \cite{Fernandez2019}, which is why for $G_\mathrm{ms}({\Psi})$ we use $p=60$. Table \ref{tab:Opti_Parameters} summarizes the parameters of the constraints $G_{\mathrm{ms}}(\Omega)$ and $G_{\mathrm{ms}}(\Psi)$ considering $[\mu^\mathrm{ero},\mu^\mathrm{int},\mu^\mathrm{dil}]=[0.75,0.50,0.25]$. 

To summarize, the user-defined values are $\PFS{r}{min,Solid}{int}$ (MinSolid), $\PFS{r}{min,Void}{int}$ (MinVoid), $\PFS{r}{max}{int}$ (MaxSolid), and $h^\mathrm{int}$ (MinGap). The minimum length scale (MinSolid and MinVoid) are imposed through the robust design approach. The maximum length scale (MaxSolid) is imposed through local volume restrictions applied on the eroded, intermediate and dilated designs. Similarly, the minimum separation distance (MinGap) is imposed through local volume restrictions applied on the eroded, intermediate and dilated designs. Finally, the formulation of the topology optimization problem with simultaneous control over the Minimum size in the Solid phase, Minimum size in the Void phase, Maximum size in the Solid phase and Minimum Gap, is as follows:
\begin{align} \label{eq:OPTI_Examples}
	\begin{split}
  		{\min_{\rho}} &\quad O_\mathrm{bj}			\\
	  	{s.t.:} &\quad  \bm{v}^{\intercal} \bm{\bar{\rho}}^{\mathrm{dil}} \leq V^*_{\mathrm{dil}} 	\\
	  		& 
			\left.	  		
	  		\begin{matrix}
	  		& \hspace{-1.5mm} G_{\mathrm{ms}}( \Omega^{\mathrm{ero}} ) \leq 0 \vspace{1mm}\\ 
	  		& \hspace{-2.0mm} G_{\mathrm{ms}}( \Omega^{\mathrm{int}} ) \leq 0 \vspace{1mm} \\ 
	  		& \hspace{-2.0mm} G_{\mathrm{ms}}( \Omega^{\mathrm{dil}} ) \leq 0 
			\end{matrix}
			\;
			\right\rbrace
			\text{\small Robust Maximum Size}		
			\\	  		
	  		&
			\left.	  		
	  		\begin{matrix}	  		
	  		& \hspace{-1.5mm} G_{\mathrm{ms}}( \Psi^{\mathrm{ero}} ) \leq 0 \vspace{1mm}\\
	  		& \hspace{-2.0mm} G_{\mathrm{ms}}( \Psi^{\mathrm{int}} ) \leq 0 \vspace{1mm}\\
	  		& \hspace{-2.0mm} G_{\mathrm{ms}}( \Psi^{\mathrm{dil}} ) \leq 0 
	  		\end{matrix}
			\;
			\right\rbrace 
			\text{\small Robust Minimum Gap}
			\\
			& \quad 0 \leq {\rho_i} \leq1
	\end{split}
\end{align}

\noindent where $O_\mathrm{bj}$ is the objective function. Therefore, in the compliance minimization problem, $O_\mathrm{bj} = \bm{f}^{\intercal} \bm{u}^{\mathrm{ero}}$, and in the force inverter design problem, $O_\mathrm{bj} = \mathrm{max} (\bm{t}^{\intercal} \bm{u}^{\mathrm{ero}} \:,\: \bm{t}^{\intercal} \bm{u}^{\mathrm{dil}})$.

\begin{figure}[t]
	\centering{}\includegraphics[width=0.95\linewidth]{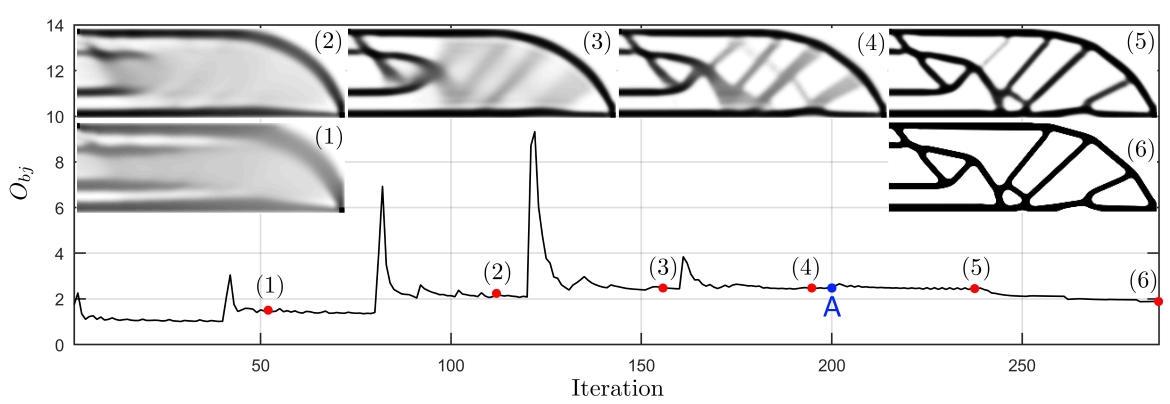}
	\vspace{-2mm}	
	\caption{Convergence curve of the 2D-MBB beam with maximum and minimum gap constraints. From iteration A, the constraints $G_{\mathrm{ms}}(\Psi)$ are deactivated. The problem is solved for $V^*_\mathrm{int}=30\%$, $\PFS{r}{min,Solid}{int}=\PFS{r}{min,Void}{int}=3$ elements, $\PFS{r}{max}{int}=5$ elements and $h^\mathrm{int}=20$ elements.}
\label{fig:Convergence_Curve}
\end{figure}

It is important to note that the chosen test region $\Psi$ for imposing MinGap is not completely exempt from the conflict between the Heaviside function and the maximum size restriction. Depending on the selected $h$, a large $\varepsilon$ can be obtained compromising the material connectivity. In such cases, we have observed that during the optimization process, when penalty parameters are still low and large gray areas are allowed, e.g.\@ for $\eta\leq 2$ and $\beta \leq 8$, material connectivity is not jeopardized. For this reason, when large $\varepsilon$ values are present in $G_{\mathrm{ms}}(\Psi)$, for instance $\varepsilon \geq 0.2$, the minimum gap constraint $G_{\mathrm{ms}}(\Psi)$ is deactivated when ${\eta \geq 2}$ and ${\beta \geq 8}$. At that moment, according to our continuation scheme, the move limit $m_l$ is $0.23$, which is small enough to prevent the distanced bars from approaching again, but large enough to rearrange the material in the bar joint. A typical convergence curve is shown in Fig.\:\ref{fig:Convergence_Curve}, where A denotes the iteration from which the set of constraints $G_{\mathrm{ms}}(\Psi)$ is deactivated.

\section{Examples and discussion}
\label{sec:5}

{\color{red}
In this section, the proposed strategy to impose Minimum size in the Solid phase (MinSolid), Minimum size in the Void phase (MinVoid), Maximum size in the Solid phase (MaxSolid) and minimum distance between solid parts (MinGap), is applied to the test cases introduced in Section \ref{sec:2}, i.e.~the force inverter, the 2D-MBB beam, and the 3D-MBB beam. In each test case we report (i) a solution with only minimum length scale control (MinSolid and MinVoid), (ii) solutions that incorporate maximum length scale control (MinSolid, MinVoid and MaxSolid), and (iii) solutions that incorporate minimum gap control (MinSolid, MinVoid, MaxSolid and MinGap). For the sake of simplicity, we refer to (i) as the reference solution, to (ii) as the maximum size constrained solutions, and to (iii) as the minimum gap constrained solutions.} 

Unless otherwise specified, the minimum length scale is defined as $\PFS{r}{min,Solid}{int}=\PFS{r}{min,Void}{int}\:$ by using the thresholds $[\mu^\mathrm{ero},\:\mu^\mathrm{int},\:\mu^\mathrm{dil}]=[0.75,\:0.50,\:0.25]$ and $r_\mathrm{fil}=2 \PFS{r}{min,Solid}{int}$. The Maximum size in the Solid phase (MaxSolid) and the minimum distance between solid members (MinGap) are imposed using the parameters listed on Table \ref{tab:Opti_Parameters} and the remaining optimization parameters are defined as in the previous sections.  

The intended length scale is graphically reported next to each solution. The objective values are reported as $O_\mathrm{bj}$ and, as is usual in density-based topology optimization, all the optimization problems in this work start from a uniform initial guess satisfying the volume constraint.

\subsection{Compliant mechanism}
\label{sec:51}

\begin{figure}[t!]
\captionsetup{font=small}
\centering    
    \begin{subfigure}[b]{0.32\linewidth}
		\centering{}\includegraphics[width=1\linewidth]{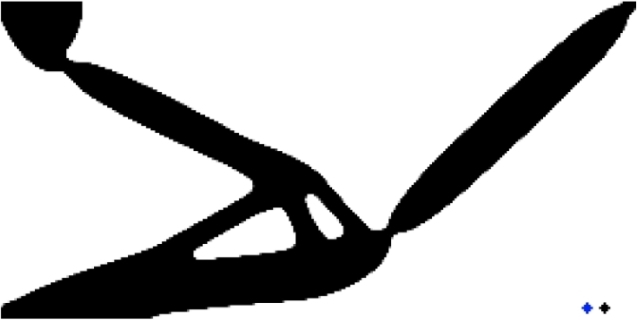}
 		\caption{$O_{\mathrm{bj}}$=$R_\mathrm{ef}$,$\;\;$ Vol=25.0$\%$}
 		\label{fig:Results_FI_a}
	\end{subfigure}
	~
	\begin{subfigure}[b]{0.32\linewidth}
		\centering{}\includegraphics[width=1\linewidth]{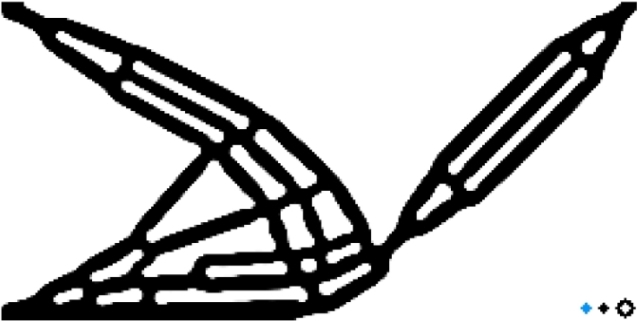}
 		\caption{{$O_{\mathrm{bj}}/R_\mathrm{ef}$=0.874,$\;\;$ Vol=25.0$\%$}}
 		\label{fig:Results_FI_b}
	\end{subfigure}
	~
	\begin{subfigure}[b]{0.32\linewidth}
		\centering{}\includegraphics[width=1\linewidth]{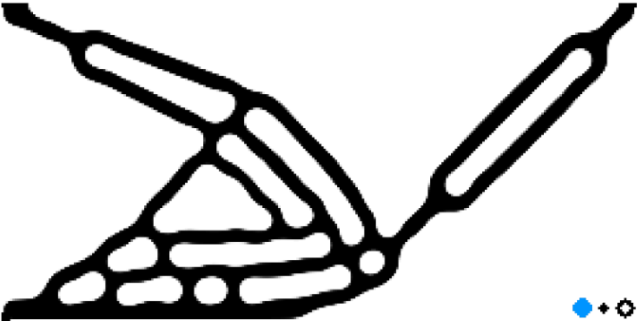}
 		\caption{{$O_{\mathrm{bj}}/R_\mathrm{ef}$=0.745,$\;\;$ Vol=24.0$\%$}}
 		\label{fig:Results_FI_X}
	\end{subfigure}
	\vspace{2mm}
	\\
 	\begin{subfigure}[b]{0.32\linewidth}
		\centering{}\includegraphics[width=1\linewidth]{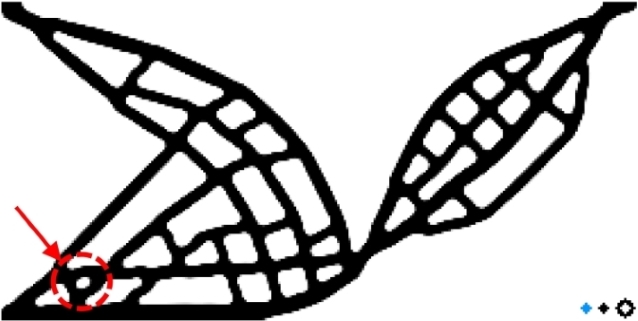}
 		\caption{$O_{\mathrm{bj}}/R_\mathrm{ef}$=0.824,$\;$ Vol=25.0$\%$}
 		\label{fig:Results_FI_c}
	\end{subfigure}
	~
	\begin{subfigure}[b]{0.32\linewidth}
		\centering{}\includegraphics[width=1\linewidth]{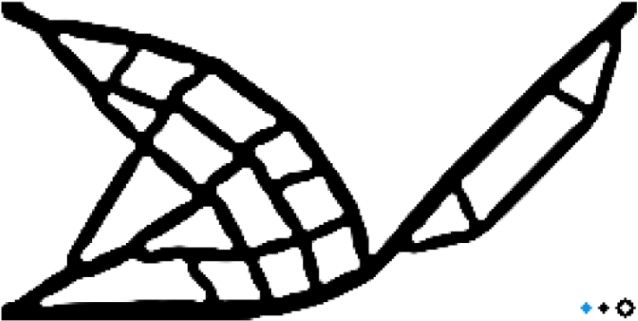}
		\caption{$O_{\mathrm{bj}}/R_\mathrm{ef}$=0.823,$\;$ Vol=24.3$\%$}
		\label{fig:Results_FI_d}
	\end{subfigure}
	~
	\begin{subfigure}[b]{0.32\linewidth}
		\centering{}\includegraphics[width=1\linewidth]{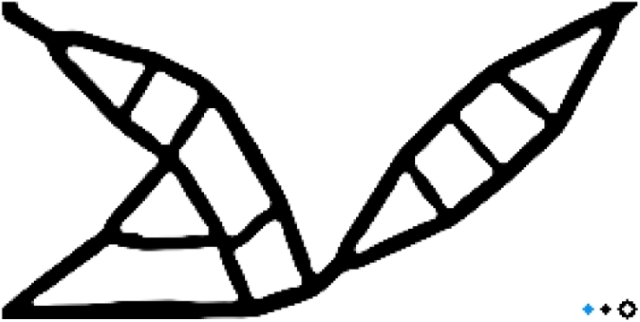}
		\caption{$O_{\mathrm{bj}}/R_\mathrm{ef}$=0.748,$\;$ Vol=23.0$\%$ }
		\label{fig:Results_FI_e}
	\end{subfigure}
	\vspace{0mm}\\
	($h^\mathrm{int}=2r_{\mathrm{max}}^\mathrm{int}$) \hspace{1.7cm} ~ \hspace{1.7cm} ($h^\mathrm{int}=4r_{\mathrm{max}}^\mathrm{int}$) \hspace{1.7cm} ~ \hspace{1.7cm}($h^\mathrm{int}=6r_{\mathrm{max}}^\mathrm{int}$)
\caption{Force inverter with length scale control. (a) MinSolid and MinVoid. (b-c) MinSolid, MinVoid and MaxSolid. (d-f) MinSolid, MinVoid, MaxSolid and MinGap. (c) is obtained using [$\PFS{\mu}{}{ero}$, $\PFS{\mu}{}{int}$, $\PFS{\mu}{}{dil}$] = [0.75, 0.65, 0.25].}
\label{fig:Results_FI}
\end{figure}

The force inverter is solved with a volume constraint of $25\%$. The minimum length scale (MinSolid and MinVoid) is set to $\PFS{r}{min,Solid}{int} = \PFS{r}{min,Void}{int} = 2$ elements and the maximum length scale (MaxSolid) is set to $r_\mathrm{max}^\mathrm{int}=4$ elements. The reference and the maximum size constrained designs are shown in Fig. \ref{fig:Results_FI_a} and \ref{fig:Results_FI_b}, respectively. It can be noticed that these designs do not contain small hinges that are likely to arise in compliant mechanism problems. In addition, the desired length scale is consistent with that intended. This affirms that maximum size constraints and the robust design approach allow effective control over MinSolid, MinVoid and MaxSolid. {\color{red} However, it can be seen that the maximum size constraint introduces large amount of closed cavities in the design, which could hinder manufacturability. Aiming to reduce the complexity of the maximum size constrained design, the Minimum size in the Void phase (MinVoid) is doubled, i.e.~$r_\mathrm{min,Void}^\mathrm{int}=4$ elements. The new design is shown in Fig. \ref{fig:Results_FI_X}. This one features less closed cavities than the former solution (Fig. \ref{fig:Results_FI_b}).  However, the structural performance is significantly reduced since the output displacement is decreased by approximately $15\%$ with respect to the former design (Fig.\:\ref{fig:Results_FI_b}). In order to reduce the geometric complexity of the maximum size constrained design without severely compromising the structural performance, we introduce the control over the minimum distance between solid members (MinGap).
}

In the following examples we consider minimum gap constrained designs. The minimum and maximum length scales are defined as in Fig. \ref{fig:Results_FI_b}, i.e.~$\PFS{r}{min,Solid}{int} = \PFS{r}{min,Void}{int} = 2$ elements and $r_\mathrm{max}^\mathrm{int} = 4$ elements. Three topology optimization problems are solved using $h^\mathrm{int}=2r_{\mathrm{max}}^\mathrm{int}$,  $h^\mathrm{int}=4r_{\mathrm{max}}^\mathrm{int}$ and $h^\mathrm{int}=6r_{\mathrm{max}}^\mathrm{int}$. Solutions are respectively shown in Figs. \ref{fig:Results_FI_c}, \ref{fig:Results_FI_d} and \ref{fig:Results_FI_e}. It can be seen that minimum gap constrained solutions contain cavities of proportional size to $h^\mathrm{int}$, which is consistent with the geometric relationship between the void fraction $\varepsilon$ and the separation distance $h$. However, as pointed in Fig. \ref{fig:Results_FI_c} with an arrow, small cavities of size $r_{\mathrm{min,Void}}^\mathrm{int}$ may be present in the optimized solution. This is because the set of constraints $G_{\mathrm{ms}}(\Psi)$ are deactivated in these examples before the optimization problem is finished. Therefore, the precise control of $h^\mathrm{int}$ is not guaranteed in these problems.

Despite increasing the distance between bars and the size of cavities, the minimum gap constraint $G_{\mathrm{ms}}(\Psi)$ with a low $h^\mathrm{int}$ does not guarantee a less complex design. For instance, compared to designs of Figs. \ref{fig:Results_FI_b} and \ref{fig:Results_FI_X}, the solution obtained with $h^\mathrm{int}=2r_{\mathrm{max}}^\mathrm{int}$ includes a greater amount of cavities. This issue may increase the complexity of the optimized design depending on the chosen manufacturing process. However, a larger $h^\mathrm{int}$ significantly decreases the quantity of cavities. In the case when $h^\mathrm{int}=6r_{\mathrm{max}}^\mathrm{int}$, the design resembles a structure composed by bars which is relatively easy to fabricate for a wide range of manufacturing processes. This suggests that, to achieve a noticeable effect on the design complexity, a large $h^\mathrm{int}$ must be used. 

{\color{red}
The difference between imposing the Minimum size in the Void phase (MinVoid) and the minimum distance between solid members (MinGap) can be noticed when comparing designs from Figs.\:\ref{fig:Results_FI_X} and \ref{fig:Results_FI_e}. Though both designs are significantly inferior in performance to the reference solution, the design in Fig.\:\ref{fig:Results_FI_e} exhibits a considerably less complex topology, featuring larger cavities and straighter bars than those in Fig.\:\ref{fig:Results_FI_X}. This is mainly due to the fact that the radius of curvature at re-entrant corners is not affected by the proposed minimum gap restriction. 
}
 
It can be seen that for large values of $h$, the optimization problem does not use all the allowed material. This is because a large distance $h$ leads to a big void fraction $\varepsilon$. In some cases, $\varepsilon$ may be greater than $1-V^*_\mathrm{dil}$ and the local volume restriction $g_i(\Psi)$ would be more restrictive than the volume constraint applied to the design. Therefore, before deactivating the constraints $G_\mathrm{ms}(\Psi)$, the maximum amount of material in the design space is defined by $1-\varepsilon$ and not by $V^*_\mathrm{dil}$. However, the excess of material, resulting from deactivating the $G_\mathrm{ms}(\Psi)$ restrictions, may not be used to create new solid parts due to the small move limits of the design variables and to the high penalty factors. The excess of material is neither used for thickening the bars because the maximum size restriction does not allow it or because the thickening negatively affects the structural performance.

\subsection{2D-MBB beam}
\label{sec:52}

\begin{figure}[t!]
\centering    
    \begin{subfigure}[b]{0.32\linewidth}
		\centering{}\includegraphics[width=1\linewidth]{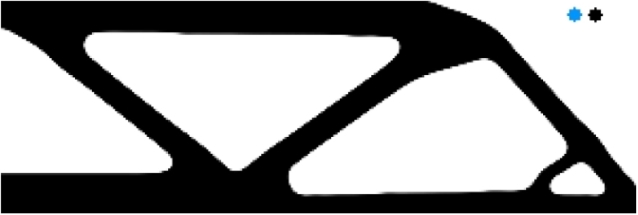}
 		\caption{$O_{\mathrm{bj}}$=$R_\mathrm{ef}$,$\;\;$ Vol=40.0$\%$}
 		\label{fig:Results_MBB_a}
	\end{subfigure}
	~
	\begin{subfigure}[b]{0.32\linewidth}
		\centering{}\includegraphics[width=1\linewidth]{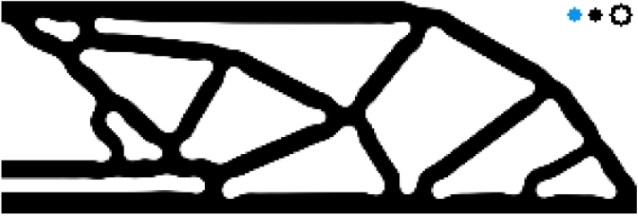}
 		\caption{$O_{\mathrm{bj}}/R_\mathrm{ef}$=1.596,$\;\;$ Vol=39.2$\%$}
 		\label{fig:Results_MBB_b}
	\end{subfigure}
	~
	\begin{subfigure}[b]{0.32\linewidth}
		\centering{}\includegraphics[width=1\linewidth]{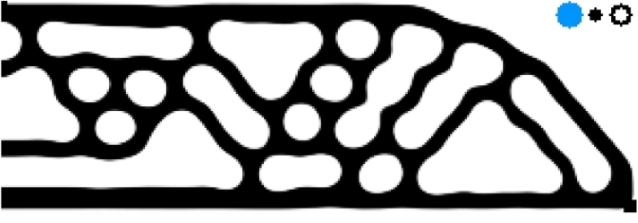}
 		\caption{$O_{\mathrm{bj}}/R_\mathrm{ef}$=2.101,$\;\;$ Vol=40.0$\%$}
 		\label{fig:Results_MBB_X}
	\end{subfigure}
	\vspace{4mm}
	\\
 	\begin{subfigure}[b]{0.32\linewidth}
		\centering{}\includegraphics[width=1\linewidth]{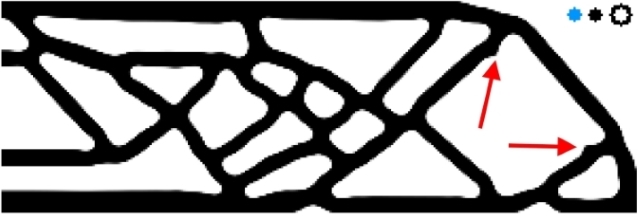}
 		\caption{$O_{\mathrm{bj}}/R_\mathrm{ef}$=1.786,$\;\;$ Vol=40.0$\%$}
 		\label{fig:Results_MBB_c}
	\end{subfigure}
	~
	\begin{subfigure}[b]{0.32\linewidth}
		\centering{}\includegraphics[width=1\linewidth]{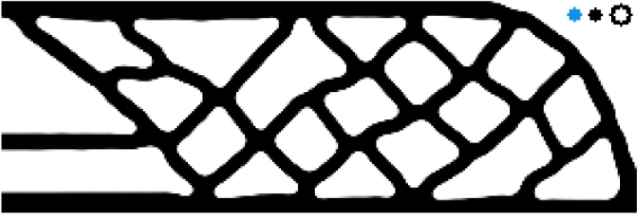}
		\caption{$O_{\mathrm{bj}}/R_\mathrm{ef}$=2.117,$\;\;$ Vol=40.0$\%$}
		\label{fig:Results_MBB_d}
	\end{subfigure}
	~
	\begin{subfigure}[b]{0.32\linewidth}
		\centering{}\includegraphics[width=1\linewidth]{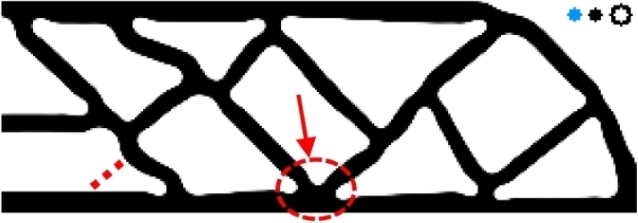}
		\caption{$O_{\mathrm{bj}}/R_\mathrm{ef}$=2.145,$\;\;$ Vol=37.8$\%$}
		\label{fig:Results_MBB_e}
	\end{subfigure}
	\vspace{0mm}\\
	($h^\mathrm{int}=2r_{\mathrm{max}}^\mathrm{int}$) \hspace{1.7cm} ~ \hspace{1.7cm} ($h^\mathrm{int}=4r_{\mathrm{max}}^\mathrm{int}$) \hspace{1.7cm} ~ \hspace{1.7cm}($h^\mathrm{int}=6r_{\mathrm{max}}^\mathrm{int}$)
\caption{2D-MBB beam with length scale control. (a) MinSolid and MinVoid. (b-c) MinSolid, MinVoid and MaxSolid. (d-f) MinSolid, MinVoid, MaxSolid and MinGap. (c) is obtained using [$\PFS{\mu}{}{ero}$, $\PFS{\mu}{}{int}$, $\PFS{\mu}{}{dil}$] = [0.75, 0.65, 0.25].}
\label{fig:Results_MBB}
\end{figure}

The 2D-MBB beam is solved with a volume constraint of $40\%$. The minimum length scale is set to $r_\mathrm{min,Solid}^\mathrm{int}=r_\mathrm{min,Void}^\mathrm{int}=3$ elements and the Maximum size in the Solid phase is set to $r_\mathrm{max}^\mathrm{int}=5$ elements. The reference and the maximum size constrained designs are shown in Figs. \ref{fig:Results_MBB_a} and \ref{fig:Results_MBB_b}, respectively. For the minimum gap constrained designs, we solve 3 optimization problems considering $h^\mathrm{int}=2r_{\mathrm{max}}^\mathrm{int}$, $h^\mathrm{int}=4r_{\mathrm{max}}^\mathrm{int}$ and $h^\mathrm{int}=6r_{\mathrm{max}}^\mathrm{int}$. Results are respectively shown in Figs. \ref{fig:Results_MBB_c}, \ref{fig:Results_MBB_d} and \ref{fig:Results_MBB_e}. In addition, for comparison purposes, we include in Fig. \ref{fig:Results_MBB_X} a solution with MinSolid, MinVoid and MaxSolid control, where the Minimum size of the Void phase is defined as $r_\mathrm{min,Void}^\mathrm{int}=6$ elements, i.e. twice as large as in the above examples.   

{\color{red} Interestingly, unlike the force inverter, the MBB beam constrained in maximum size becomes more complex when enlarging the minimum size of the cavities, since Fig. \ref{fig:Results_MBB_X} features more bars and closed cavities than Fig. \ref{fig:Results_MBB_b}. Therefore, for a modest increase of the Minimum size in the Void phase (MinVoid), it is not guaranteed that the number of closed cavities will decrease. Regarding the minimum gap constrained designs, it can be seen that the separation distance $h$ does not affect the radius of curvature at re-entrant corners. However, as with the force inverter, a large minimum gap $h$ is required to significantly reduce the amount of cavities.} 

In compliance minimization, bar thickening enhances structural performance. This is why in these problems the amount of unused material is less compared to the force inverter problem. Even if the design is stuck in a local optimum unable to incorporate new structural features, the present bars can be thickened to the size tolerated by the maximum size constraint. This can be seen in the solution of Fig. \ref{fig:Results_MBB_e}, where most of the structural members are of size $r_\mathrm{max}^\mathrm{int}$. In fact, in this solution there are regions whose size exceeds $r_\mathrm{max}^\mathrm{int}$, as the one shown by the arrow. This is a typical behaviour of the \textit{p-mean} aggregation function, which underestimates the most critical local constraint (see \citep{Fernandez2019} for details).   However, it can be noticed that in compliance minimization problems, maximum size and minimum gap constraints highly reduce the structural performance. This is reasonable as the geometrical constraints avoid the concentration of material at the lower and upper zone of the design domain, which reduces the moment of inertia and therefore the bending stiffness of the structure.

Two arrows in Fig. \ref{fig:Results_MBB_c} show structural features that suggest the solution is a local optimum. These pieces of material belonged to a bar that connected them in iterations preceding the ending of the optimization problem. These salient portions of material were not removed because the small move limit and the small number of remaining iterations were not sufficient to complete the removal. This illustrates a drawback of the adopted continuation method, which can be terminated without a fully converged solution. Similarly, the design in Fig. \ref{fig:Results_MBB_e} presents a curved bar at the bottom left. Its shape is not logical from a structural point of view as it reduces its axial stiffness. The curvature of this bar is due to the fact that in interactions previous to deactivating the set of constraints $G_\mathrm{ms}(\Psi)$, there was a bar that connected the horizontal bar at the bottom with the curved bar, as shown schematically by a dotted line in Fig. \ref{fig:Results_MBB_e}. The connecting bar had a high amount of intermediate densities, which is why it is not present in the final solution. These features that are distinctive of a local optimum can be avoided by changing the continuation strategy, for instance, by increasing the maximum number of iterations or by changing the updating strategy of $m_\mathrm{l}$ to be less conservative.

\subsection{3D-MBB beam}
\label{sec:53}

This problem is solved with a volume constraint of $40\%$. The chosen minimum and maximum length scales are $r_\mathrm{min,Solid}^\mathrm{int}=r_\mathrm{min,Void}^\mathrm{int}=3$ elements and $r_\mathrm{max}^\mathrm{int}=5$ elements. The reference solution is shown in the first row of Table \ref{tab:Results_3DV04}. To ease graphical interpretation of the design, 6 cross sections are shown next to the solution. The second row shows the solution obtained when the maximum size restriction is applied only in the intermediate design. Here, it can be seen that by restricting the intermediate design only, cavities smaller than $r_\mathrm{min,Void}^\mathrm{int}$ are obtained. This is consistent with the observation made for the 2D case in Section 3, where it is graphically shown that at least the dilated design must be restricted in order to obtain cavities of size $r_\mathrm{min,Void}^\mathrm{int}$. 

The third row in Table \ref{tab:Results_3DV04} shows the solution obtained with the maximum size constraint applied in all the three designs. From the solution it can be seen that the maximum size restriction considerably increases the complexity of the design. The restriction promotes the introduction of plates, which are placed parallel to the lower and upper part of the design space. These plates remain very close to each other to increase the area moment of inertia and thus the bending stiffness.

\begin{table}[htpb]
	\caption{Quarter 3D-MBB beam with volume constraint of $40\%$.} 
	\centering
		\begin{tabular}{c c c}
				 \toprule
				 & $\bm{\bar{\rho}}^{int}$ & Cross Sections  \\
				 \cmidrule(r){2-2} \cmidrule(r){3-3}
				 \multirow{9}{*}{\begin{minipage}[t]{0.18\textwidth}
				    \endgraf
				    \center{Reference}
				    \endgraf \vspace{-2mm}
					\center{$O_\mathrm{bj}=R_\mathrm{ef}$.}
        			\end{minipage}}
				 & 
				 \multirow{9}{*}{\includegraphics[width=0.33\linewidth]{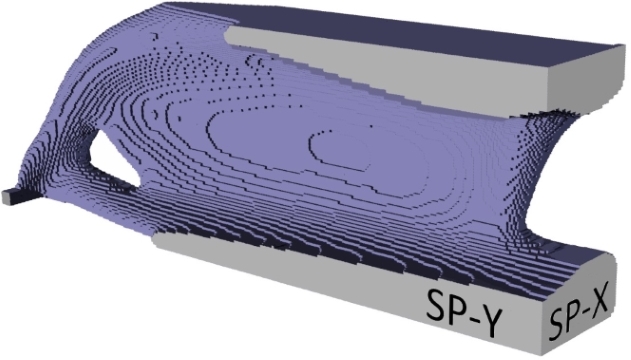} }
				 & 
				 \multirow{9}{*}[0 mm]{\includegraphics[width=0.33\linewidth]{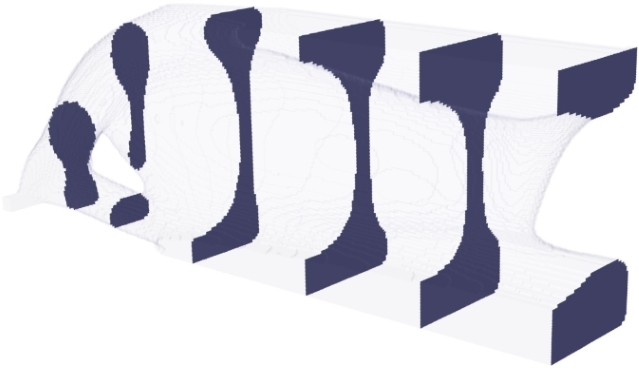} }
				 \vspace{-1mm}\\ \\ \\ \\ \\ \\ \\ \\ 
				 \multirow{9}{*}{\begin{minipage}[t]{0.14\textwidth}
				    \endgraf
				    \center{Only $G_{\mathrm{ms}}(\Omega^\mathrm{int})$}
					\endgraf \vspace{-2mm}
          			\center{$O_{\mathrm{bj}}/R_\mathrm{ef}={1.182}$}
        			\end{minipage}}
				 & 
				 \multirow{9}{*}[0 mm]{\includegraphics[width=0.33\linewidth]{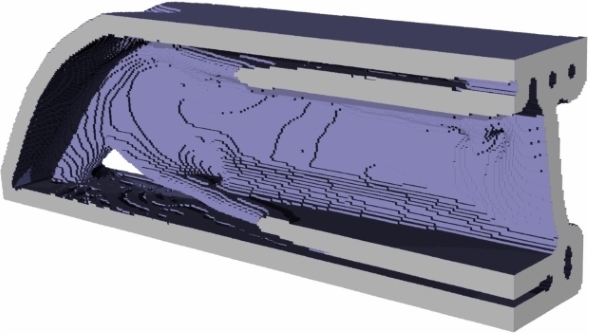}}
				 &
				 \multirow{9}{*}[0 mm]{\includegraphics[width=0.33\linewidth]{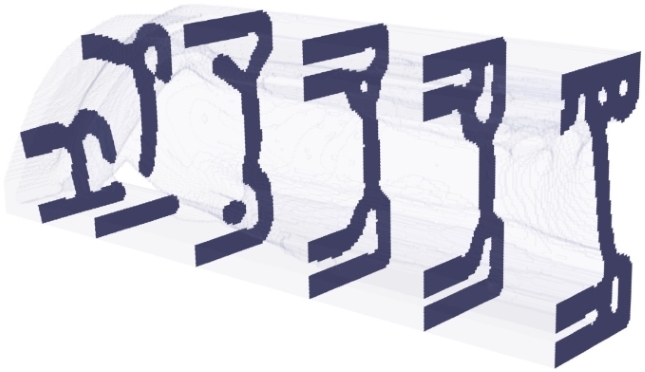}}
				 \vspace{-1mm} \\ \\ \\ \\ \\ \\ \\ \\ 
				 \multirow{9}{*}{\begin{minipage}[t]{0.14\textwidth}
				    \endgraf
				    \center{Robust $G_{\mathrm{ms}}(\Omega)$}
					\endgraf \vspace{-2mm}
					\center{$O_{\mathrm{bj}}/R_\mathrm{ef}={1.232}$}
        			\end{minipage}}
				 & 
				 \multirow{9}{*}[0 mm]{\includegraphics[width=0.33\linewidth]{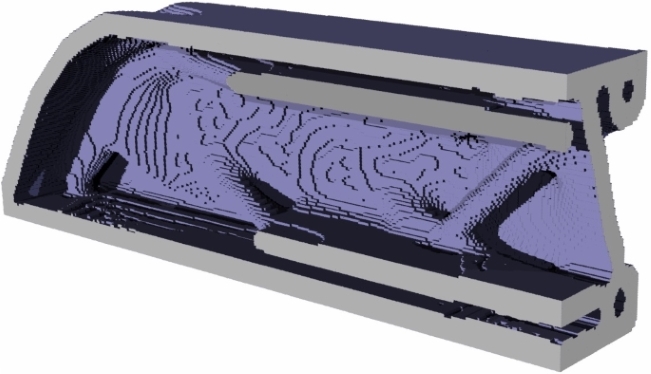}}
				 &
				 \multirow{9}{*}[0 mm]{\includegraphics[width=0.33\linewidth]{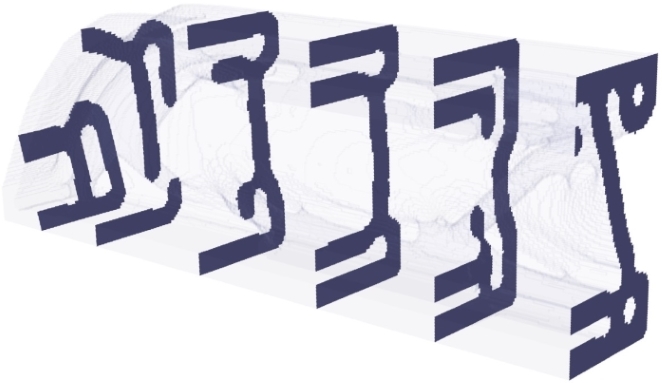}}
				 \vspace{-1mm} \\ \\ \\ \\ \\ \\ \\ \\ 
				 \multirow{9}{*}{\begin{minipage}[t]{0.14\textwidth}
				 	\endgraf
				    \center{Robust $G_{\mathrm{ms}}(\Omega)$}
				    \endgraf \vspace{-2mm}
				    \center{MinVoid $\times$ 2}
					\endgraf \vspace{-2mm}
          			\center{$O_{\mathrm{bj}}/R_\mathrm{ef}=1.321$}
        			\end{minipage}}
				 & 
				 \multirow{9}{*}[0 mm]{\includegraphics[width=0.33\linewidth]{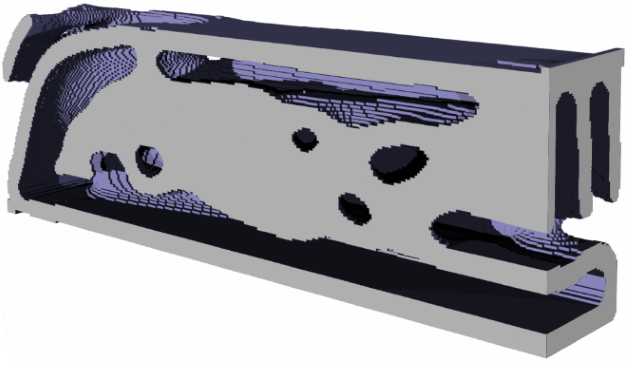}}
				 &
				 \multirow{9}{*}[0 mm]{\includegraphics[width=0.33\linewidth]{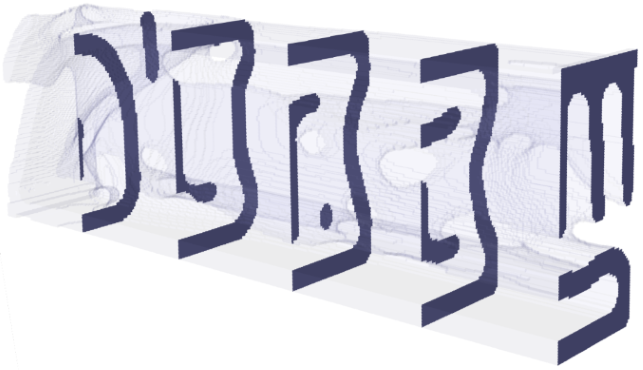}}
				 \vspace{-1mm} \\ \\ \\ \\ \\ \\ \\ \\ 
				 \multirow{9}{*}{\begin{minipage}[t]{0.14\textwidth}
				    \endgraf
				    \center{Robust $G_{\mathrm{ms}}(\Omega)$}
				    \endgraf \vspace{-2mm}
				    \center{Robust $G_{\mathrm{ms}}(\Psi)$}
				    \endgraf \vspace{-2mm}
				    \center{$h^\mathrm{int}=2r_\mathrm{max}^\mathrm{int}$}
					\endgraf \vspace{-2mm}
          			\center{$O_{\mathrm{bj}}/R_\mathrm{ef}=1.239$}
        			\end{minipage}}
				 & 
				 \multirow{9}{*}[0 mm]{\includegraphics[width=0.33\linewidth]{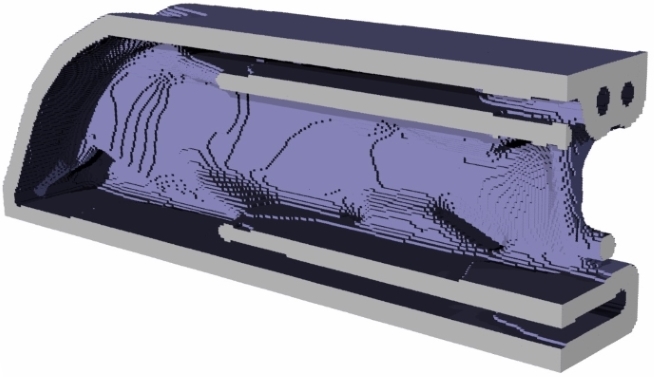}}
				 &
				 \multirow{9}{*}[0 mm]{\includegraphics[width=0.33\linewidth]{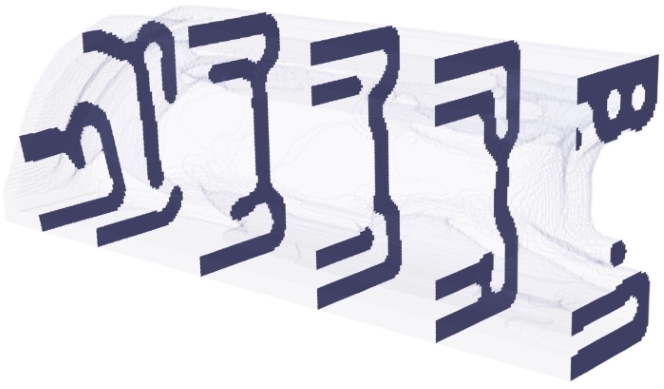}}
				 \vspace{-1mm} \\ \\ \\ \\ \\ \\ \\ \\
				 \multirow{9}{*}{\begin{minipage}[t]{0.14\textwidth}
				    \endgraf
				    \center{Robust $G_{\mathrm{ms}}(\Omega)$}
				    \endgraf \vspace{-2mm}
				    \center{Robust $G_{\mathrm{ms}}(\Psi)$}
				    \endgraf \vspace{-2mm}
				    \center{$h^\mathrm{int}=4r_\mathrm{max}^\mathrm{int}$}
					\endgraf \vspace{-2mm}
					\center{$O_{\mathrm{bj}}/R_\mathrm{ef}=1.313$}
        			\end{minipage}}
				 & 
				 \multirow{9}{*}[0 mm]{\includegraphics[width=0.33\linewidth]{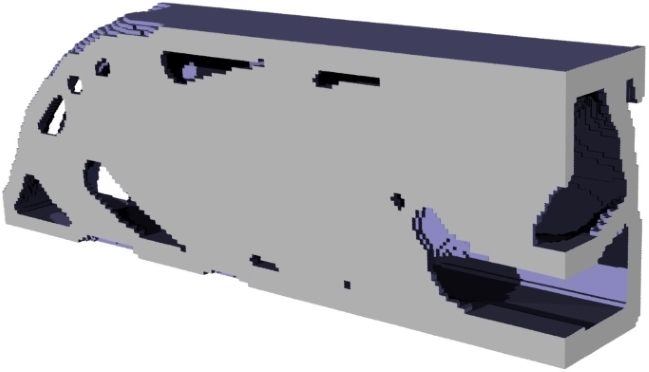}}
				 &
				 \multirow{9}{*}[0 mm]{\includegraphics[width=0.33\linewidth]{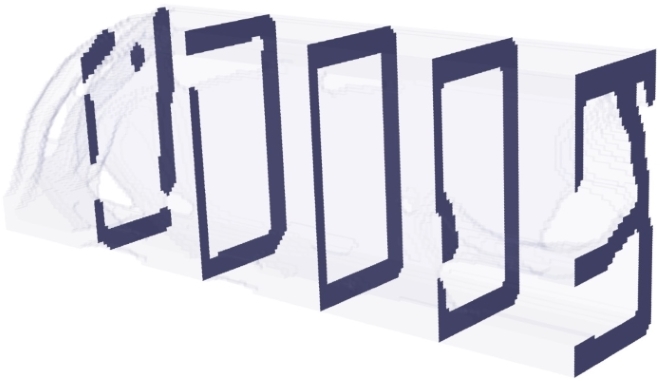}}
				 \\ \\ \\ \\ \\ \\ \\ \vspace{2mm} \\ 
				 \bottomrule
		\end{tabular}
	\label{tab:Results_3DV04}
\end{table}

{\color{red}
Aiming to reduce the complexity of the maximum size constrained design, the Minimum size in the Void phase (MinVoid) is doubled in the following example, i.e. $\PFS{r}{min,Void}{int} = 6$ elements (using the thresholds [0.75, 0.65, 0.25]). The new maximum size constrained design is shown in the fourth row of Table \ref{tab:Results_3DV04}. As expected, this solution features larger cavities and a larger radius of curvature at re-entrant corners. Interestingly, by enlarging the minimum size in the void phase (MinVoid), solid parts are placed at the upper left corner of the design. This place is commonly avoided by the optimizer as it is a zone of low strain energy. The material placed in this zone suggests that the MinVoid is too large to allow the placement of the available material at the interior of the beam, or that the solution is stuck in a local optimum.
}

\begin{figure}[t!] 
\captionsetup[subfigure]{labelformat=empty}
\centering   
    \begin{subfigure}[b]{0.44\linewidth}
		\centering{}\includegraphics[width=1\linewidth]{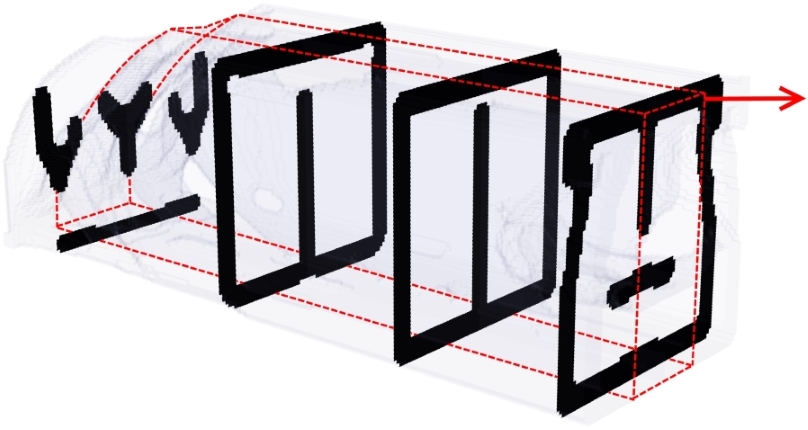}		
		\caption{}
		\label{fig:Results_3DV04_Gap_a}
	\end{subfigure}
	~
	\begin{subfigure}[b]{0.35\linewidth}
		\centering{}\includegraphics[width=1\linewidth]{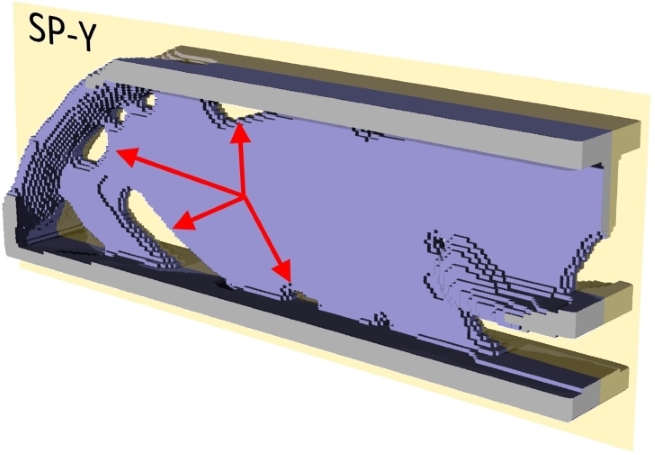}
		\caption{}
		\label{fig:Results_3DV04_Gap_b}
	\end{subfigure}
	\vspace{-8mm}\\
	(a) \hspace{6cm} (b) \\
	\vspace{-1mm}
\caption{Half 3D-MBB beam design from Table \ref{tab:Results_3DV04}, fifth row. In (a) the mirrored solution with respect to SP-Y. In (b) the central portion shows a thin 3D-MBB beam. The arrows point cavities smaller than the defined $h$.}
\label{fig:Results_3DV04_Gap}
\end{figure}

The fifth row of Table \ref{tab:Results_3DV04} shows a minimum gap constrained solution obtained with $h^\mathrm{int}=2r_\mathrm{max}^\mathrm{int}$. It can be seen that the separation distance between plates is bigger than in the maximum size constrained design (row 3). However, the complexity of the design is not significantly reduced. As noted in the previous problems, large values of $h$ must be used to get a major effect on the topology. The drawback is that increasing $h$ enlarges the $\Psi$ regions and reduces the sparsity of the $\mathrm{\mathbf{D}}_{\mathrm{II}}\left( \Psi \right)$ matrices, which inevitable increases memory requirements. As regard to this issue and considering the chosen discretization of 1.3 million elements, it was not possible to create with our hardware such matrices when $h>2r_\mathrm{max}$. Therefore, to solve the optimization problem with $h^\mathrm{int}=4r_\mathrm{max}^\mathrm{int}$, the discretization is reduced to $0.4$ million elements. The solution is shown in the sixth row of Table \ref{tab:Results_3DV04}. To facilitate the interpretation of this solution, different cross sections of the optimized design including the mirrored part with respect to SP-Y are shown in Fig. \ref{fig:Results_3DV04_Gap_a}. It can be noticed that the topology is remarkably different from the preceding solutions. The material tends to be placed at the boundary of the design space and no small cavities between parallel plates are observed. Thus, our end design seems less complex to manufacture than the previous solutions.

\begin{table}[b!]
	\caption{Quarter 3D-MBB beam with volume constraint of $20\%$.} 
	\centering
		\begin{tabular}{c c c}
				 \toprule
				  & $\bm{\bar{\rho}}^{\mathrm{int}}$ & Cross Sections  \\
				 \cmidrule(r){2-2} \cmidrule(r){3-3}
				 \multirow{9}{*}{\begin{minipage}[t]{0.15\textwidth}
				 	\endgraf
					\center{Reference}
				    \endgraf
					\center{$O_{\mathrm{bj}}=R_\mathrm{ef}$}
        			\end{minipage}}
				 & 
				 \multirow{9}{*}{\includegraphics[width=0.365\linewidth]{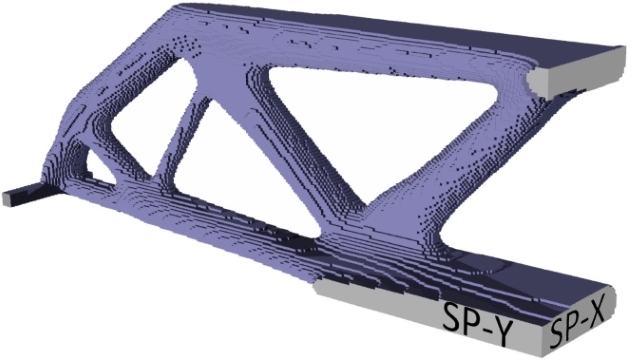} }
				 & 
				 \multirow{9}{*}[0 mm]{\includegraphics[width=0.365\linewidth]{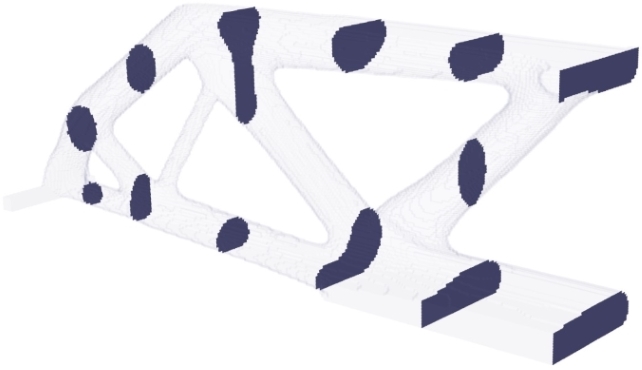} }
				 \vspace{2mm}\\ \\ \\ \\ \\ \\ \\ \\ 
				 \multirow{9}{*}{\begin{minipage}[t]{0.17\textwidth}
					\endgraf				    
				    \center{Robust $G_{\mathrm{ms}}(\Omega)$}
				    \endgraf \vspace{-2mm}
					\center{$O_{\mathrm{bj}}/R_\mathrm{ef}={1.176}$}
        			\end{minipage}}
				 & 
				 \multirow{9}{*}[0 mm]{\includegraphics[width=0.365\linewidth]{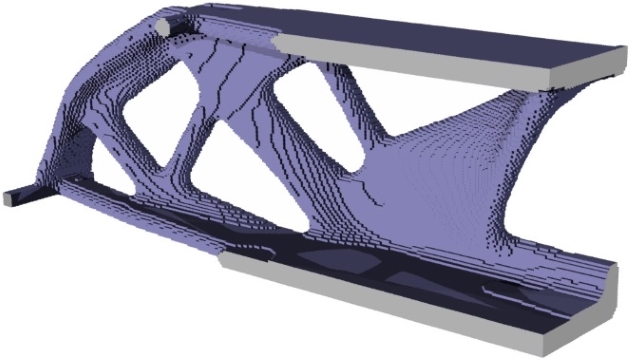}}
				 &
				 \multirow{9}{*}[0 mm]{\includegraphics[width=0.365\linewidth]{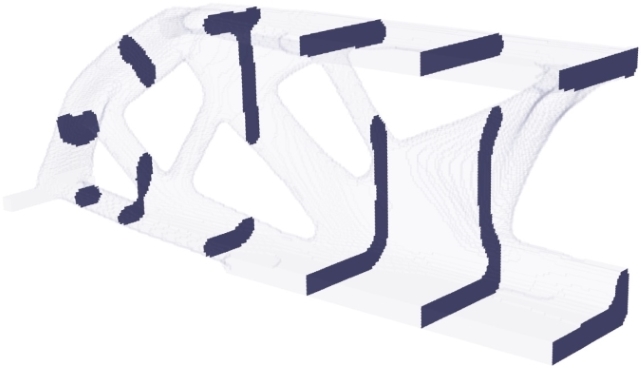}}
				 \\ \\ \\ \\ \\ \\ \\ \vspace{2mm} \\ 
				 \bottomrule
		\end{tabular}
	\label{tab:Results_3DV02}
\end{table}

It is interesting to note that the reduction in structural performance, associated to maximum and minimum size constraints, is less in the 3D-MBB beam  than in the 2D-MBB beam. Indeed, the third dimension allows the material to be rearranged more efficiently for the sake of the area moment of inertia. A good example of this fact can be seen in the 3D-MBB beam with $h^\mathrm{int}=4r_\mathrm{max}^\mathrm{int}$. There, the third dimension allows the placement of a thin beam along the SP-Y, similar to the principle of spaced trusses. The thin beam is shown separately in Fig. \ref{fig:Results_3DV04_Gap_b} since it is hidden inside the structure when the beam is reflected with respect to SP-Y. In the same figure, some cavities are pointed out which have a small size in comparison to the chosen $h^\mathrm{int}$. This is due to two drawbacks of the proposed strategy. First, the geometric relationship between $h$ and $\varepsilon$ is obtained for two parallel plates, therefore under other conditions the distance $h$ will not be controlled. Second, the minimum gap constraints are deactivated before the optimization problem is finished, therefore the precise control of the minimum gap distance cannot be guaranteed.

It is worth mentioning that the maximum size restriction does not necessarily increase the complexity of the design. Hence, depending on the chosen length scale and the volume constraint, the minimum gap constraint may not always be needed. For instance, Table \ref{tab:Results_3DV02} shows the reference and maximum size constrained solutions considering the length scales of the previous example, but using a volume constraint of $20\%$. It can be seen that the maximum size constrained design does not contain small cavities, making it just as complex as the reference design.

\begin{figure}[t!] 
\centering   
\vspace{3mm}
    \begin{subfigure}[b]{0.75\linewidth}
		\centering{}\includegraphics[width=1\linewidth]{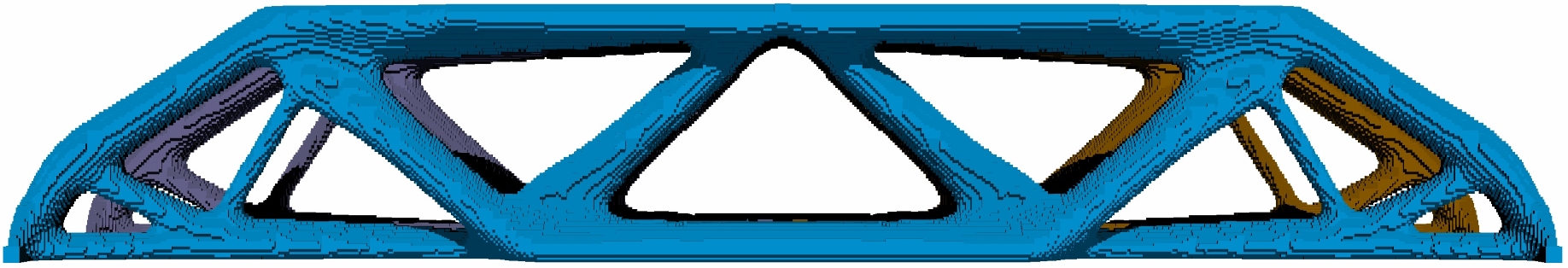}
	\end{subfigure}
	\hspace{3mm} ~
	\begin{subfigure}[b]{0.17\linewidth}
		\centering{}\includegraphics[width=1\linewidth]{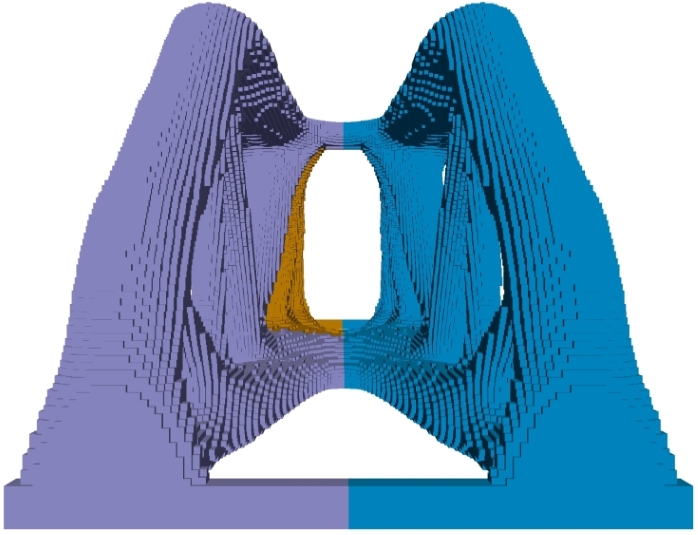}
	\end{subfigure}
	\\ \vspace{2mm}
	\hspace{-30mm}
	\begin{subfigure}[b]{0.7\linewidth}
		\centering{}\includegraphics[width=1\linewidth]{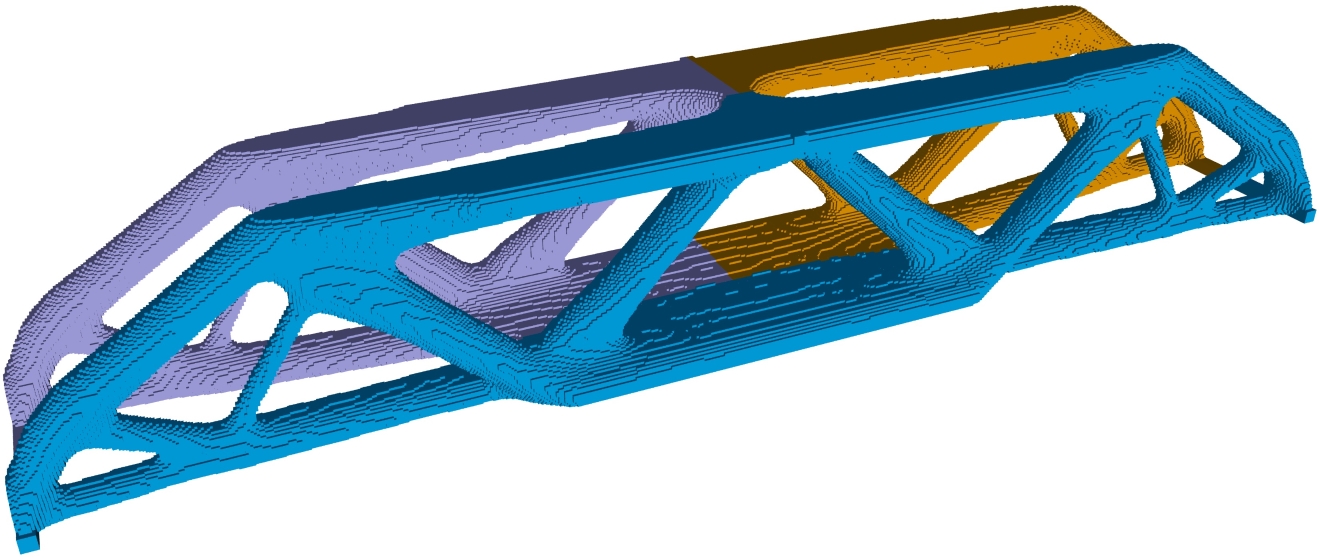}
	\end{subfigure}
\caption{Front, side and perspective view of the full 3D-MBB beam. Structure obtained after two reflections of the reference solution shown in Table \ref{tab:Results_3DV02}.}
\label{fig:REF_PLANS}
\end{figure}

We close this section by mentioning that results shown in this article are nearly discrete. These, on average, have a gray level indicator of $3\%$ (indicator defined by \citet{Sigmund2007}). In addition, as can be seen in the reported solutions, the length scale is satisfied at the boundaries of the design space by means of the numerical treatment applied on the filtering process and on the maximum size constraint. For instance, the 3D solutions satisfy the imposed length scale even after applying the 2 reflections (with respect to SP-X and SP-Y). In order to not overextend the document, only the reference solution shown in Table \ref{tab:Results_3DV02} is used to recover the full 3D-MBB beam. The full optimized design is shown in Fig. \ref{fig:REF_PLANS}. The colors on the structure indicate the reflection process. 

\section{Conclusions}
\label{sec:6}

This paper focuses on density-based topology optimization and proposes a combined strategy to simultaneously impose the Minimum size in the Solid phase (MinSolid), the Minimum size in the Void phase (MinVoid), the Maximum size in the Solid phase (MaxSolid) and the minimum separation distance between solid parts (MinGap). MinSolid and MinVoid are imposed through the robust design approach based on the eroded, intermediate and dilated designs. MaxSolid and MinGap are imposed through local volume restrictions applied on the eroded, intermediate and dilated designs. The strategies are evaluated on 2D and 3D test cases, which allow us to conclude the following:

\begin{itemize}

\item[\textendash] The maximum size constraint and the robust design approach allow effective length scale control (MinSolid, MinVoid and MaxSolid). For this, the maximum size restriction must be scaled with respect to the design in which it is applied (eroded, intermediate or dilated), and at least the dilated design must be restricted. However, this strategy may require a continuation method with large number of iterations to avoid topological features typical of a local optimum, such as curved bars or small hinges. 

\item[\textendash] For low volume restrictions, the maximum size constraint does not necessarily increase the complexity of the design, since the introduction of closed cavities is unlikely to happen. When a large amount of material is allowed in the design space, the MaxSolid constraint considerably increases the number of small cavities in the design. In such cases, the complexity of the design can be reduced by increasing the minimum size of the cavities through combination of parameters of the robust design approach.

\item[\textendash] Since the Minimum size in the Void phase (MinVoid) imposed by the robust design approach defines the radius of curvature at re-entrant corners, an excessively large MinVoid can affect structural performance as the bars composing the structure lose straightness and stiffness.

\item[\textendash] The local volume restrictions can also be used to impose a minimum distance between two solid members, or a Minimum Gap (MinGap). This formulation allows, indirectly, to reduce the number of cavities in the design and increase the size of those remaining, notably, without affecting the radius of curvature at re-entrant corners. Therefore, simultaneous control of MinSolid, MinVoid, MaxSolid and MinGap, can be useful to facilitate the manufacturability of maximum size constrained designs.

\item[\textendash] In a regular mesh, it is possible to numerically include the treatment of the filter at the boundaries of the design space without using a real extension of the finite element mesh. Similarly, the maximum size constraint can also be adapted to simulate an extension of the mesh outside the design space.   
\end{itemize}

\section*{Acknowledgement}
This work was supported by the project AERO+, funded by the Plan Marshall 4.0 and the Walloon Region of Belgium. Computational resources have been provided by the \textit{Consortium des \'{E}quipements de Calcul Intensif} (C\'{E}CI), funded by the Scientific Research Fund of Belgium (F.R.S.-FNRS).

\bibliographystyle{elsarticle-num-names} 
\bibliography{elsarticle-template-1-num} 

\end{document}